\documentclass[fleqn,usenatbib]{mnras}

\usepackage{newtxtext,newtxmath}
\usepackage[T1]{fontenc}
\usepackage{ae,aecompl}

\usepackage{graphicx}	% Including figure files
\usepackage{amsmath}	% Advanced maths commands
\usepackage{amssymb}	% Extra maths symbols
\usepackage{xcolor}

\usepackage[utf8]{inputenc}  			% pour lire les accents en francais	
\usepackage[T1]{fontenc}				% pour lire les accents en francais

\usepackage{adjustbox}

\usepackage{pdflscape}
\usepackage{caption}

\DeclareCaptionFormat{cont}{#1 (cont.)#2#3\par}
\usepackage[xcolor={divpdf,grey},authormarkup=none]{changes} % À enlever dans la version à soumettre
\definechangesauthor[name={RF},color=red]{R}

\newcommand{\I}{\,{\sevensize I}}
\newcommand{\II}{\,{\sevensize II}}
\newcommand{\III}{\,{\sevensize III}}

\title[Optical Identification and Confirmation of Supernova Remnants in NGC\,3344]
{3D Optical Spectroscopic Study of NGC\,3344 with SITELLE: I.~Identification and Confirmation of Supernova Remnants}
\author[I. Moumen et al.]{I. Moumen$^{1,2}$\thanks{E-mail: ismael@cfht.hawaii.edu (IM)},
C. Robert$^{1}$, D. Devost$^{2}$, R. P. Martin$^{3}$, L. Rousseau-Nepton$^{2,3}$,
\newauthor
L. Drissen$^{1}$, and T. Martin$^{1}$
\\
$^{1}$Département de physique, de génie physique et d’optique, Université Laval, and \\ 
Centre de Recherche en Astrophysique du Québec (CRAQ), Québec, QC, G1V 0A6, Canada\\
$^{2}$Canada-France-Hawaii Telescope, Kamuela, HI, 96743, USA\\
$^{3}$Department of Physics and Astronomy, University of Hawaii at Hilo, Hilo, HI, 96720, USA
}

\date{Accepted XXX. Received YYY; in original form ZZZ}

\pubyear{2018}

\begin{document}
\label{firstpage}
\pagerange{\pageref{firstpage}--\pageref{lastpage}}
\maketitle

\begin{abstract}
We present the first optical identification and confirmation of a sample of supernova remnants (SNRs) in the nearby galaxy NGC\,3344. Using high spectral and spatial resolution data, obtained with the CFHT imaging Fourier transform spectrograph SITELLE, we identified about 2200~emission line regions, many of which are H\II\ regions, diffuse ionized gas regions, and also SNRs. Considering the stellar population and diffuse ionized gas background, that are quite important in NGC\,3344, we have selected 129~SNR candidates based on four criteria for regions where the emission lines flux ratio [S\II]/H$\alpha$\,$\ge$\,0.4. Emission lines of [O\II]$\lambda$3727, H$\beta$, [O\III]$\lambda\lambda$4959,5007, H$\alpha$, [N\II]$\lambda\lambda$6548,6583, and [S\II]$\lambda\lambda$6716,6731 have been measured to study the ionized gas properties of the SNR candidates. We adopted a self-consistent spectroscopic analysis, based on Sabbadin plots and BPT diagrams, to confirm the shock-heated nature of the ionization mechanism in the candidates sample. With this analysis, we end up with 42~Confirmed SNRs, 45~Probable SNRs, and 42~Less likely SNRs. Using shock models, the Confirmed SNRs seems to have a metallicity ranging between LMC and 2$\times$solar.  
We looked for correlations between the size of the Confirmed SNRs and their emission lines ratios, their galaxy environment, and their galactocentric distance: we see a trend for a metallicity gradient among the SNR population, along with some evolutionary effects.
\\

\end{abstract}


%%%%%%%%%%%%%%%%%%%%%%%%%%%%%%%%%%%%%%%%%%%%%%%%%%
%%%%%%%%%%%%%%%%%%%%%%%%%%%%%%%%%%%%%%%%%%%%%%%%%%
\section{Introduction}
%%%%%%%%%%%%%%%%%%%%%%%%%%%%%%%%%%%%%%%%%%%%%%%%%%

Supernova remnants (SNRs) are key objects to study the chemical enrichment of the interstellar medium (ISM) and stellar evolution, while SNRs produced by core-collapse supernovae are tightly linked to star-formation sites in galaxies. Supernova explosions represent the tragic end of the life of many types of stars, while they interact with the ISM by injecting enriched material and energy into their environment, often being at the origin of superwinds expanding far away from the galaxies \citep{doi:10.1146/annurev.aa.15.090177.001135}. Depending on the shock velocity and the density of the surrounding ISM, the high-energy shock waves resulting from SN explosions are proposed to either trigger or repress star formation (e.g. \citep{2009MNRAS.399.2183N, 2006MNRAS.373..811M, 2013NatSR....E1738B}.

The strength of their radiation and the energy dissipated by the shocks are such that a vast bubble of ionized gas is generated and the expansion shell is particularly bright. For instance, they are a non-negligeable contributor to the observed H$\alpha$ emission line in galaxies, which can be easily mixed with the emission from H\II\ regions often used in more distant objects to measure the star formation rate (SFR). Nevertheless, the exact contribution from SNRs to the H$\alpha$ flux is not well known and could vary from one galaxy to another, 
e.g. \cite{2015MNRAS.446..943V} found a flux contribution of 5 $\pm$ 5 \%. 

The morphology of SNRs depends on different factors like their evolutionary phase, the input energy released by the explosion, the mass loss history of the progenitor, and the density of the surrounding ISM. The SNR evolution can be summarized by a four-stage model \citep{1972ARA&A..10..129W}. During the first stage, called the free expansion phase, the supernova ejecta sweeps up ISM material as it expands freely until its mass becomes equal to the mass of the swept up material. Theoretically, this phase can last between 90 and 300 years and is characterized by constant temperature and a constant expansion velocity of the shell. The second evolutionary stage, called the adiabatic phase or Sedov-Taylor phase, is described by the Sedov-Taylor self-similar solution \citep{1959sdmm.book.....S, 1950RSPSA.201..159T}. In this phase, the quantity of the material swept up from the circumstellar and interstellar medium becomes so important that the expanding shell starts to slow down and cool. In general, the estimated duration of this phase is between 100 and 10\,000 years but it can also be missed completely in some cases. However, if the expansion is happening in a hot interstellar medium, it will last longer and the SNR diameter may reach 180\,pc \citep{1977ApJ...218..148M}. During the third stage, called the radiative phase or the snowplow phase, the mass of the swept up material is dramatically increased, which forces the velocity of the shock front to decrease down to $\sim$200\,km\,s$^{-1}$. The temperature behind the shock front drops to $\sim$10$^{5}$\,K and the energy lost due to recombination becomes significant, creating a cooling region behind the shock front and producing shock-heated collisionally ionized species (such as [S\II] and [O\III]) along with hydrogen recombination lines. In this phase, the SNR diameter expands from about 14\,pc to 50\,pc in a typical ISM with a density of 1.0\,cm$^{-3}$ \citep{1990ASIC..305....1C}. During this phase, which can last up to 20\,000 years, the SNR becomes visible in the optical band. The last evolutionary stage is a dissipative process during which the velocity of the shock reaches the sound speed of the ambient ISM and the SNR is dispersed. The size of the SNR during this phase is from 50\,pc to over 100\,pc. 
In the optical,  the remnant diameter can vary from a few parsecs up to more than a 100\,pc \citep{2009SerAJ.179...55C}. With larger diameters, superbubbles are another type of emission region with shocks related emission lines, which may involve massive stars and/or SN explosions. 
Superbubbles have typical velocity shock below 100\,km\,s$^{-1}$ \citep{2015yCat..74460943V}.

Because of their proximity, SNRs observed in the Milky Way allow a detailed study of their physical properties and interactions with the surrounding ISM. According to the latest version of the Galactic SNR catalogs \citep{2017yCat.7278....0G,2012AdSpR..49.1313F}\footnote{http://www.physics.umanitoba.ca/snr/SNRcat}, our galaxy is hosting at least 294~remnants. But these catalogs remain limited because of the high absorption in the Galactic plane and the large uncertainty associated with the distance measurement. On the other hand, these constraints are often less important in the case of extragalactic SNRs in nearly face-on galaxies.

Extragalactic SNRs are very important to understand the physics of the SNRs, as individual objects and as a component of the ISM and galaxies. First, a distance measurement (and SNR flux and size) is less of a burden as it may be given by other indicators for the galaxy.  Second, the advancement of  astronomical instruments allows us to observe simultaneously complete samples of SNRs which help to take  into account observational selection effects. Historically, the first extragalactic SNRs were identified in the Magellanic Clouds using radio and optical observations \citep{1963Natur.199..681M, 1973ApJ...180..725M}. Today, 25~nearby galaxies are known to host almost 1200~extragalactic SNRs \citep{2015MNRAS.446..943V}. It is interesting to highlight that with the exception of SNRs in the SMC and LMC, almost all the extragalactic SNRs were first identified in the optical. The number of observed SNRs in a galaxy, from the optical, varies from a small number to a few hundred. 25~SNRs have been identified in the SMC \citep{2012A&A...545A.128H}, 26 in NGC\,6946 \citep{1997ApJS..112...49M}, 53 in the LMC \citep{2016A&A...585A.162M}, 93 in M101 \citep{2012AJ....143...85F}, 217 in M33 \citep{2010ApJS..187..495L, 2018ApJ...855..140L}, and 296 in M83 \citep{2014ApJ...788...55B}. \cite{2013MNRAS.429..189L} detected more than 400 SNRs in six nearby galaxies (NGC\,2403, NGC\,3077, NGC\,4214, NGC\,4395, NGC\,4449, and NGC\,5204) based on optical photometric and spectroscopic observations.

The identification and confirmation of extragalactic SNRs are mainly done using data from the radio \citep{1997ApJS..109..417L, 2001ApJ...560..719L, 2001ApJ...551..702H}, visible \citep{2010A&A...517A..91S, 2009A&A...493.1061S, 2004ApJS..155..101B, 1998ApJS..117...89G, 1997ApJS..113..333M, 1997ApJS..112...49M}, and X-Ray \citep{2005AJ....130..539G, 2001ApJ...561..189P} wavelength ranges. Only a few extragalactic SNRs have been identified in the infrared using the [Fe\II]\,1.64\,$\micron$ emission line \citep{1997ApJ...476..105G}. 
The radiation of SNRs in different wavelength ranges is under the influence of biases as they cover different aspects of the ISM environment and SNR age and evolution \citep{2000ApJ...544..780P, 2010ApJ...725..842L, 2012Ap&SS.337..573S, 2017hsn..book.2005L}. 
In the radio, only SNR candidates associated with H\II\ regions are identified, which means that radio SNR samples are biased towards star-forming regions. 
In the X-Ray, candidates are selected if they display a soft spectrum and if they are associated with an H\II\ region, which means that X-Ray samples have the same bias than the radio samples and are also biased against SNR candidates with hard spectra and no optical counterparts. In the optical, samples are under the influence of biases privileging the identification of SNRs located in low density environments. A multi-wavelength (X-Ray, optical, and radio) study of SNRs in NGC\,300 by \cite{2000ApJ...544..780P} revealed 16 new SNRs, 2 in the radio and 14 in the X-Ray, in addition to the 28 SNRs previously identified in the optical. The lack of new optical detection is explained by the fact that optical SNRs can only be detected when they represent relatively low confusion with other H$\alpha$ emission sources. The optical SNRs found here are  generally located well away from star-forming regions. Consequently, SNR samples identified optically are often not complete. 
Another technique, based on the search for Large-Velocity-Width Sources (LVWS), is used to identify SNRs with optical spectroscopy \citep{1986ApJ...311...85C, 1990ApJ...365..164C}. While H\II\ regions show low-velocity dispersion 
\citep[$\sigma_{v} \le~30$~km\,s$^{-1}$;][]{1977ApJ...213...15M, 1983ApJ...274..141G}, 
broad emission line widths observed in the LVWS can be caused  by stellar winds or supernova. Using this technique, \citet{1986ApJ...311...85C}  discovered four LVWS inside the Giant H\II\ regions NGC\,5471 A,B,C and NGC\,5461 in the nearby galaxy M101.

In the visible, a criterion often used for the identification of extragalactic SNRs is based on a value of the emission lines ratio [S\II]$\lambda\lambda$6716,6731/H$\alpha$\,$\geq$\,0.4 \citep{1973ApJ...182..697M}. 
This criterion was used, for example, to produce the catalogue of \cite{2015yCat..74460943V}. 
It allows differentiation between photoionized nebulae, like H\II\ regions, and SNRs shock-heated nebulae \citep{2004ApJS..155..101B}.  Physically, in a typical  H\II\ region,  where the photoionization of the gas by hot stars is dominating, the second ionization state of the Sulfur, S++,  is favored over S+, and the expected theoretical ratio [S\II]/H$\alpha$ is between 0.1 and 0.3 \citep{2009A&A...493.1061S}. In an SNR, shock waves are propagating into the ISM after the supernova explosion. The material cools behind these shock waves and it increases the quantity of S+. Consequently, the ratio [S\II]/H$\alpha$ becomes higher than 0.3. Regions of diffuse ionized gas (DIG), as seen in many galaxies \citep{1999ApJ...523..223H, 2009ApJ...704..842B} also presents a relatively strong [S\II]/H$\alpha$ ratio (as well as a strong [N\II]/H$\alpha$ ratio; \citep{1994ApJ...431..156W}; \citep{2005ApJ...632..277E}. These are regions with temperatures similar to H\II\ regions ($\sim$10$^4$\,K), lower densities (0.1\,cm$^{-3}$), and slightly lower ionization states. The origin of the DIG is not well understood yet and seems to be complex; it is often suspected of being caused by ionizing photons escaping H\II\ regions and traveling large distances, but it can also be related to a generation of post-AGB stars, fast shocks, a weak AGN, and dust diffusion \citep[e.g. ][]{2011MNRAS.415.2182F, 2012A&A...544A..57H, 2012ApJ...758..109S, 2014MNRAS.444.3961D, 2017ApJ...834...40H}

Physical and chemical properties of SNRs can also be determined using emission line diagnostics in the visible; for instance, one can get the electron density with [S\II]$\lambda$6716/[S\II]$\lambda$6731, the shock velocity with [O\III]$\lambda$5007/H$\beta$, and the chemical abundances with [N\II]$\lambda\lambda$6548,6583/H$\alpha$ \citep[e.g.][]{2017arXiv170107840L}. Some emission lines ratios have been found to display various correlations which may be useful to understand the SNR evolution and environment. For example, a study of the SNRs in M33 by \citet{1998ApJS..117...89G} shows a galactocentric gradient for [N\II]$\lambda\lambda$6548,6583/H$\alpha$, no relation between [N\II]$\lambda\lambda$6548,6583/H$\alpha$ and [S\II]$\lambda$6716/[S\II]$\lambda$6731, a weak correlation between the SNRs diameter and  
[N\II]$\lambda\lambda$6548,6583/H$\alpha$ or  
[S\II]$\lambda\lambda$6716,6731/H$\alpha$, a strong correlation between [N\II]$\lambda\lambda$6548,6583/H$\alpha$ and  
[S\II]$\lambda\lambda$6716,6731/H$\alpha$, and also between [O\I]$\lambda$6300/H$\alpha$ and [S\II]$\lambda\lambda$6716,6731/H$\alpha$. 
More correlations have been seen in the case of SNRs in M31;  \citet{1981ApJ...247..879B, 1982ApJ...254...50B} found a clear galactocentric gradient for the ratio [N\II]$\lambda\lambda$6548,6583/H$\alpha$, a little evidence of a gradient for  [S\II]$\lambda\lambda$6716,6731/H$\alpha$, [O\III]$\lambda\lambda$4959,5007/H$\beta$, and 
[O\III]$\lambda\lambda$4959,5007/[O\II]$\lambda$3727.

NGC\,3344 is an isolated galaxy \citep{1973AISAO...8....3K}, located at 6.1\,Mpc \citep{2013AJ....146...86T}, and classified as a (R)SAB(r)bc by \cite{1991rc3..book.....D}. 
Seen nearly face-on, NGC\,3344 presents a bright inner ring with multiple spiral arms, and numerous H\II\ regions. A weak bar can be seen inside the ring. 
Its optical diameter is about 8$^{\prime}$, while 
its H\I\ diameter is 18$^{\prime}$
\citep{1983AJ.....88..272H}. Recently, a Type Ibc supernova (SN 2012fh) was discovered at RA = 10:43:34.05 and DEC = +24:53:29.0 \citep{2012CBET.3263....1N}.

As part of a SITELLE study of the ionized gas emission in NGC\,3344, this first paper focuses on the identification and confirmation of the galaxy's SNR population. The paper is organized as follows: Section~2 describes the observation of NGC\,3344 using SITELLE while section~3 presents the data reduction process and the technique used to measure the emission lines. In Section~4, we describe the method applied to identify the emission line regions, the method developed to subtract the galaxy stellar populations contribution along with the DIG component, the equations used to calculate the internal extinction, and the selection criteria for the SNR candidates. 
Section~5 presents our method, based on the emission lines ratios along with Sabbadin plots \citep{1977A&A....60..147S} and Baldwin, Phillips \& Terlevich (BPT) diagrams \citep{1981PASP...93....5B}, used to classify the candidates into three categories, including the Confirmed SNRs. The physical proprieties of the Confirmed SNR are discussed in Section~6 considering, among others, shock models. Conclusions from this work are then presented in Section~7. The detailed study of the H\II\ regions of NGC\,3344, along with a discussion of the  galaxy SFR and chemical evolution, using the same SITELLE data, will follow in a upcoming paper (Moumen et al. in prep.). 

% Table 1
\begin{table*}
	\centering
	\caption{Observing parameters and sky conditions for NGC\,3344}
	\label{tab:param_obs_table}
	\begin{tabular}{lccc} % four columns, alignment for each
		\hline
		Filter & SN1 & SN2 & SN3 \\
		 		& 365-385 nm & 480-520 nm & 651-685 nm \\
		\hline\hline
		Date 					& 2016 March 07  & 2016 March 06 & 2016 March 05 \\
		Seeing 					& 0.8$^{\prime\prime}$ & 0.8$^{\prime\prime}$ & 0.8$^{\prime\prime}$ \\
		Sky quality         	& Clear & Clear & Clear \\
		Folding order 					& 8  & 6 & 8 \\
		Step size [nm] 				& 1647 & 1680 & 2943 \\
		Number of steps 				& 70 & 134 & 258 \\
        Spectral resolution (R) 	& 400 & 600 & 1500 \\
		Exposure time per step	[s]	& 138.0 & 79.5  & 34.0 \\
		Total observing time 	[h] 		& 2.8 & 3.1 & 2.7 \\
		\hline
	\end{tabular}
\end{table*}

%%%%%%%%%%%%%%%%%%%%%%%%%%%%%%%%%%%%%%%%%%%%%%%%%%
%%%%%%%%%%%%%%%%%%%%%%%%%%%%%%%%%%%%%%%%%%%%%%%%%%
\section{OBSERVATION}
%%%%%%%%%%%%%%%%%%%%%%%%%%%%%%%%%%%%%%%%%%%%%%%%%%

NGC\,3344 was observed in March 2016 (RUNID: 16AH41; PI: R.P.~Martin) with the imaging Fourier transform spectrometer (iFTS) SITELLE \citep{2010SPIE.7735E..0BD, 2019MNRAS.485.3930D} installed on the 3.6-m Canada-France-Hawaii Telescope (CFHT).

With a large 11$^{\prime}$ $\times$ 11$^{\prime}$ field of view (FoV) and a 
seeing-limited spatial resolution sampled at 0.32$^{\prime\prime}$, SITELLE allows us to obtain more than 4 million spectra in the wavelength range from 350 to 900~nm. In the case of NGC\,3344, the seeing was typically 0.8$^{\prime\prime}$. We selected the three filters SN1 (365-385\,nm), SN2 (480-520\,nm), and SN3 (651-685\,nm) in order to get the strong emission lines useful to study the ionized gas properties ([O\II]$\lambda$3727, [O\III]$\lambda\lambda$4959,5007, H$\beta$, [N\II]$\lambda\lambda$6548,6584, H$\alpha$, and [S\II]$\lambda\lambda$6716,6731). 
Observation with each filter provides a wavelength calibrated data cube.
The spectral resolution and observing time for each filter, along with other observing parameters are listed in Table~\ref{tab:param_obs_table}. A higher spectral resolution was used for the SN3 filter to allow us to properly separate the emission lines of [N\II]$\lambda\lambda$6548,6583 and H$\alpha$, and to study the ionized gas kinematics. 

%%%%%%%%%%%%%%%%%%%%%%%%%%%%%%%%%%%%%%%%%%%%%%%%%%
%%%%%%%%%%%%%%%%%%%%%%%%%%%%%%%%%%%%%%%%%%%%%%%%%%
\section{DATA REDUCTION AND LINES FITTING}
\label{sec:datareduction}
%%%%%%%%%%%%%%%%%%%%%%%%%%%%%%%%%%%%%%%%%%%%%%%%%%

The data reduction was performed with ORBS \citep[v. 4.0-DR1-beta;][]{2015ASPC..495..327M, 2016MNRAS.463.4223M}. ORBS is an automatic data reduction software designed for iFTS data obtained with SITELLE freely available on the web\footnote{https://sourceforge.net/projects/orb-orbs/}. ORBS follows in a few steps to transform the interferogram cubes (from the two cameras) into a spectral datacube for each filter \citep[see][]{2019MNRAS.485.3930D}. First, the standard CCD image calibrations are applied, including the bias subtraction, the flat-field correction, and the cosmic rays removal. Second, due to a slight optical misalignment between the cameras, an alignment of the interferogram cubes is performed before combining the two cubes and correcting for atmospheric variations (airmass and clouds). Finally, a discrete Fourier transform is applied, along with the phase correction, on each pixel of the interferogram cube to create the spectral datacube. The wavelength calibration is carried out by ORBS using a high-resolution HeNe data cube collected during the observing run.  Images of the standard star GD71 were available for the flux calibration in the three filters.

Given that the spatial resolution is seeing limited, the spaxels have been spatially binned 3$\times$3~pixels to increase the signal-to-noise ratio (S/N) of the fainter regions. The Galactic extinction for NGC\,3344 is negligible ($A_{V} = 0.091$; NED\footnote{NASA/IPAC Extragalactic Database: http://ned.ipac.caltech.edu}) so we ignore it here and in the following analysis.

\begin{figure}
	\includegraphics[width=\columnwidth]{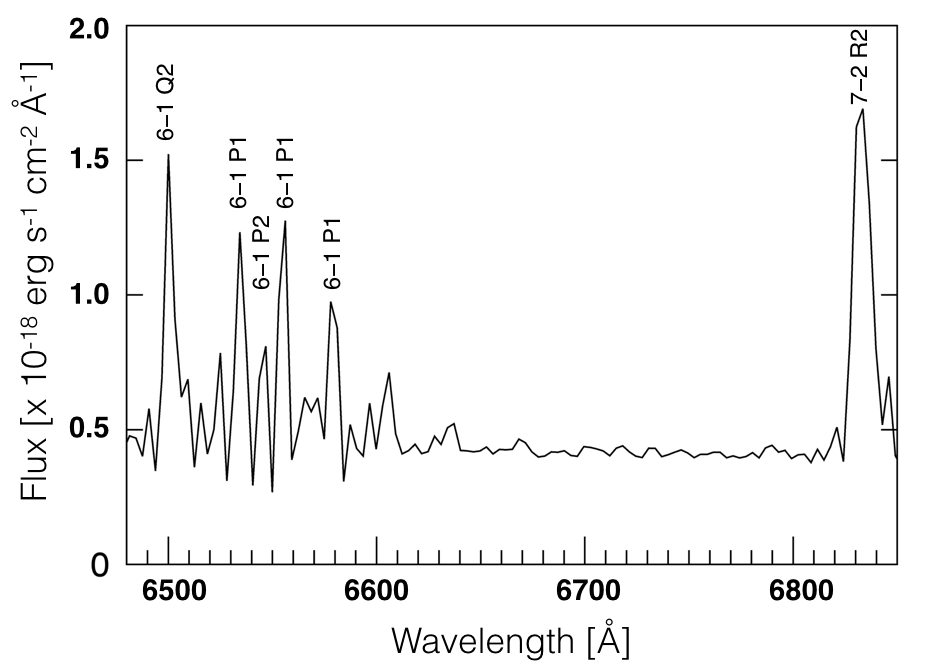}
		\caption{An example of the sky background spectrum in the case of the SN3 filter. The strong OH lines seen in this wavelength range are identified.}
	    \label{fig:skyLineSN3}
\end{figure}

The sky background was subtracted using the median spectrum built from more than 2$\times$10$^{4}$ pixels in the FoV, located away from the galaxy. As an example, Figure~\ref{fig:skyLineSN3} shows the sky spectrum obtained for the SN3 filter. Many OH lines are seen in this wavelength range. These lines are actually used to refine the wavelength calibration in the SN3 filter \citep[i.e. to take into account the flexure in the instrument during the observation, while the HeNe datacube used for the first order calibration is obtained with the telescope at the zenith; for more details, see][]{2019MNRAS.485.3930D}.

%Figure 1

\begin{figure*}
	\includegraphics[width=\textwidth]{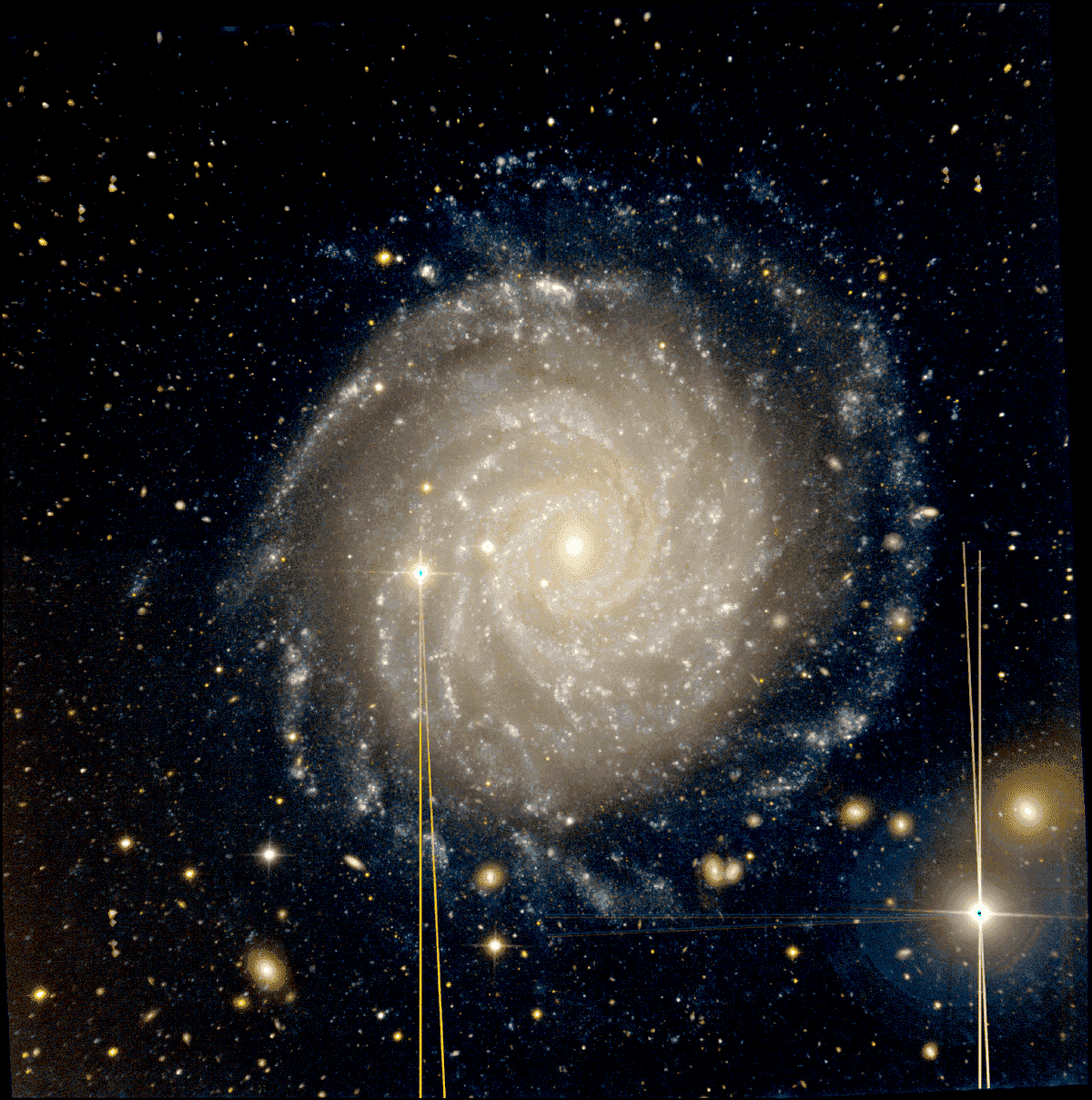}
    \caption{The SITELLE image of NGC\,3344 obtained from the combination of the  SN1 (blue) and SN3 (red) deep images. The image is centred on the galaxy (at RA\,=\,10h43m31.15s and DEC\,=\,+24$^\circ$55$^{\prime}$20.0$^{\prime\prime}$) and fills the FoV of SITELLE (11$^{\prime}$ $\times$ 11$^{\prime}$). Bright lines are artefacts caused by saturated stars.  North is up and East is left.}
    \label{fig:figure_deep}
\end{figure*}

The three datacubes have been spatially aligned relative to each other using a homemade Python program that enables a sub-pixel accuracy. Using the datacubes deep image (i.e. the image obtained after summing all the interferograms collected in one filter, prior to the application of the Fourier transform), this program calculates the shifts along the spatial axis $x$ and $y$ using the function \texttt{chi2\_shift} in the Astropy Python package \texttt{image\_registration} and it applies a correction for the misalignment using the function \texttt{shiftnd} from the same package. The SN3 deep image was used as the reference since it is the deepest one. We measured a shift [$dx$, $dy$] of [$-$0.41, 0.94] for SN1 and [0.48, $-$0.54] for SN2. The image of NGC\,3344 obtained from the combination of two of the filter deep images is presented in Figure~\ref{fig:figure_deep}.

% Figure 2

\begin{figure*}
	\includegraphics[width=0.8\textwidth]{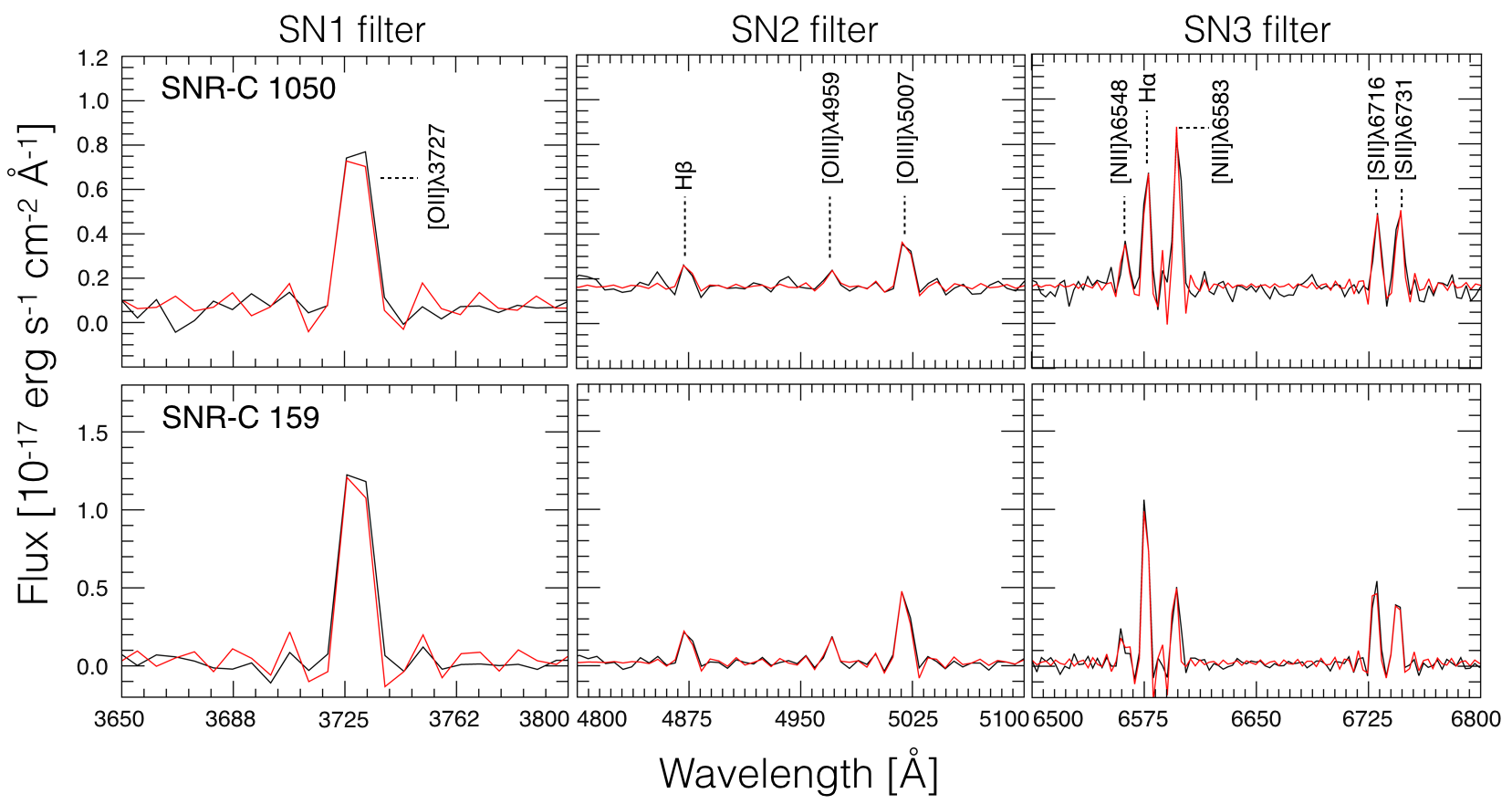}
		\caption{
		Examples of SITELLE spectra for one pixel within the SNR candidates SNR-C\,1050 (top) and SNR-C\,159 (bottom). The segments for the three filters are shown including the emission lines, from right to left:	SN1 (R = 400), SN2 (R = 600), and SN3 (R = 1500). In black, the observed spectrum and in red, the fit obtained with ORCS. The strong emission lines are identified. The S/N$_{{\rm H}\alpha}$\,=\,19 for SNR-C\,1050 and 80 for SNR-C\,159.} 
	    \label{fig:snr_spectra}
\end{figure*}

SITELLE has an instrumental profile best reproduced by a cardinal sine (sinc) function. ORCS \citep[v.~1.0.1;][]{2015ASPC..495..327M, 2016MNRAS.463.4223M}, a data extraction software developed specifically for SITELLE used to fit simultaneously all the lines in each datacube. ORCS returned maps, for each emission line, of the amplitude (intensity), FWHM, continuum height, flux, and velocity (based on one or multiple lines centroid). More maps are also returned for the parameters uncertainty. For example, Figure~\ref{fig:snr_spectra} presents the spectrum of one pixel within two SNR candidates studied in this paper, along with their fit obtained with ORCS.

% Figure 3

\begin{figure}
	\includegraphics[width=\columnwidth]{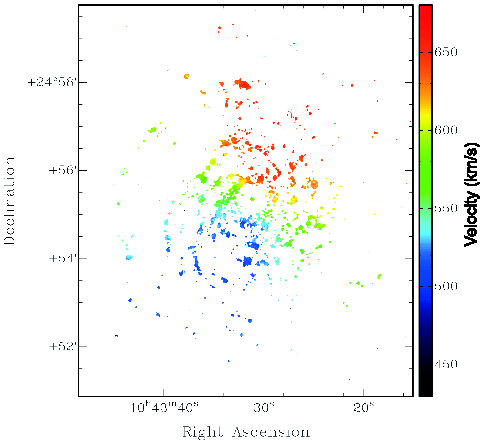}
		\caption{NGC\,3344 velocity map. This map was obtained considering
        only the spaxels with S/N$_{{\rm H}\alpha}$\,$\geq$\,3. 
        }
	    \label{fig:exVel_map}
\end{figure}

In the Appendix~\ref{Flux_maps_A}, Figure~\ref{fig:figure_fluxmaps} shows maps of the flux for all the lines used in this work. It is interesting to notice how the [O\III]$\lambda\lambda$4959,5007 emission lines are absent in the galaxy inner ring, while this structure is well marked by the other emission lines, including [O\II]$\lambda$3727. Figure~\ref{fig:exVel_map} presents the ionized gas velocity map gathered considering only spaxels with S/N$_{{\rm H}\alpha} \geq 3$. With a small inclination ($i = 18.7^\circ$; HyperLeda\footnote{http://leda.univ-lyon1.fr/}), NGC\,3344 displays a  small velocity gradient, of the order of 200~km~s$^{-1}$, between its receding and approaching side. The systemic velocity measured at the galaxy centre (579$\pm$18~km~s$^{-1}$) is in agreement with the value of NED (580$\pm$1~km~s$^{-1}$).

%Figure 4

%%%%%%%%%%%%%%%%%%%%%%%%%%%%%%%%%%%%%%%%%%%%%%%%%%
%%%%%%%%%%%%%%%%%%%%%%%%%%%%%%%%%%%%%%%%%%%%%%%%%%
\section{SNR CANDIDATES}
%%%%%%%%%%%%%%%%%%%%%%%%%%%%%%%%%%%%%%%%%%%%%%%%%%

%%%%%%%%%%%%%%%%%%%%%%%%%%%%%%%%%%%%%%%%%%%%%%%%%%
\subsection{Automatic Detection of Emission Regions}
%%%%%%%%%%%%%%%%%%%%%%%%%%%%%%%%%%%%%%%%%%%%%%%%%%
\label{sec:auto}

\begin{figure}
	\includegraphics[width=\columnwidth]{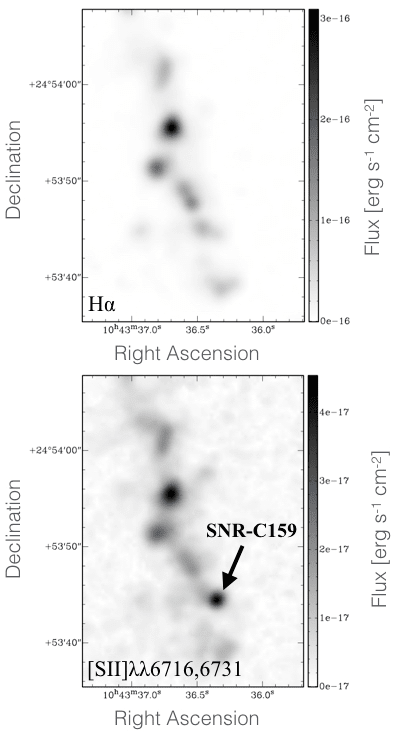}
		\caption{H$\alpha$ (top) and [S\II]$\lambda$6716+[S\II]$\lambda$6731 (bottom) flux maps near the SNR candidate SNR-C\,159. While the scale of the images was selected so the brightness of the H\II\ regions in the H$\alpha$ and the [S\II] maps is comparable, SNR-C\,159 appears clearly in the [S\II] map which is known to be a good tracer of shock-heated gas. 
		}
	    \label{fig:exsnr159}
\end{figure}

In order to identify the SNR candidates in a non-subjective way, we used the automated identification technique for ionized gas regions described by \citet{2018MNRAS.477.4152R}. This technique was initially created to study the star-forming regions in the nearby, nearly face-on spiral galaxy NGC\,628 where more than 4200 H\II\ region candidates have thus been  identified. We used this technique, although with some adaptations for our purpose to find SNRs, following the same main three steps:

\noindent (i) The identification of the emission peaks is done using a combination of the [S\II]$\lambda$6716 and [S\II]$\lambda$6731 continuum-subtracted images. \citet{2018MNRAS.477.4152R} considered the three emission lines H$\alpha$, H$\beta$, and [O\III]$\lambda\lambda$4959,5007 for the identification of star-forming regions in NGC\,628. In the case of SNRs, compared to H\II\ regions, the [S\II] emission lines are rather strong, relative to H$\alpha$, and may be related to a different emission process (i.e. shocks vs. photoionization), and are therefore considered here as better suited to define the SNR's boundary. As an example, Figure~\ref{fig:exsnr159} presents the H$\alpha$ and [S\II] maps near one of our best SNR candidates, SNR-C\,159; the difference is striking with a relatively stronger [S\II] flux for the SNR compared to the surrounding H\II\ regions. Therefore, using the combined [S\II] continuum-subtracted image, an emission peak is confirmed if its intensity is greater than the intensity of the 5 surrounding pixels (this number allows us to reduce the amount of false detection due to the noise, without missing fainter peaks in crowded regions) and if the total intensity of a box centred on the emission peak is higher than the adopted detection threshold. The size of the box is 3$\times$3~pixels and the threshold adopted, which is slightly different from a CCD quadrant to another, vary from 2.5$\times$10$^{-19}$ in faint regions to 8$\times$10$^{-18}$~erg~s$^{-1}$~cm$^{-2}$~\AA$^{-1}$ in the brightest regions. We also add a spectral constraint which consists in considering only pixels with S/N\,$>$\,3 for both of the [S\II] lines.

% Figure 5

\noindent (ii) The determination of the zone of influence (an area influenced by the ionizing photons and shocks from an emission peak) around each emission peak is done by studying the distance between each pixel and the surrounding peaks. A maximum radius of 30 pixels was adopted (i.e. at a distance of 6.1~Mpc for NGC\,3344, this radius corresponds to 285~pc), much larger than the maximum size known for SNRs \citep{2009SerAJ.179...55C} but allowing to detect more extended objects like superbubbles and DIG regions. In general a pixel is associated to its closest peak. If more than one peak is located at the same distance, the pixel is associated to the brightest peak. Finally, the position of the most distant pixels defines the boundaries of the zone of influence.

\begin{figure}
	\includegraphics[width=\columnwidth]{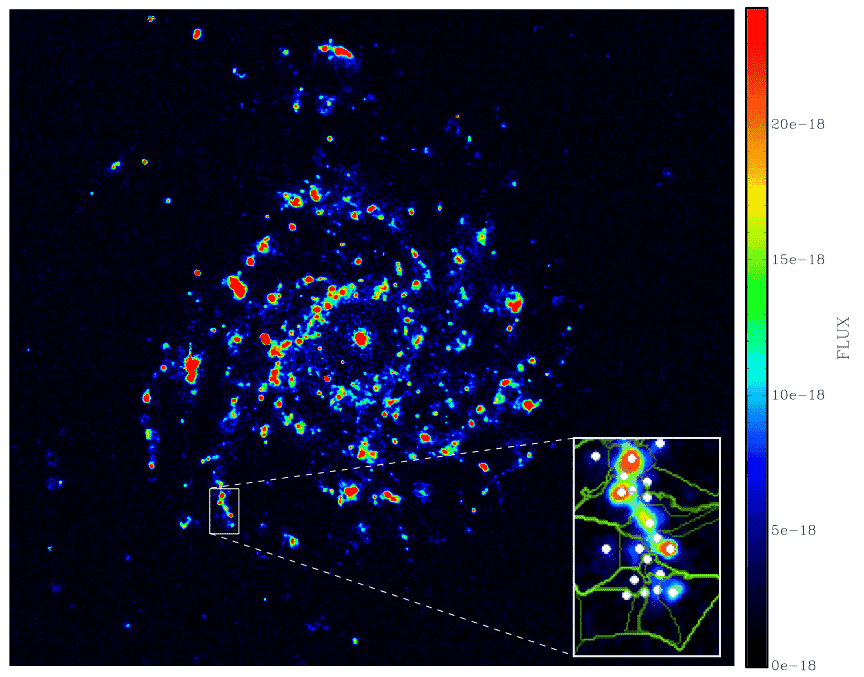}
		\caption{
		The [S\II] map near the SNR candidate SNR-C\,159, where the outer limit for various emission regions are shown.}
	    \label{fig:exzoom-outerlim}
\end{figure}

\noindent (iii) The outer limit of an emission region is defined as the distance where the slope of the flux profile decreases by less than 2\% within the zone of influence. The flux profile is calculated by summing the flux of pixels located in circular annuli 25~pc thick centred on the peak. Figure~\ref{fig:exzoom-outerlim} presents the [S\II] map near the SNR candidate SNR-C\,159, where the domain, i.e. the outer limit, for a sample of emission regions is shown. 

%Figure 6

\begin{figure}
	\includegraphics[width=\columnwidth]{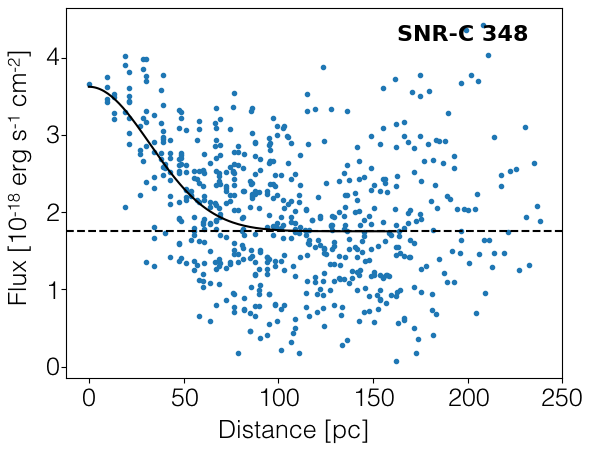}
	\includegraphics[width=\columnwidth]{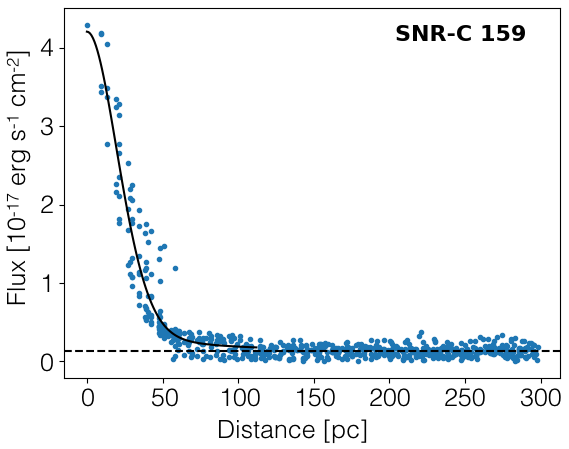}
		\caption{Pseudo-Voight profile of the diffused region SNR-C\,348 (most probably a DIG region) and the compact region SNR-C\,159 (one of the best SNR candidates), as obtained with the technique of Rousseau-Nepton et al. (2018) adapted to SNR candidates.
		}
	    \label{fig:exProfil_snr}
\end{figure}

For each emission region thus defined, the code then fits a radial profile to the flux. The theoretical flux profile adopted is a pseudo-Voight profile (i.e. a combination of a Gaussian and Lorentzian function; see Eq.~3 of \citet{2018MNRAS.477.4152R}), allowing to accurately fit a wide variety of morphology for the emission regions (i.e. from highly peaked to very diffused). The code then returns for each emission region a list of parameters for the best profile fit: the peak intensity ($A$), the FWHM of the profile, the correlation coefficient ($R$) between the profile fit and the flux points, and the background level. Figure~\ref{fig:exProfil_snr} shows examples of the pseudo-Voight profiles obtained for two emission regions: the extended and more diffused region SNR-C\,348 (most probably a DIG region) and the more  compact region SNR-C\,159 (one of the best SNR candidates).

The number of emission regions found, using the technique of  Rousseau-Nepton et al. (2018) but with the combined [S\II] map, 
is rather large; 2192 emission regions are found. At this point, it is clear that most of these regions are not SNRs, but H\II\, regions (almost 3000 emission regions have been found using the same technique but with the H$\alpha$+H$\beta$+[O\III] map; Moumen et al. in prep.) and also DIG regions. 

%Figure 7

%%%%%%%%%%%%%%%%%%%%%%%%%%%%%%%%%%%%%%%%%%%%%%%%%%
\subsection{Removal of the Stellar Populations and Diffuse Ionized Gas Background}
%%%%%%%%%%%%%%%%%%%%%%%%%%%%%%%%%%%%%%%%%%%%%%%%%%
\label{sec:bg}

In order to isolate and study the emission lines associated with SNRs, it is important to take into account the light on the line of sight from other sources in the galaxy. NGC\,3344 is rather massive \citep{2013AJ....145..101K}, it is therefore expected that its disk and bulge stellar populations will be responsible for absorption lines that will be superimposed to the emission lines (mainly the Balmer lines). H\II\ regions are numerous in NGC\,3344 -- large complexes can be seen in its ring for example -- and may be the place where numerous SN explosions occur. Therefore, H\II\ regions are as well expected to pollute the spectrum of many SNRs while: 1) the ionized gas produces additional emission lines, and 2) the young stellar population adds another absorption component underneath the emission lines. In this massive star-forming galaxy, one can also anticipate DIG emission lines to be mixed with the SNR signatures. 

We have considered two different methods to create the galaxy background spectrum to be subtracted from each emission region. For the first method, a Local Galaxy Background (LGB) was obtained for each region considering a circular aperture centred on each region. The LGB internal radius corresponds to the outer limit of the region, while its external radius is given when its spectrum reaches S/N\,$\simeq$\,25 in the continuum  (a value larger than the best signal-to-noise ratio obtained for the integrated region spectrum). Prior to summing individual spectra in order to create the LGB spectra, a mask was applied to the datacubes to avoid all the other emission regions identified (up to their outer limit) and also all the H\II\ region candidates found independently using the technique presented in Section~\ref{sec:auto} but, in this case, with the H$\alpha$+H$\beta$+[O\III] image.

For the second method, we considered a Global Galaxy Background (GGB) obtained by creating median spectra in rings centred on the galaxy. The spaxels position and galactocentric radius have been first calculated considering the galaxy inclination $i = 18.7^\circ$ (Hyperleda) and position angle $PA = 150^\circ$ (NED). The width of the different rings is variable as it aims for S/N\,$\simeq$\,25 in the continuum of the GGB spectra (except at large galactocentric radius, where the noise becomes too important as discussed below). Prior to the summation of the GGB spectrum, all the SNR and H\II\ region candidates have been masked and all spaxels have been shifted to the galaxy velocity rest frame using the velocity map shown in Figure~\ref{fig:exVel_map}.

\begin{figure}
	\includegraphics[width=\columnwidth]{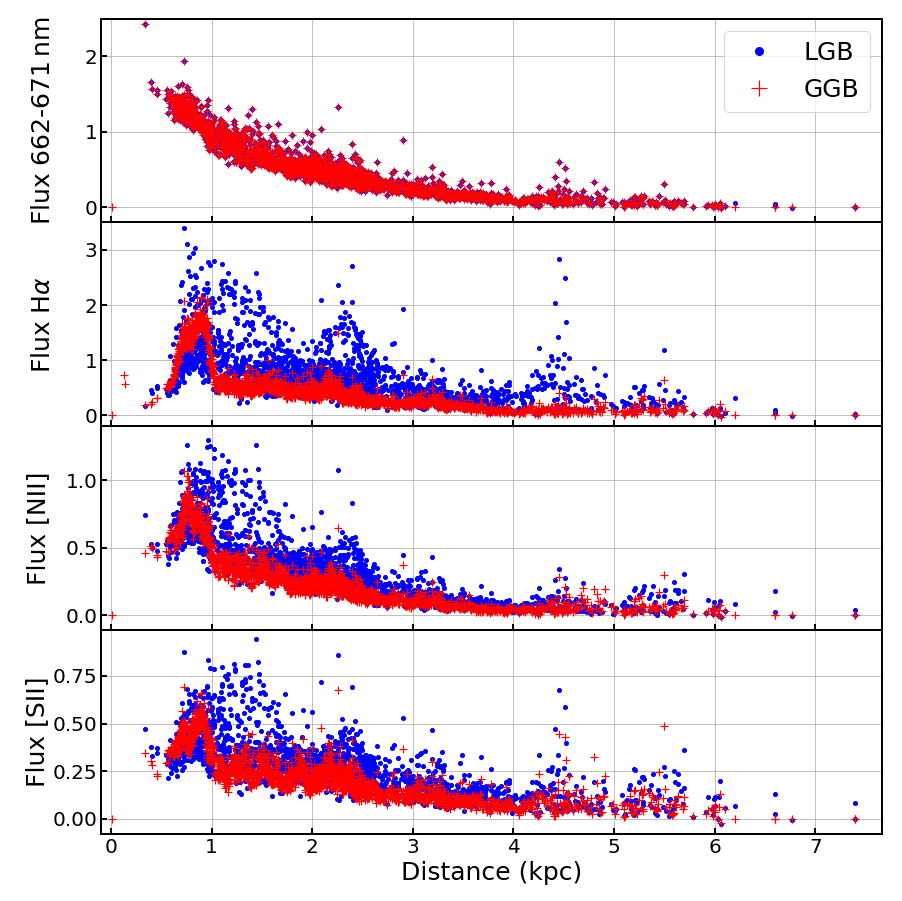}
    \caption{Comparison of the SN3 emission lines' flux in the galaxy background spectra created with the local LGB (in blue) and global GGB method (in red). There is one data point for all the emission regions, corresponding to the position (i.e. the galactocentric distance) of the emission region peak. All the background spectra have been scaled to the continuum level of the regions prior to line measurements. The galaxy inner ring is located  at a galactocentric distance of $\sim$1\,kpc.
    % refaire avec GCD
    }
    \label{fig:bg3}
 \end{figure}

We find that both methods used to calculate the galaxy background for the different emission regions have pros and cons. While the LGB takes more easily into account the local distribution of the emission regions, it is not ideal where a high density of emission regions forces a much larger external background radius. For example, in the inner ring, the LGB can select spaxels which are quite far, up to 350\,pc, from the region itself. On the other hand, the GGB may represents better the overall behaviour of the old stellar populations and the DIG at a given galactocentric radius, but it does not take into account the local distribution of the surrounding emission regions. Figure~\ref{fig:bg3} shows, as an example, the flux of the emission lines H$\alpha$, [N\II], and [S\II], along with the SN3 continuum level, measured in the galaxy background spectra obtained with both methods. First, the two background methods display lines which are only in emission, i.e. pure absorption lines of H$\alpha$ or H$\beta$ from the galaxy stellar populations are never seen in the background spectra. The DIG emission is important all over the galaxy disk.  The background H$\alpha$ emission line has a maximum flux  corresponding to $\sim$12.56\%  of the average H$\alpha$ flux for all the emission regions. Second, both methods indicate stronger emission lines in the inner ring (at a galactocentric distance of $\sim$1\,kpc), with a decreasing intensity toward the outer disk. The LGB displays on average stronger lines and more variations in the spiral arms. In this paper, where we aim at identifying SNRs in an automatic and systematic way, we will favor the GGB method. For some specific purposes, i.e. to discuss the extinction and the effect of the environment on the identification of the SNRs, a comparison using the LGB method will also be done. 

\begin{figure}
	\includegraphics[width=\columnwidth]{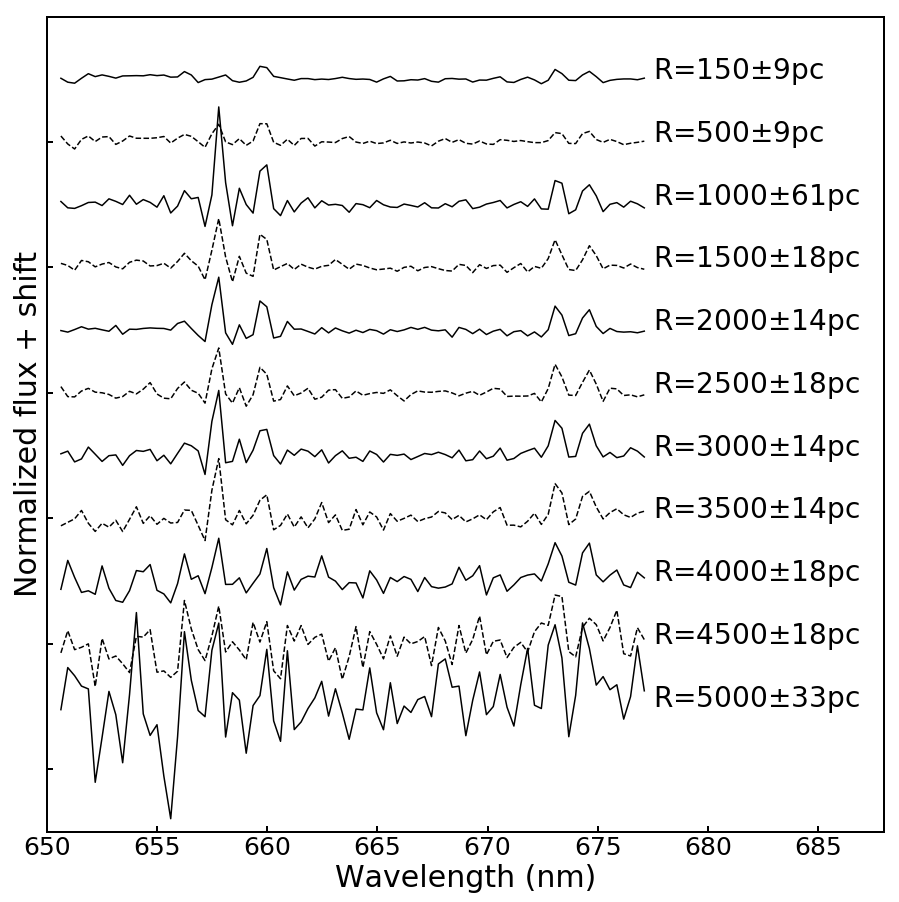}
    \caption{SN3 global galaxy background (GGB) spectra for different galactocentric distances. For each spectrum, the central radius of the background annulus is indicated on the plot, along with the annulus half width. NGC\,3344 shows an important DIG contribution that dominates these spectra. At a galactocentric radius of $\geq$\,4\,kpc, the noise level becomes important. 
    % refaire avec GCD
    }
    \label{fig:bg_sp}
\end{figure}

\begin{figure*}
    \includegraphics[width=0.8\textwidth]{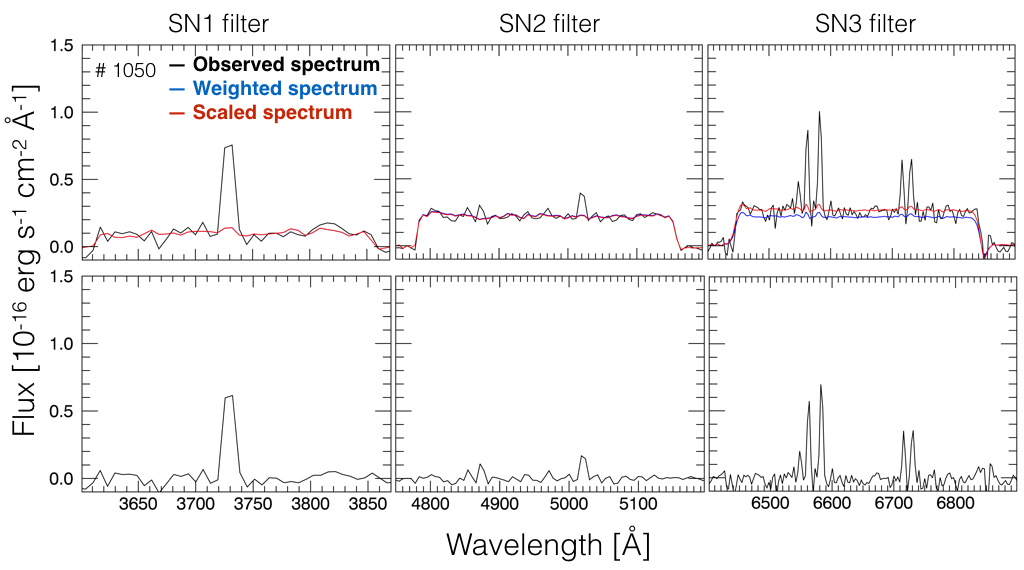}
    \includegraphics[width=0.8\textwidth]{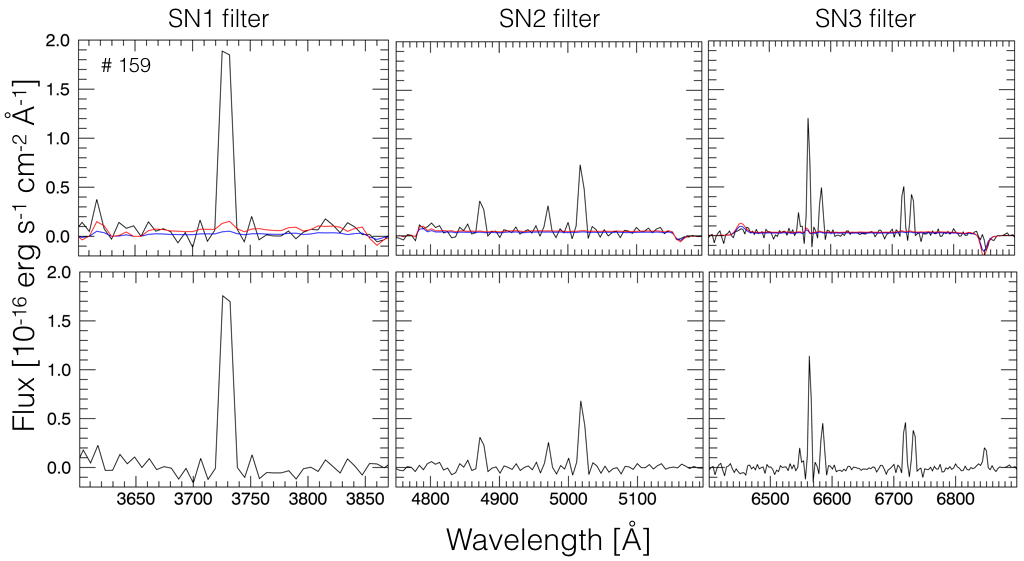}
    \caption{Spectra of two SNR candidates after the subtraction of the GGB background. SNR-C\,1050 is located inside the galaxy inner ring while SNR-C\,159 has a lower background level at a distance of 3.25\,kpc from the galaxy centre. The weighted spectrum is the background spectrum assigned to the region, before it is scaled to the continuum level of the region.
    }
    \label{fig:spectra_bg}
\end{figure*}

Figure~\ref{fig:bg_sp} presents a sequence of normalized GGB SN3 spectra plotted, for comparison, as a function of their galactocentric distance. As already indicated by Figure~\ref{fig:bg3},  these GGB spectra display a significant change in the emission lines strength with their position; the H$\alpha$ emission is quite strong in the galaxy inner ring (at $\sim$1\,kpc). It is clear that the noise becomes very important at a very large radius (even though we tried to reach S/N\,$\simeq$\,25 by increasing the width of the ring). This is mainly explained by the fact that the continuum flux from the disk stellar populations drops quickly. Because of the high level of noise seen at larger radius, and because the spectra are quite similar between 2.5 and 4\,kpc, we simply used the GGB spectrum at 3.5\,kpc for all the region located further away. In the following, unless stated otherwise, for each of the emission region we have subtracted its corresponding GGB spectrum, i.e. for each of the spaxel in an emission region, a GGB spectrum is assigned, scaled to the continuum level of the spaxel, and subtracted. The scaling factor is in general very close to one and assures us that the continuum level of the emission region is never negative. Figure~\ref{fig:spectra_bg} shows the background corrected spectrum of two SNR candidates, one away in the disk and one inside the inner ring, along with the scaled GGB spectrum used.

%Figure 8

%Figure 9

%Figure 10

%%%%%%%%%%%%%%%%%%%%%%%%%%%%%%%%%%%%%%%%%%%%%%%%%%
\subsection{SNR Candidate Selection}
%%%%%%%%%%%%%%%%%%%%%%%%%%%%%%%%%%%%%%%%%%%%%%%%%%
\label{sec:iden-snr}

In order to select the SNR candidates among the emission regions identified in the previous subsections, we used four criteria. These criteria often involve the size (i.e. diameter) of the region. This size parameter is represented by $\Sigma$ through all the paper and is simply set equal to the FWHM of the pseudo-Voight profile obtained for each region. This definition may underestimate the real size of a region, but it allows a secure integrated flux measurement for the region emission lines while avoiding an uncertainty due to the immediate background contamination. The selection criteria are:

\noindent 1) The integrated spectrum of a region must have a line ratio ([S\II]$\lambda$6716+[S\II]$\lambda$6731)/H$\alpha$ $\geq$\,0.4. This is supported by the work of \citet{1963Natur.199..681M} and  \citet{1973ApJ...180..725M}. A total of 1058 emission regions satisfy this criterion alone.

\noindent 2) The spectrum of each region must have S/N\,$\geq$\,5 for the three emission lines H$\alpha$, [S\II]$\lambda$6716, and [S\II]$\lambda$6731. The number of emission regions that meet this criterion alone is 778.

\noindent 3) The size $\Sigma$ of a region should not exceed 120\,pc. This value is slightly larger than the expected diameter of 100\,pc for Galactic SNRs \citep{2009SerAJ.179...55C}, to take into account mainly the galaxy distance uncertainty. In total, 1030 emission regions satisfy this criterion alone.

\noindent 4) The pseudo-Voight profile of a region must present a correlation coefficient $R \geq 0.5$. In general, emission regions with a lower correlation coefficient are seen in the images as extended DIG regions or blended emission regions of all kinds. A total of 1637 emission regions meet this requirement alone.

\begin{figure}
	\includegraphics[width=\columnwidth]{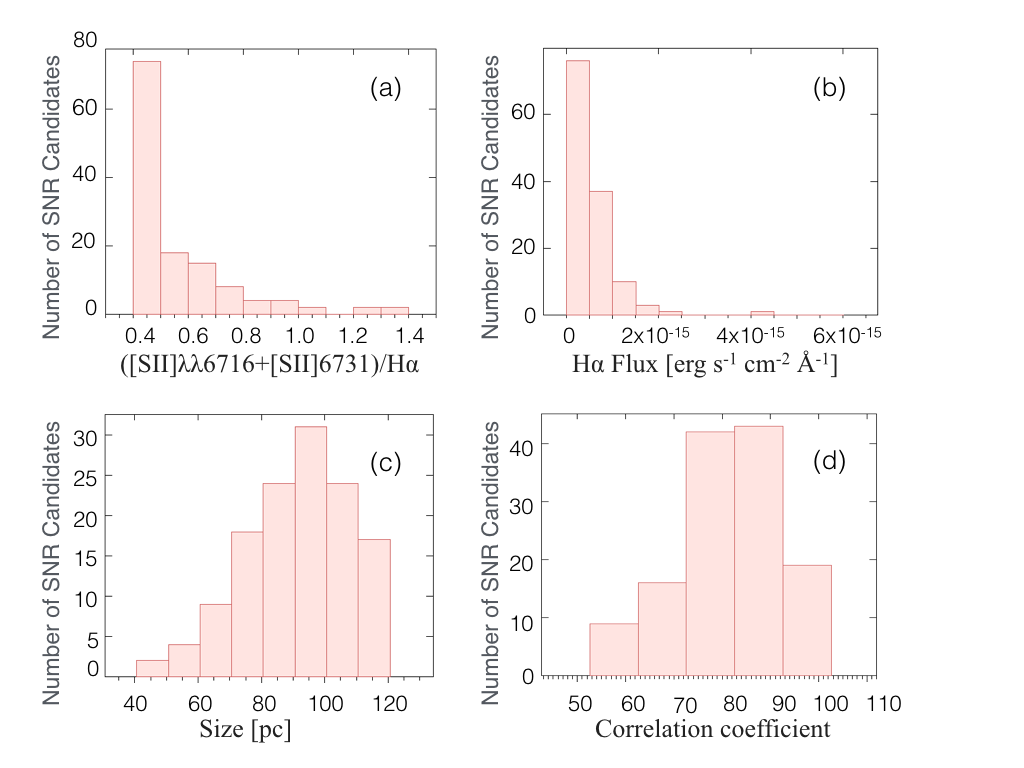}
    \caption{Histograms of the selected SNR candidates as a function of: (a)~the [S\II]/H$\alpha$ flux ratio, (b)~the H$\alpha$ Flux, (c)~the size $\Sigma$, and (d)~the profile correlation coefficient $R$. 
    All these candidates satisfy the selection requirements given in Section~\ref{sec:iden-snr}. 
    }
    \label{fig:r2c_histo}
\end{figure}

Combining these four criteria, we obtain a list of 129 SNR candidates. Figure~\ref{fig:r2c_histo} shows that our selected candidates display a wide range of [S\II]/H$\alpha$ and H$\alpha$ flux.
The median H$\alpha$ flux measured for the selected SNR candidates is 3.90$\times$10$^{-16}$\,erg\,s$^{-1}$\,cm$^{-2}$\,\AA$^{-1}$. On average, the size $\Sigma$ of the regions is near 90\,pc, and the profile correlation coefficient $R$ is $\sim$\,80\%. We find no SNR candidate with $\Sigma$\,$<$\,40\,pc, but this limit corresponds to the seeing at the time of the observations. 

In Appendix~\ref{cat_snr_cand}, Figure~\ref{fig:figure_snrs2Ha} presents the H$\alpha$+[S\II] image with a close-up of the 129 SNR candidates. The candidates appear as green sources due to their relatively stronger [S\II] emission. Table~\ref{tab:example_table_snr} gives the SNR candidates properties: their identification number, coordinates, integrated H$\alpha$ flux and ([S\II]$\lambda$6716+[S\II]$\lambda$6731)/H$\alpha$ flux ratio, region size $\Sigma$, pseudo-Voight profile correlation coefficient $R$, and galactocentric distance. All the emission lines integrated flux for the 129 candidates are listed in Table~\ref{tab:table_emission_lines_flux}. 

%Figure 11

%%%%%%%%%%%%%%%%%%%%%%%%%%%%%%%%%%%%%%%%%%%%%%%%%%
\subsection{Intrinsic Extinction}
%%%%%%%%%%%%%%%%%%%%%%%%%%%%%%%%%%%%%%%%%%%%%%%%%%
\label{sec:ext}

In general, studies of SNRs make use of emission line ratios that are built from lines near each other in the spectrum, and therefore consider that an extinction correction is not required (e.g. \citealt{2018ApJ...855..140L}). As done in previous optical studies of SNRs, we will consider similar line ratios, but we will also use the [O\II]$\lambda$3727/H$\beta$ ratio, since it is available with our SITELLE data. In this case, a reliable extinction  correction is necessary. The study of the galaxy intrinsic extinction is done here considering the Balmer decrement. Our sample of emission line regions detected with our technique (\S~\ref{sec:auto}) have different physical gas conditions, but since we are concentrating on the study of the SNRs (and most of the other type of emission regions will be put aside during our selection and confirmation process, as presented below), we adopt a theoretical flux ratio $(F_{{\rm H}\alpha}/F_{{\rm H}\beta})_{\rm theo}$\,=\,3.0, as prescribed for the physical properties of SNRs \citep{2006agna.book.....O}. A similar approach was also taken by \cite{1985ApJ...289..582B} to study the SNRs in M33. The color excess in the SNR candidates obtained in our study is then given by the equation: 
\begin{equation}
\label{eq_ebv}
E(B-V) = \frac {2.5} {1.07} \log \bigg( 
\frac { ( F_{{\rm H}\alpha} / F_{{\rm H}\beta} )_{\rm obs} } {3.00} \bigg),
\end{equation}
where  $E(\beta - \alpha)/E(B-V)$\,=\,1.07 (as extracted from the analysis of \citealt{1989ApJ...345..245C}), and where $({F_{{\rm H}\alpha}/F_{{\rm H}\beta}})_{\rm obs}$ is the observed flux ratio of the two main Balmer lines.

As shown in the Appendix~\ref{fig:figure_fluxmaps}, the H$\alpha$ map is deep enough to study the whole galaxy pixel per pixel. However, the H$\beta$ map presents far less pixels (only 12\% of the whole sample of the H$\alpha$ pixels) with S/N\,$\geq$\,3. In this case, instead of calculating the extinction for each pixel, we are using the integrated flux of each emission region (i.e. by combining the spectrum of the pixels located within an aperture centred on the emission peak with a diameter equal to $\Sigma$, i.e. the FWHM of the region pseudo-Voight profile). When considering the integrated flux, 80\% of the regions have an H$\beta$ detection with S/N\,$\geq$\,3. 

\begin{figure*}
	\includegraphics[width=\textwidth]{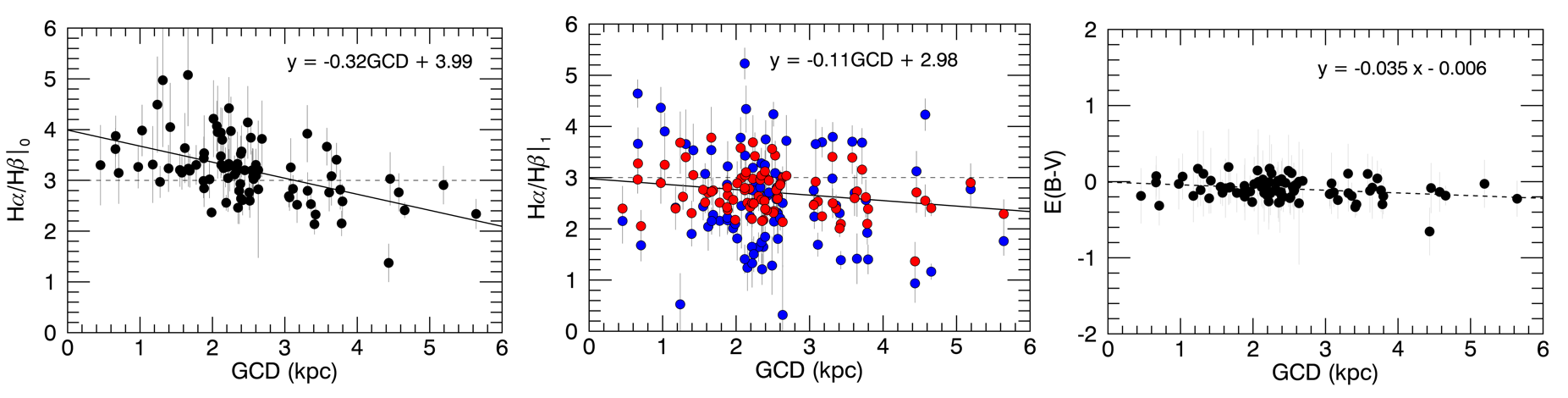}
	\caption{The H$\alpha$/H$\beta$ ratio for all the SNR candidates. Left: before the background subtraction. Centre: after the background subtraction using the LGB (blue) and the GGB (red) measurements. 
	Right: the galactocentric variation of the color excess obtained using Equation~\ref{eq_ebv} and considering only the SNR candidates with $S/N > 5$ for all the emission lines H$\beta$, H$\alpha$, [SII]$\lambda$6716, and [SII]$\lambda$6731.
    }
    \label{fig:hbhamap}
\end{figure*}

Figure~\ref{fig:hbhamap} shows the flux ratio of the H$\alpha$ line over H$\beta$ observed (before and after the subtraction of the galaxy background -- GGB and LGB) for each region as a function of their galactocentric distance (GCD). Prior to the background subtraction, the average ratio is larger than the selected theoretical value of 3 in the galaxy inner 2-3\,kpc, while it drops below this theoretical value at a larger distance. Different technical effects have been investigated to explain values below the theoretical ratio: datacube misalignment, aperture size vs atmospheric refraction, and flux calibration; but no clear correlation has appeared. On the other hand, the region with the lowest ratio, SNR-C2161 with a ratio H$\alpha$/H$\beta = 1.37\pm0.37$, is actually rather extended and a physical explanation involving an important contamination from the DIG is proposed to explain values below the theoretical ratio. After the background subtraction, the H$\alpha$/H$\beta$ ratio is on average near 3 (considering the uncertainties; see Fig.~\ref{fig:hbhamap}) through the whole disk, but again an important scatter is seen. As shown in the figure, the subtraction of the local background (LGB) produce more scatter in the H$\alpha$/H$\beta$ ratio than the subtraction of the global background (GGB). As suspected, the LGB is more easily influenced by the non uniformity of the emission from surrounding nebulae. 

Since the background subtracted H$\alpha$/H$\beta$ ratio is on average very close to the theoretical value, one might conclude that the galaxy internal extinction is rather low (values of $E(B-V)$ are also shown in Fig.~\ref{fig:hbhamap}). But this is not so clear considering the importance of the dispersion and the complex nature of the galaxy background. As discussed in Section~\ref{sec:bg}, the background contribution is a combination of the underlying stellar populations absorption lines and DIG emission lines. As it can be seen with the examples of Figure~\ref{fig:spectra_bg}, these two effects seems to cancel out well in the H$\beta$ line compared to H$\alpha$, resulting in a relatively weaker H$\alpha$ emission line and a ratio H$\alpha$/H$\beta$ closer to the theoretical value. On the other hand, the background is more complex than presented here, as its components are spatially variable, relative to each other and relative to the SNRs, as the DIG present distinct physical properties (e.g. electron temperatures and densities) compared to SNRs, and as these components are mixed with non-homogeneous layers of dust. We therefore suspect that the scatter seen in the background subtracted H$\alpha$/H$\beta$ ratio is related to the complexity of the background. With the data in hand, we cannot separate these multiple components and their extinction (as mentioned previously, we never get to observe simple Balmer absorption lines in the galaxy disk, which would help us for example to isolate the effect of the stellar populations). It is therefore difficult to estimate here the level at which the background subtraction affects the extinction measurement

Considering that the background corrected H$\alpha$/H$\beta$ ratio is on average near 3, no extinction correction was actually applied to the data. In the remaining of the paper, only the background (GGB) correction was made.

%Figure 12

%%%%%%%%%%%%%%%%%%%%%%%%%%%%%%%%%%%%%%%%%%%%%%%%%%
%%%%%%%%%%%%%%%%%%%%%%%%%%%%%%%%%%%%%%%%%%%%%%%%%%
\section{SNR Optical Confirmation}  
%%%%%%%%%%%%%%%%%%%%%%%%%%%%%%%%%%%%%%%%%%%%%%%%%%

\begin{figure*}
	\includegraphics[width=0.8\textwidth]{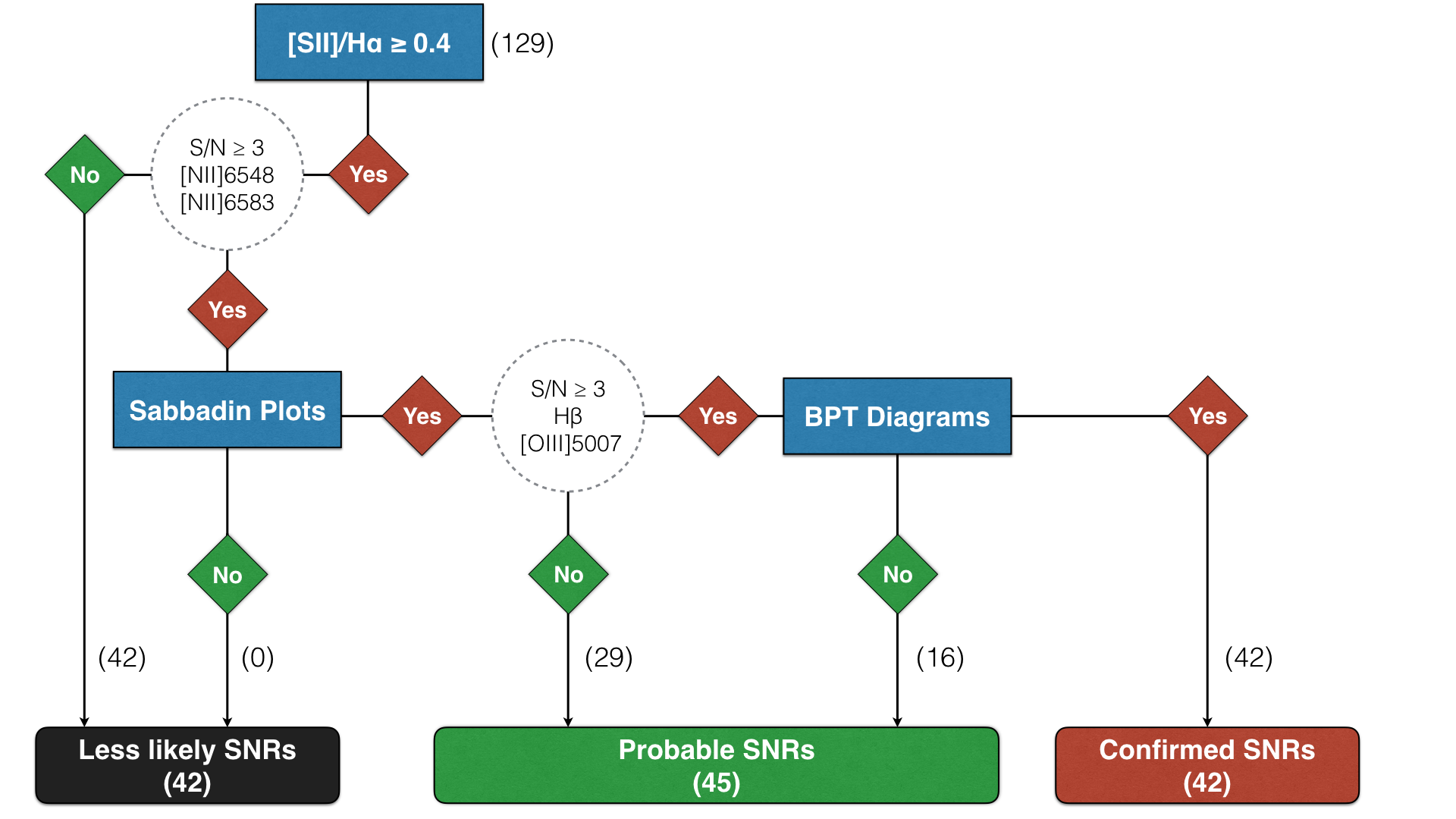}
    \caption{
    Tree of decision used to confirm the SNRs in NGC\,3344. 
    Sabbadin plots and BPT diagrams, based on emission line ratios, offer constrains for the ionization mechanism. 
    By the end of this analysis, the 129 SNR candidates fall into 
    one of the three categories: Confirmed (42), Probable (45), or Less likely SNRs (42).
    }
    \label{fig:decision_figure}
\end{figure*}

The emission lines available in the SITELLE data are here considered 
in Sabbadin plots (\S~\ref{sec:sabaddin}) and BPT diagrams (\S~\ref{sec:bpt}), 
in order to identify the ionization mechanism at play in the regions and confirm the best SNRs in our sample of 129~candidates. Figure~\ref{fig:decision_figure} presents our tree of decisions used to classify the SNR candidates into three categories: Confirmed SNRs, Probable SNRs, and Less likely SNRs. The following subsections describe in detail our analysis and classification scheme for the SNR candidates. 

%Figure 13

%%%%%%%%%%%%%%%%%%%%%%%%%%%%%%%%%%%%%%%%%%%%%%%%%%
\subsection{Sabbadin Plots} 
%%%%%%%%%%%%%%%%%%%%%%%%%%%%%%%%%%%%%%%%%%%%%%%%%%
\label{sec:sabaddin}

\citet{1977A&A....60..147S} performed a spectroscopic and photometric investigations on the nearby Planetary Nebula (PN) Sharpless 176 and compared its emission line ratios with those of SNRs, H\II\ regions, and other PNe. In their plots, where they compare emission line ratios of H$\alpha$/([N\II]$\lambda$6548+[N\II]$\lambda$6583), H$\alpha$/([S\II]$\lambda$6716+[S\II]$\lambda$6731), and [S\II]$\lambda$6716/[S\II]$\lambda$6731, SNRs, H\II\ regions, and PNe are distinct phenomena as they occupy different zones. 
While the electron density is expected to be higher in SNRs during the radiative phase (compared to H\II\ regions),  a smaller ratio [S\II]$\lambda$6716/[S\II]$\lambda$6731 and larger ratios of a line involving electron collisions (like [N\II] and [S\II]) over the H$\alpha$ resonance line is expected. Following the work of Sabbadin et al., more emission regions from our galaxy and from nearby galaxies have been considered so that it confirmed the behavior originally found. For example, a detailed version of the Sabbadin plots considering many SNRs identified in the optical was produced by \citet{2013MNRAS.429..189L}. 

The first rule in confirming the SNR nature of our 129 candidates is to find them in the SNR zone of at least one of the three Sabbadin plots. 
The choice for only one plot is justified by the fact that some physical factors (like the contamination by an H\II\ region and/or by the DIG) may move the SNR away from the shock region in some of the Sabbadin plots.
Figure~\ref{fig:sabaddin1} shows the Sabbadin plots, where our 129 candidates are represented by filled circles and where the domain for Galactic SNRs, H\II\ regions, and PNe are delimited by shaded zones and lines. The color of the  circles gives the final classification as discussed in Section~\ref{sec:fc}. Circles with a `$+$' symbol indicate SNR candidates with S/N\,$<$\,3 for the [N\II] lines (the criterion used to select the SNR candidates already implied S/N\,$\geq$\,5 for the H$\alpha$ and [S\II] lines).  The 42 circles with a `$+$' are in black as they are automatically considered Less likely SNRs. The remaining candidates (87/129), fall within (considering the line ratio uncertainties) the SNR zone in at least one of the three Sabbadin plots. 

In the Sabbadin plots, three candidates, SNR-C\,233, 332, and 1135, display a large value of the [S\II]$\lambda$6716/[S\II]$\lambda$6731 ratio, but with a large uncertainty due to weaker lines. Although older SNRs may have, in theory, a ratio larger that 1.43, we believe that these candidates, because of their weak lines, are better explained by very low density DIG regions. The candidate with a very high density (low [S\II]$\lambda$6716/[S\II]$\lambda$6731 ratio) and the lowest H$\alpha$/[N\II] ratio is SNR-C\,1050. This candidate is the only one located inside the galaxy inner ring. Within the inner ring, the DIG emission is important, with strong nitrogen emission, but we are confident that along with the galaxy background subtraction, the DIG contribution was removed properly. No  star-forming regions are surrounding SNR-C\,1050 which may indicate that it was produced more recently by a type Ia supernovae.

%Figure 14
 \begin{figure*}
	\includegraphics[width=\textwidth]{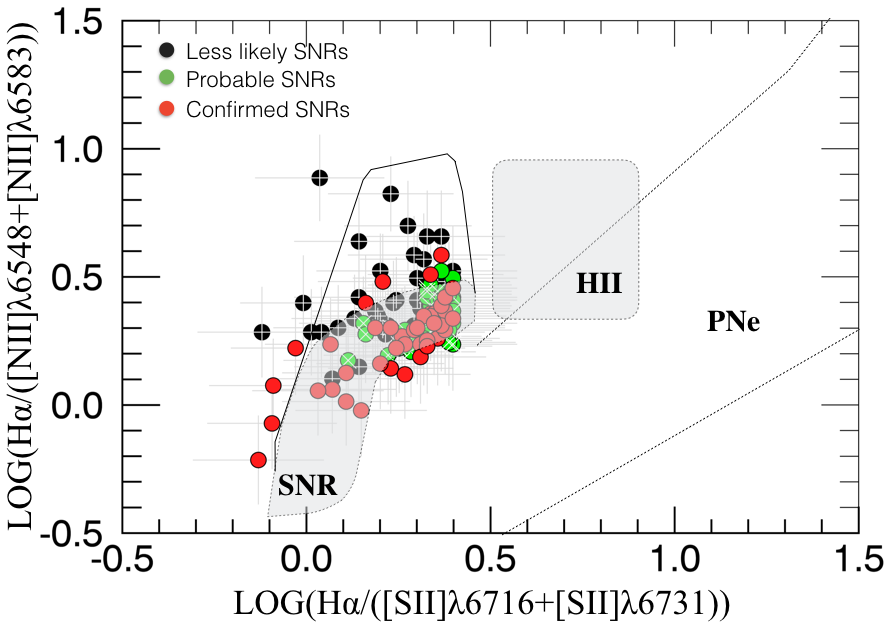}
    \caption{Sabbadin plots with the 129 SNR candidates found in NGC\,3344.
    The final classification of the candidates is shown using different colors: red for the Confirmed SNRs, green for the Probable SNRs, and black for the Less likely SNRs. Shaded zones and lines are used to separate the position of the SNRs, H\II\ regions, and planetary nebulae according to \citet{1977A&A....60..147S}.
    Circles with a `$+$' symbol indicate SNR candidates with S/N\,$<$\,3 for the [N\II]$\lambda\lambda$6548,6583 lines, while circles with a `$\times$' symbol indicate candidates with S/N\,$<$\,3 for the H$\beta$ or [O\III] lines. 
}
    \label{fig:sabaddin1}
\end{figure*}

 \begin{figure*}
	\includegraphics[width=\textwidth]{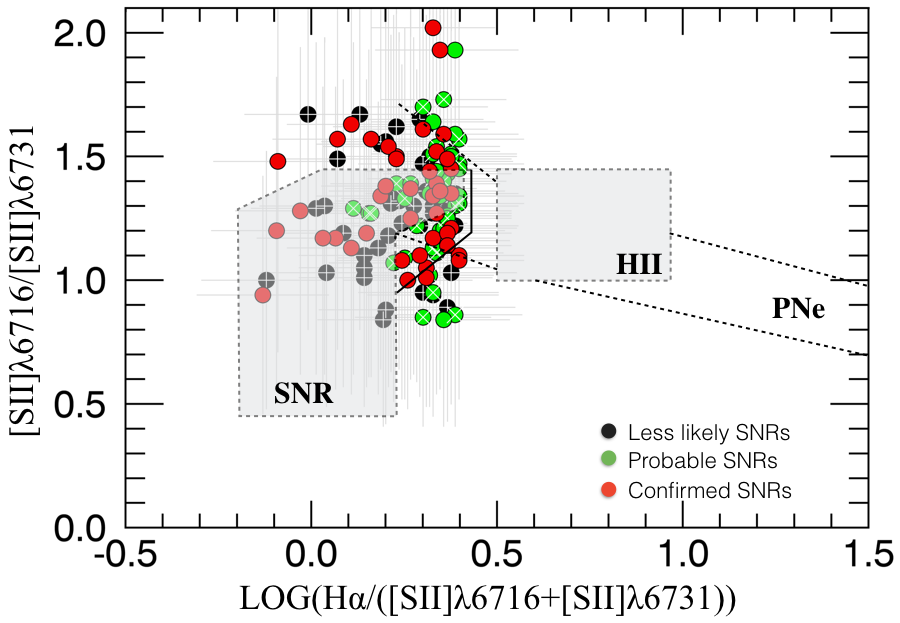}
    \contcaption{}
\end{figure*}

 \begin{figure*}
	\includegraphics[width=\textwidth]{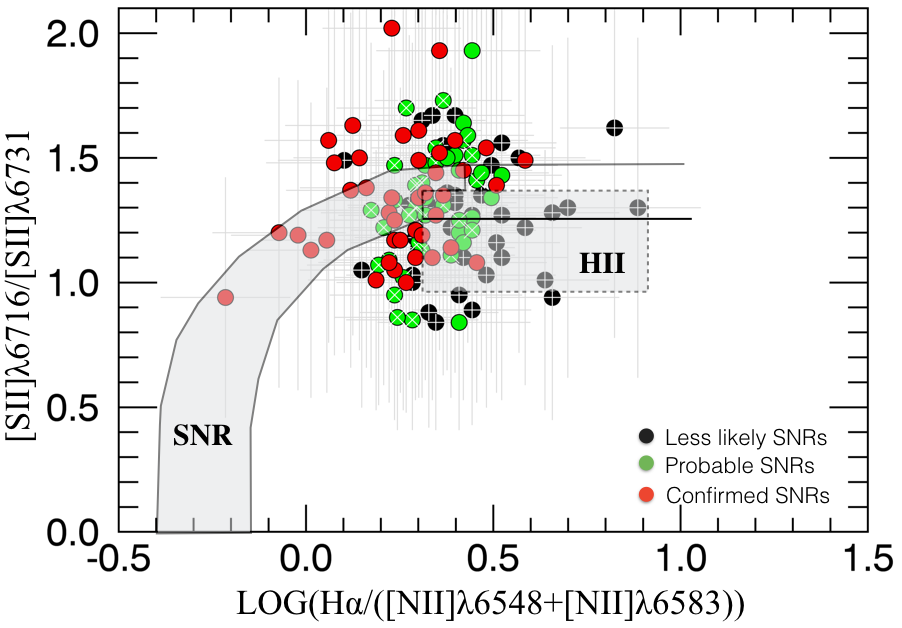}
    \contcaption{}
\end{figure*}

%%%%%%%%%%%%%%%%%%%%%%%%%%%%%%%%%%%%%%%%%%%%%%%%%%
\subsection{BPT Diagrams} 
%%%%%%%%%%%%%%%%%%%%%%%%%%%%%%%%%%%%%%%%%%%%%%%%%%
\label{sec:bpt}

The BPT diagrams, named after Baldwin, Phillips, and Terlevich \citep{1981PASP...93....5B}, represent another tool to identify the principal excitation mechanism in the nebular gas. Baldwin et al. showed empirically that several combinations of optical emission lines can be used to separate objects into four categories according to the principal excitation mechanism. These categories are: normal H\II regions, planetary nebulae, objects photoionized by a power-law continuum, and objects excited by shock-wave heating. Using these emission lines, Baldwin et al. created three diagrams: BPT-NII with [N\II]$\lambda$6583/H$\alpha$ vs [O\III]$\lambda$5007/H$\beta$;  BPT-SII with [S\II]$\lambda\lambda$6716,6731/H$\alpha$ vs [O\III]$\lambda$5007/H$\beta$; and  BPT-OI with [O\I]$\lambda$6300/H$\alpha$ vs [O\III]$\lambda$5007/H$\beta$. All the emission lines used in the BPT diagrams are available with SITELLE except the [O\I]$\lambda$6300 line. 

\begin{figure}
	\includegraphics[width=\columnwidth]{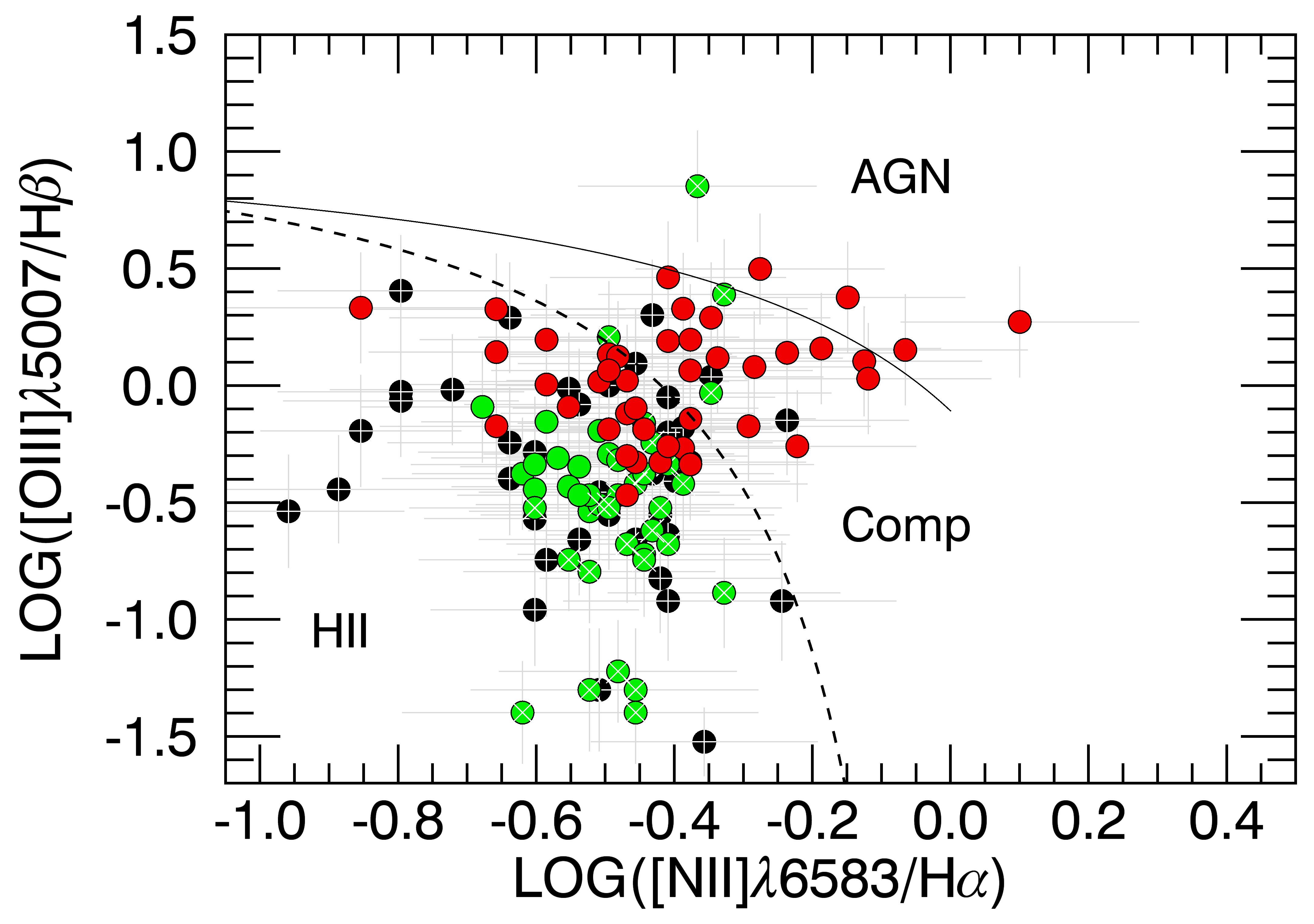}
	\includegraphics[width=\columnwidth]{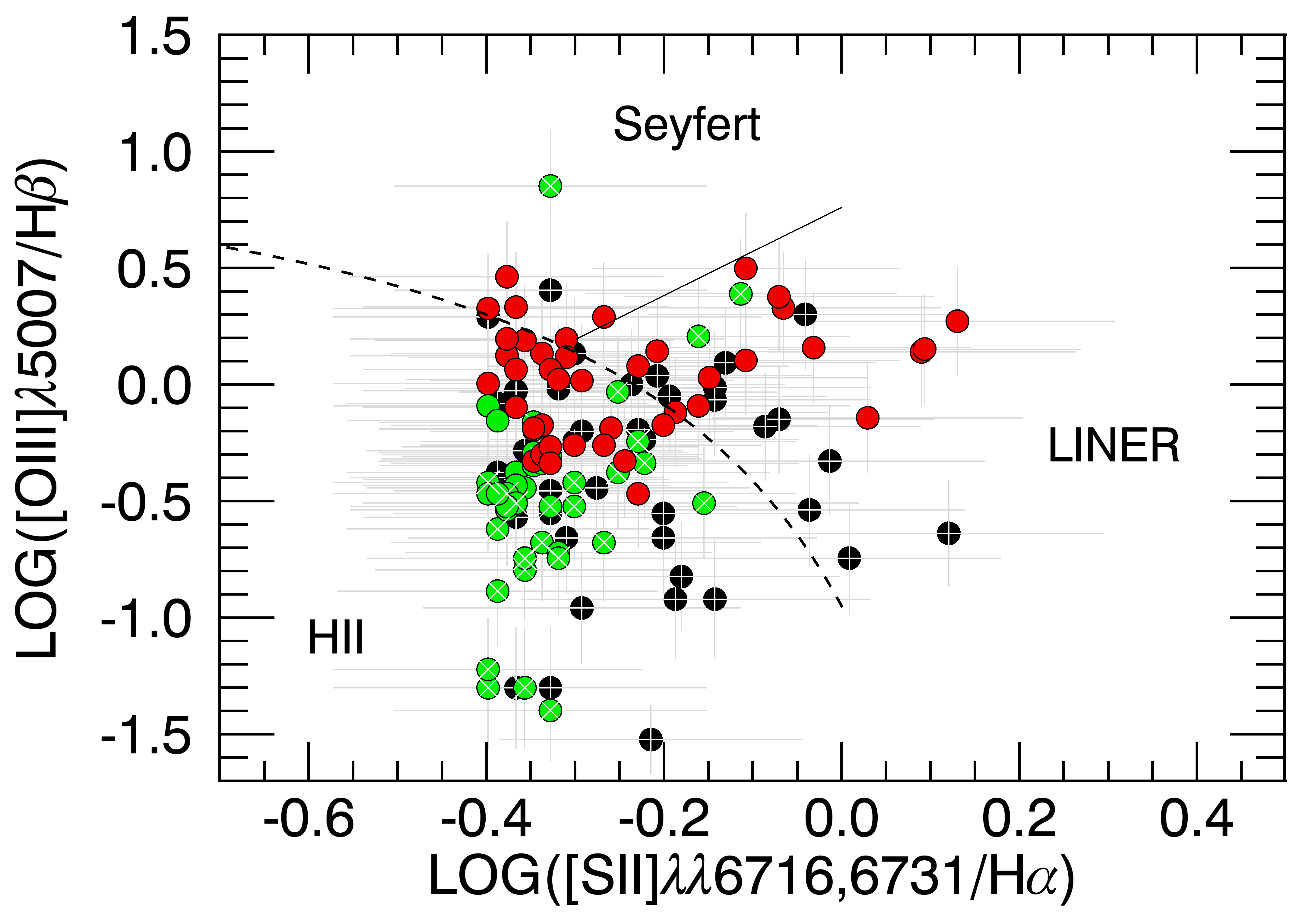}
    \caption{BPT-NII and BPT-SII diagrams of the 129 SNR candidates found in NGC\,3344. Colors and symbols are the same as presented in Figure~\ref{fig:sabaddin1}. The domain for the H\II\ regions (dotted line), the composite/transition regions, and pure shock regions labelled LINERS, Seyfert, and AGN (continuous line) are indicated based on the work of \citet{2003MNRAS.346.1055K} and \citet{2001ApJ...556..121K}, respectively. 
    }
    \label{fig:BPT1}
\end{figure}

The second rule in confirming the nature of the SNR candidates is for them to fall outside the H\II\ region zone in at least one of the BPT diagrams that we can draw with the available data.
Again, if a background residue is present, it may move the
SNR away from the shock region and, therefore, the constraint is put for only one BPT diagram.
Figure~\ref{fig:BPT1} shows the BPT-NII and BPT-SII diagrams, 
where the filled circles (with different colors for the final classification as discussed in \S~\ref{sec:fc}) represent our 129 SNR candidates. In these diagrams, the domains for the H\II\ regions, shocks, LINERS, and Seyfert are indicated based on the work of \citet{2003MNRAS.346.1055K} and \citet{2001ApJ...556..121K}. As in  Section~\ref{sec:sabaddin},  circles with a `$+$' symbol indicate candidates with S/N\,$<$\,3 for the [N\II] lines. Circles with a `$\times$' symbol are used in the case of candidates with S/N\,$<$\,3 for the [O\III]$\lambda$5007 or H$\beta$ lines (and the selection criterion for the SNR candidates already implied S/N\,$\geq$\,5 for the H$\alpha$ and [S\II] lines). Along with the 42 regions with a `$+$' symbol, there  are 29 regions with a `$\times$' symbol. 
The remaining 42 candidates (with S/N\,$\geq$\,3 for H$\beta$, [O\III], and [N\II]), fall (considering the line ratio uncertainties) inside the zone for shock regions in, at least, one of the two BPT diagrams. No foreground AGN is known in the field of view which could correspond to any of our SNR candidates.

%%%%%%%%%%%%%%%%%%%%%%%%%%%%%%%%%%%%%%%%%%%%%%%%%%
\subsection{Final Classification}
%%%%%%%%%%%%%%%%%%%%%%%%%%%%%%%%%%%%%%%%%%%%%%%%%%
\label{sec:fc}

If both the Sabbadin plots and BPT diagrams related classification rules are satisfied, the candidate falls in the category of Confirmed SNRs, and it is represented by a filled circle in red in Figures~\ref{fig:sabaddin1} and \ref{fig:BPT1}. If the candidate is within an SNR domain in at least one of the three Sabbadin plots and is not satisfying the rule for the BPT diagrams, it is considered as a Probable SNR and it is shown in green in the figures. If the candidate is outside the SNR domain in all five plots (with the required signal-to-noise ratio for the lines), it is a Less likely SNR and it is represented in black in the figures. From the start, all 129 candidates have a ratio [S\II]/H$\alpha$\,$\ge$\,0.4, but only 42 end as Confirmed SNRs, 45 as Probable SNRs, and 42  are called Less likely SNRs.

Because of their poor signal-to-noise ratio, many candidates (with `$+$' or `$\times$' symbols in Fig.~\ref{fig:sabaddin1} and \ref{fig:BPT1}) have been put in the category of Less likely SNRs and Probable SNRs, despite the fact that some of them are located in the SNR zones in Sabbadin plots and BPT diagrams. A deeper observation could help confirm their category as, for example, an old SNR may indeed present a very weak [O\III]$\lambda$5007 line.

On the other hand, Less likely SNRs that fall in the domain for H\II\ regions in the BPT diagrams (Fig.~\ref{fig:BPT1}) and in the Sabbadin plots (Fig.~\ref{fig:sabaddin1}) raise the question of some remaining contamination by the immediate environment: are we looking at SNR superposed with an H\II\ region, are we confusing an H\II\ region bathing into the DIG with an SNR, or are we seeing clumps in the DIG?
For the three SNR categories, we actually find no significant difference between the average distance ($\sim$\,$7.5 \pm 2.5$~pixels) between the SNR and the closest emission peak (in H$\alpha$ and [S\II]) or between the average number ($\sim$\,$8 \pm 3$~pixels) of emission peaks surrounding an SNR within a radius of 15 pixels.

For the line ratios presented in Figures~\ref{fig:sabaddin1} and \ref{fig:BPT1}, we have subtracted a global galaxy background (GGB) which may not be ideal to consider the immediate environment of the regions, but, as we have shown in Figure~\ref{fig:bg3}, it displayed  less dispersion compared to the local galaxy background (LGB).

%%%%%%%%%%%%%%%%%%%%%%%%%%%%%%%%%%%%%%%%%%%%%%%%%%
%%%%%%%%%%%%%%%%%%%%%%%%%%%%%%%%%%%%%%%%%%%%%%%%%%
\section{SNR Properties} %enlever le s a SNR
%%%%%%%%%%%%%%%%%%%%%%%%%%%%%%%%%%%%%%%%%%%%%%%%%%

%%%%%%%%%%%%%%%%%%%%%%%%%%%%%%%%%%%%%%%%%%%%%%%%%%
\subsection{Shock Models}
%%%%%%%%%%%%%%%%%%%%%%%%%%%%%%%%%%%%%%%%%%%%%%%%%%

Shock models have been developed by various teams and are often used to study the physical properties of SNRs. The models from \cite{2008ApJS..178...20A}  and \cite{1984ApJ...276..653D} have been of particular interest for us.  \cite{2008ApJS..178...20A} produced a library of fully radiative shock models calculated with the 
photoionization and shocks code MAPPINGS III. This code produces grids
of models taking into account different emission line ratios with shock velocities in the range [100-1000]~km~s$^{-1}$ and the magnetic parameter \textit{B/n$^{1/2}$} between 10$^{-4}$ to 10~$\mu$G cm$^{3/2}$ for five different atomic abundance sets and for a preshock density of 1.0~cm$^{-3}$. \cite{1984ApJ...276..653D} derived an extensive grid of models for fixed physical conditions appropriate to a typical evolved SNR.

\begin{figure}
	\includegraphics[width=\columnwidth]{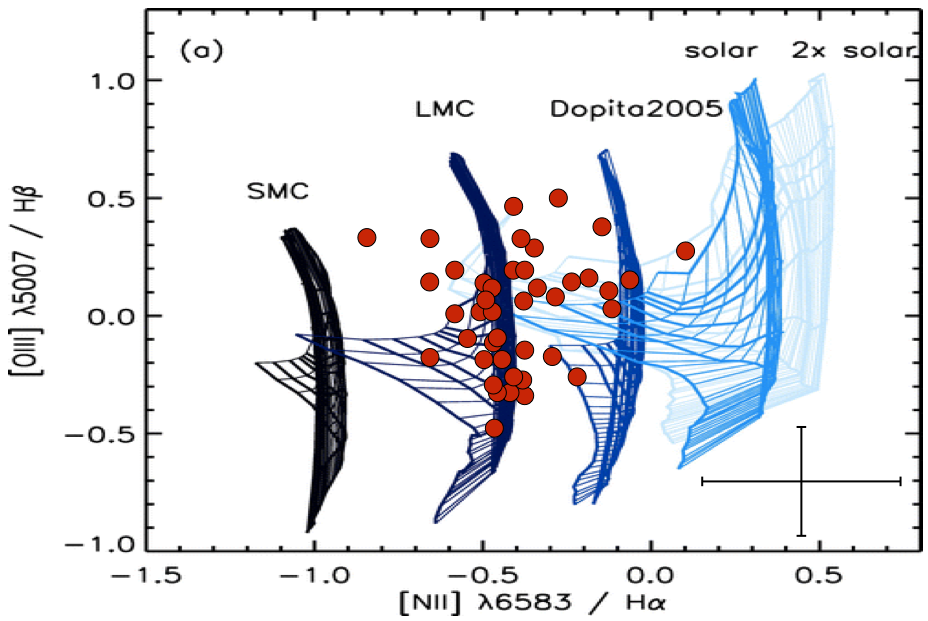}
	\includegraphics[width=\columnwidth]{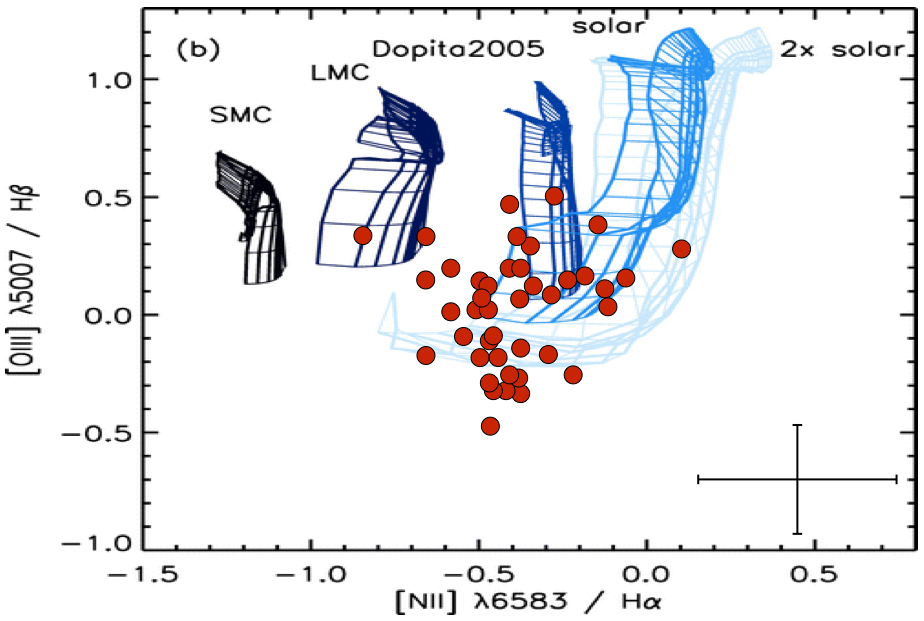}
    \caption{[O\III]$\lambda$5007/H$\beta$ versus [N\II]$\lambda$6583/H$\alpha$ based on  
    (a)~the shock-only and (b)~the shock+precursor models of \citet{2008ApJS..178...20A} 
    for five abundances (SMC, LMC, Dopita2005, solar, and \mbox{2 $\times$ solar}, 
    as indicated on the plots) with the density $n = 1$~cm$^{-3}$. 
    Vertical lines represent different values of 
    the magnetic parameter, from 10$^{-4}$ to 10 $\mu$G
    cm$^{3/2}$,
    while horizontal lines represent the shock velocity, from 200 to 1000~km\,s$^{-1}$, % with a step of 50~km\,s$^{-1}$,
    } 
    from the top to the bottom for the shock-only model and from the bottom to the top for the shock+precursor model. 
    In both models, while different values of shock velocity and magnetic field may be considered, a very low metallicity (i.e. SMC type) is excluded.  
    \label{fig:ModelsAllen}
\end{figure}

In Figure~\ref{fig:ModelsAllen}, we present grids from \citet{2008ApJS..178...20A} for the emission line ratios [O\III]$\lambda$5007/H$\beta$ versus [N\II]$\lambda$6583/H$\alpha$ for the shock-only and shock+precursor models considering different abundances.
The first plot (shock-only) represents the physical extreme case of the absence of ionized gas ahead of the shock front,  while the second plot (shock+precursor) represents the case of having an extensive, radiation bounded, precursor region ahead of the shock. For these grids, the vertical lines represent different values of the magnetic field, while the horizontal lines represent the shock velocity. This figure indicates a strong dependence of the emission lines ratio [N\II]$\lambda$6583/H$\alpha$ with the metallicity, while the [O\III]$\lambda$5007/H$\beta$ is more sensitive to the shock velocity. Our sample of Confirmed SNRs have been superposed on these plots. 
In both models, while different values of the magnetic field may be considered, a very low metallicity (i.e. SMC type) is excluded and a low value of the shock velocity is favored. More particularly in the case of the shock+precursor model, a shock velocity below 250~km\,s$^{-1}$ is prescribed for the Dopita2005, solar, and 2$\times$solar metallicity.  With the shock+precusor model, many Confirmed SNRs are best reproduced with Dopita2005, solar, the 2$\times$solar metallicity grid which is consistent with the value found in literature (12 + LOG(O/H) = 8.72) for H\II\  regions \citep{2014AJ....147..131P}.
The model of \citet{2008ApJS..178...20A} also predicts a weak  [O\III]$\lambda$5007/H$\beta$ line ratio when the shock velocity is small.
The absence of the [O\III]$\lambda$5007 as a result of shocks with a low velocity is
also supported by the model of \citet{1987ApJ...316..323H}. Therefore some of the candidates in the Probable SNRs category due to a week [O\III] line with a poor signal-to-noise ratio, should be reconsidered with deeper observations. 
Finally, shock waves in SNRs evolve from being non-radiative to radiative where their velocity becomes lower. For Galactic SNRs, such as Cygnus Loop, proper motions of the filaments show a complicated structure and one can observe both non-radiative and radiative shocks depending on the environment in which these filaments expand \citep{2014ApJ...791...30M, 2009ApJ...702..327S}. Using  optical data, we can distinguish radiative from non-radiative shocks by looking at their H$\alpha$ and [S\II] emission. While non-radiative shocks would be seen only in H$\alpha$, radiative shocks would show up in both filters. In our sample, all the SNR candidates seem to be in the radiative shock phase since they have emission in both H$\alpha$ and [S\II] emission with a signal-to-noise higher than 5.

%Figure 16
 \begin{figure}
	\includegraphics[width=\columnwidth]{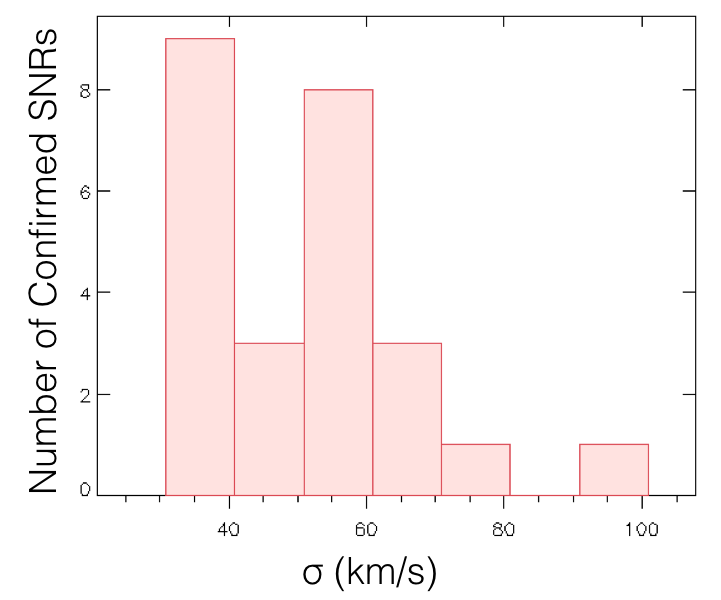}
    \caption{
    Histogram of the H$\alpha$ velocity dispersion $\sigma_v$ for the Confirmed SNRs}
    \label{fig:HistoDisp}
\end{figure}

 \begin{figure*}
 	\includegraphics[width=\textwidth]{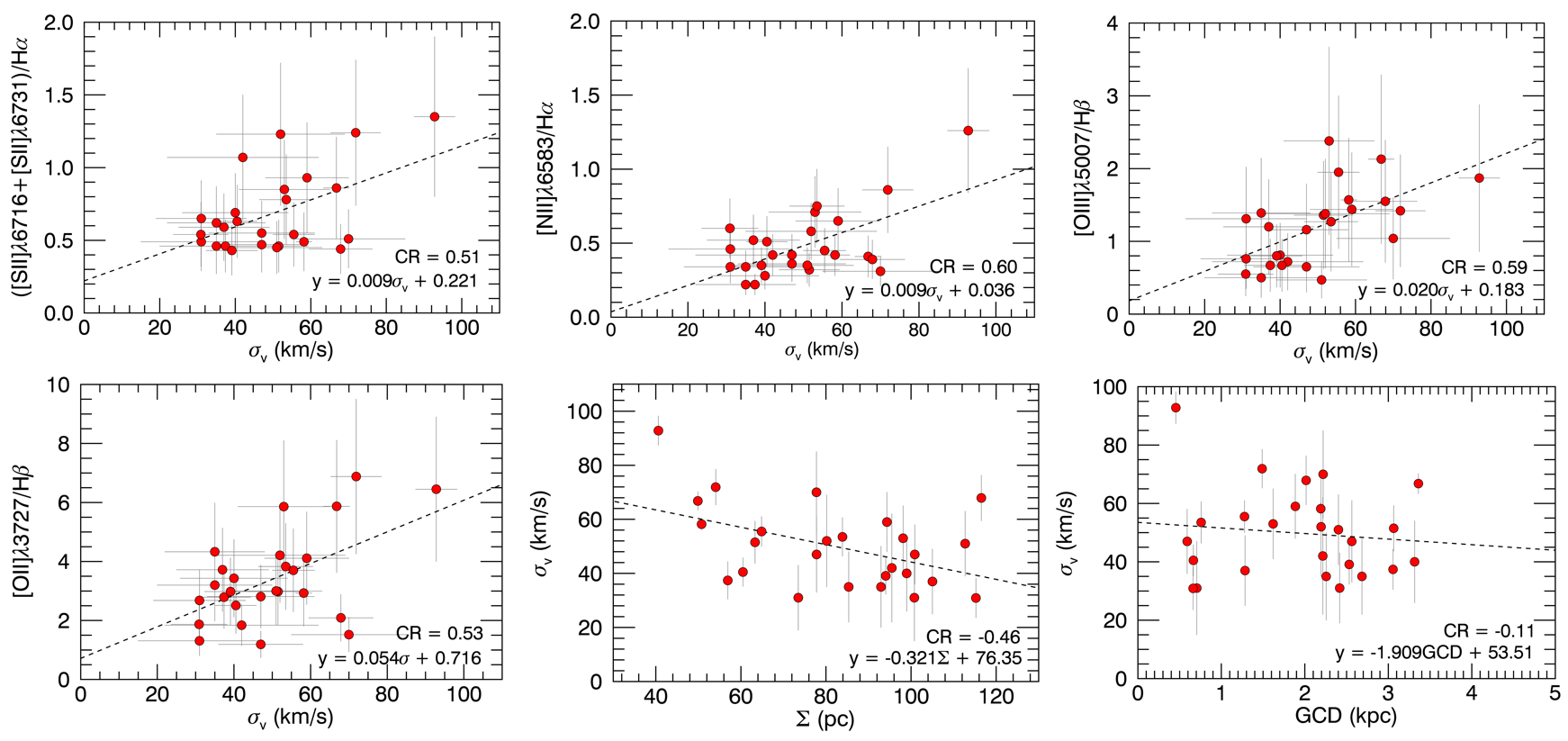}
    \caption{Relations between the H$\alpha$ velocity dispersion $\sigma_v$ of the Confirmed SNRs and their emission lines ratios, size~($\Sigma$) and galactocentric distance (GCD). The parameters of the linear fit (dotted line), along with the correlation coefficient ($CR$), are indicated for each relation.}
    \label{fig:Ha-sigma}
\end{figure*}

%Figure 16

\begin{figure*}
	\includegraphics[width=0.7\textwidth]{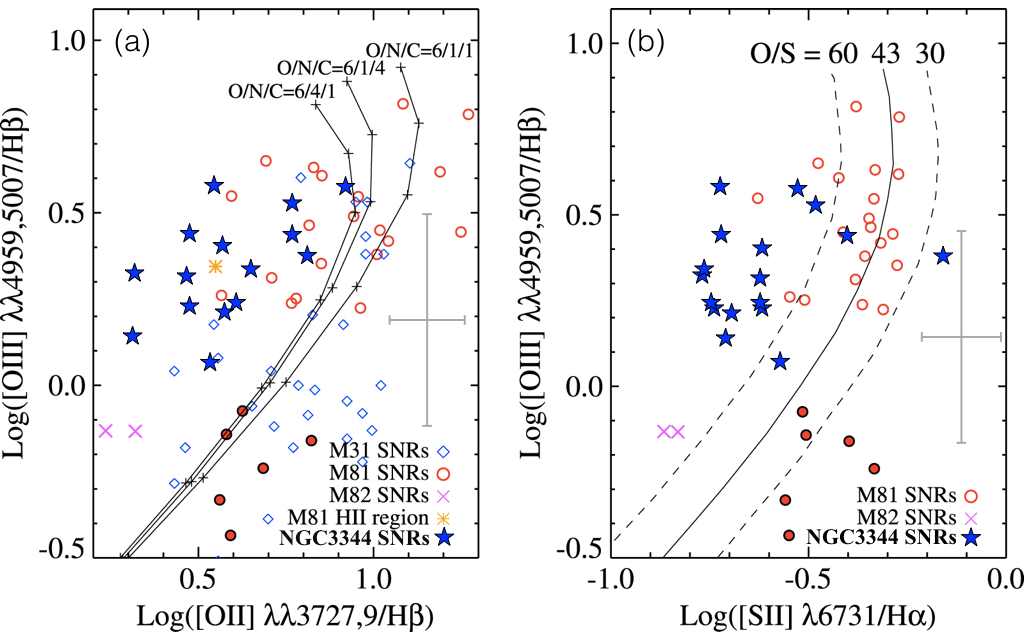}\\
	\includegraphics[width=0.7\textwidth]{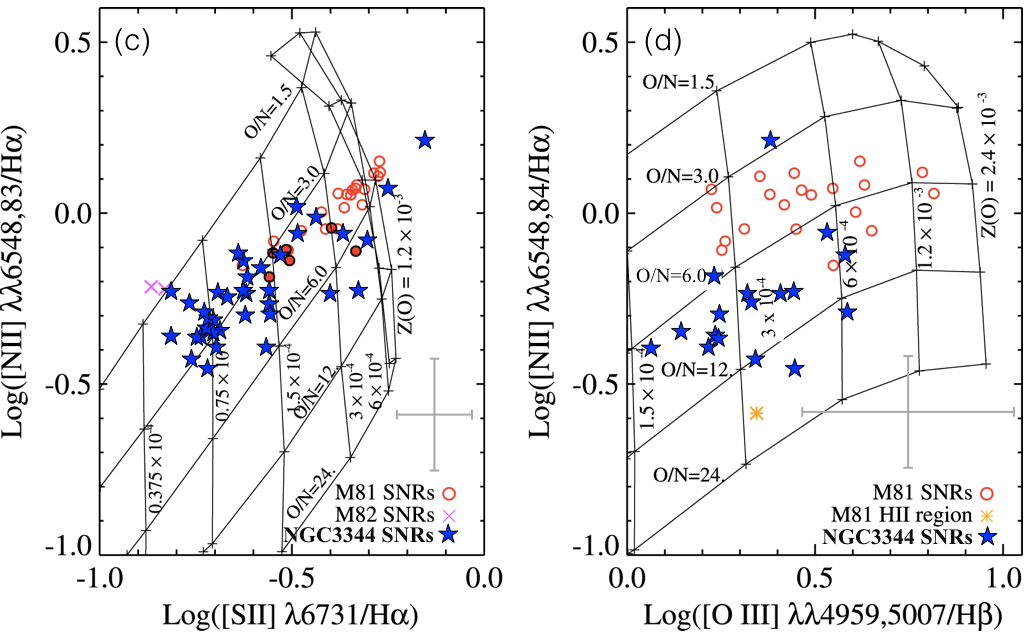}
    \caption{
    Emission line ratios based on the model of Dopita et al. (1984) for a shock velocity 
    of 106~km\,s$^{-1}$. (a) for a fixed abundance ratio O/S=42.8 and different abundances O/N/C as indicated. (b) for different ratios O/S as indicated. (c) and (d) for a fixed abundance 
    ratio O/S=42.8 and different values of O/N and Z(O) as indicated. SNRs from M81 and M82 
    (Lee et al. 2015) along with SNRs from M31  (Galarza et al. 1999) and our 
    Confirmed SNRs in NGC\,3344 are superposed to the curves and grids.
    }
    \label{fig:dopita}
\end{figure*}

Figure~\ref{fig:dopita} shows curves and grids of different emission lines ratios by \cite{1984ApJ...276..653D} in the case of a weak shock velocity of 106~km\,s$^{-1}$. 
The curves and grids take into account different abundances O/N/C, O/S, and O/N, and metallicities Z(O). The first plot makes use of the [O\II]$\lambda$3727 line. For comparison, SNRs identified in M81 and M82 \citep{2015ApJ...804...63L} and in M31 \citep{1999AJ....118.2775G} have been superposed on the different plots, when possible, along with our Confirmed SNRs. Our SNRs sample do not match the curves in the top plots (where a fixed abundance ratio O/S=42.8 was considered). As this is also the case for the M31 SNRs,
\citet{2015ApJ...804...63L} proposed that [N\II]/H$\alpha$ vs. [O\III]/H$\beta$  diagram may be better for the estimation of the SNR abundances. As shown in the bottom plots of Figure~\ref{fig:dopita}, this may be the case since most of the objects falls within the model grids, allowing to derive the O/N abundance and Z(O). In this case, using [N\II]/H$\alpha$ vs. [O\III]/H$\beta$, our Confirmed SNRs have an O/N abundance between  6 and 12 and Z(O) varying from 1.5$\times$10$^{-4}$ to 6$\times$10$^{-4}$.  The O/N abundance seems to be higher than the central O/N obtained from \citep[][O/N $\simeq$ 4.36]{2014AJ....147..131P}. However, this value is consistent with values found using [N\II]/H$\alpha$ vs. [S\II]$\lambda$6731/H$\alpha$ (O/N between 1.5 and 6). 

Finally, other shock models by \cite{1985ApJ...298..651C} and \cite{1987ApJ...316..323H} 
proposed that a value of the ratio [O\III]$\lambda$5007/H$\beta$ bellow 6 indicates shocks with complete recombination zones and values higher than 6 are expected to be found for shocks with incomplete recombination zones \citep[which means that the entire postshock cooling zone is not fully established;][]{1988ApJ...324..869R}. As we can see in the Figure~\ref{fig:BPT1}, all the Confirmed SNRs in our sample have a ratio [O\III]$\lambda$5007/H$\beta$ bellow 6 (the highest [O\III]$\lambda$5007/H$\beta$ ratio is $\sim$\,1) which means we are seeing shocks with complete recombination zones.

Figure \ref{fig:HistoDisp} presents the distribution of the velocity dispersion $\sigma_{v}$ measured with the H$\alpha$ line for the confirmed SNRs.  
Values of $\sigma_{v}$ are ranging from 31 to 93 km\,s$^{-1}$ with two peaks at 35 and 55 km\,s$^{-1}$. In Figure \ref{fig:Ha-sigma}, we explored the relation between $\sigma_{v}$ and the galactocentric distance of the regions, their size $\Sigma$, and their different emission lines ratios.
The correlation coefficient for these relations are rather small.  The most significant relations are found with the emission line ratios %[S\II]/H$alpha$,
[N\II]$\lambda$6583/H$\alpha$ and [O\III]$\lambda$5007/H$\beta$.

%%%%%%%%%%%%%%%%%%%%%%%%%%%%%%%%%%%%%%%%%%%%%%%%%%
\subsection{SNR sizes} 
%%%%%%%%%%%%%%%%%%%%%%%%%%%%%%%%%%%%%%%%%%%%%%%%%%

\begin{figure}
\centering
    \includegraphics[width=0.7\columnwidth]{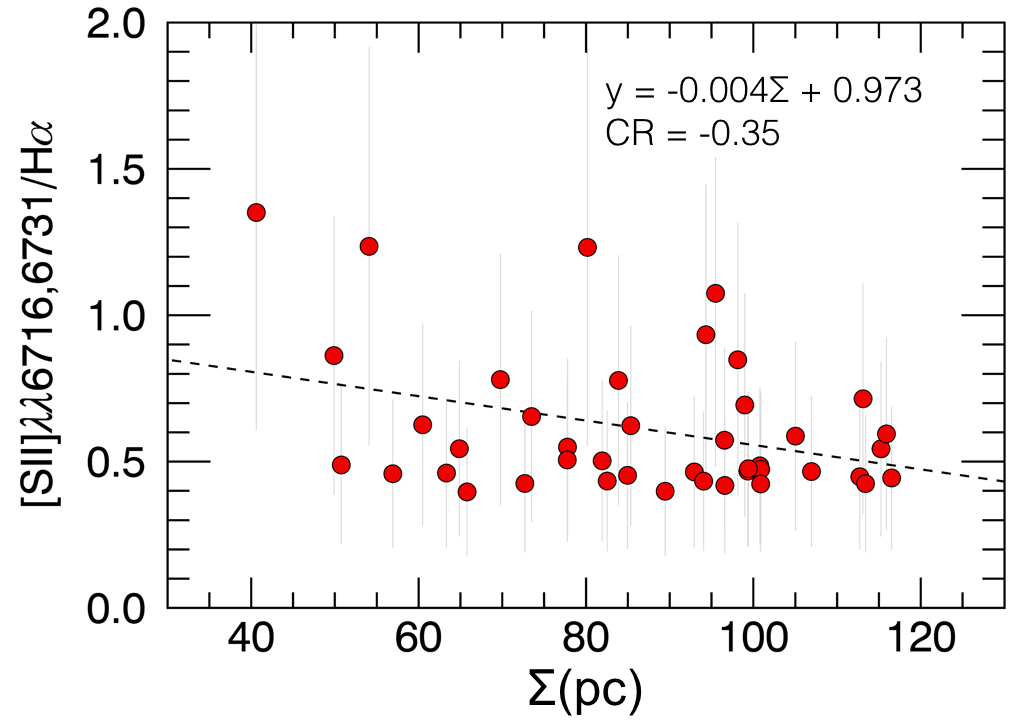}
    \includegraphics[width=0.7\columnwidth]{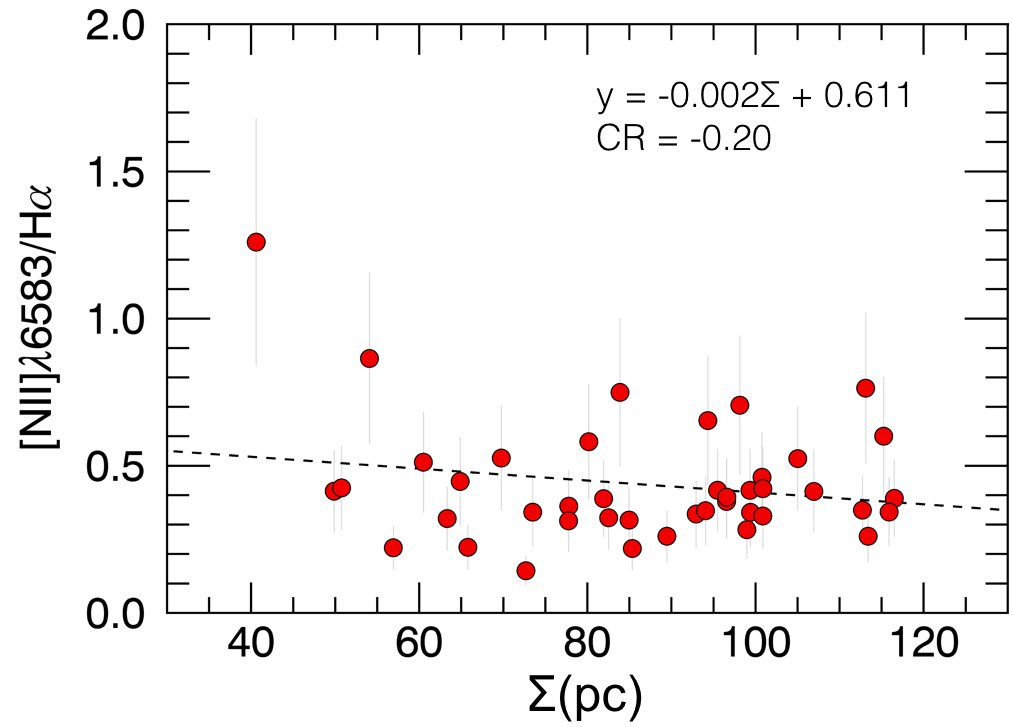}
    \includegraphics[width=0.7\columnwidth]{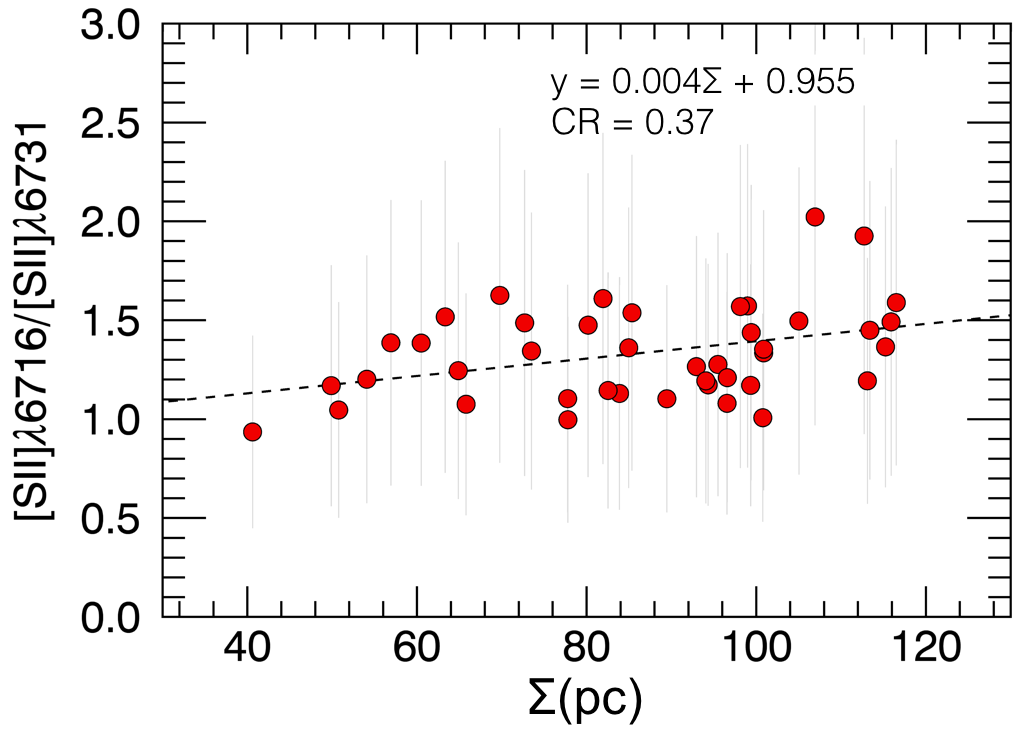}
    \includegraphics[width=0.7\columnwidth]{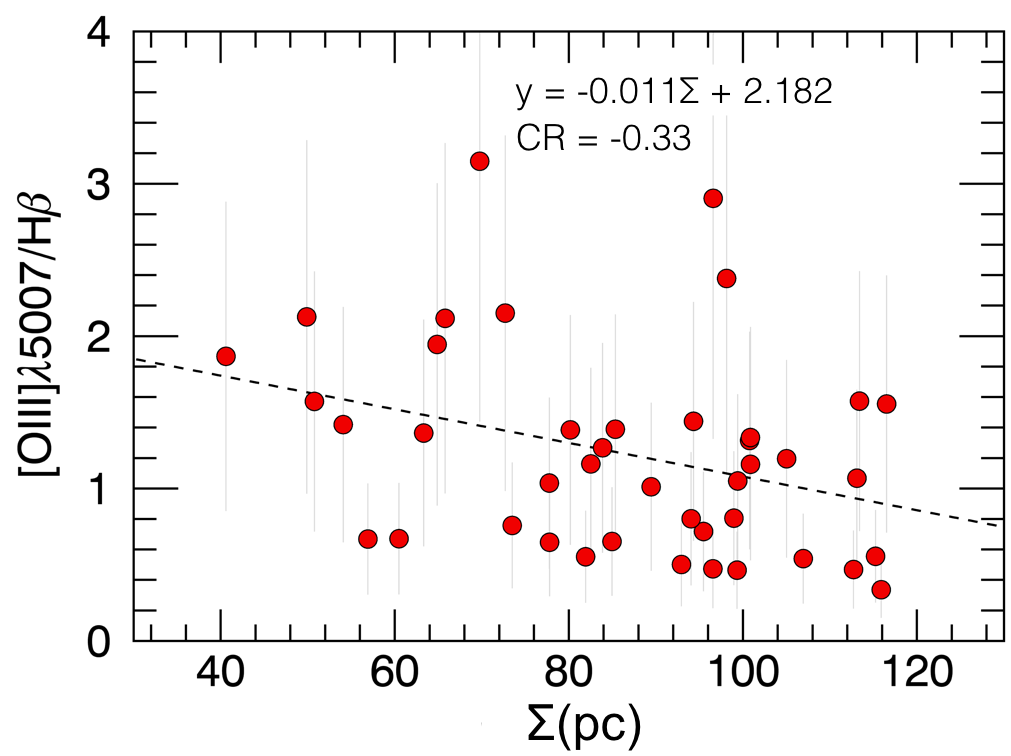}
    \includegraphics[width=0.7\columnwidth]{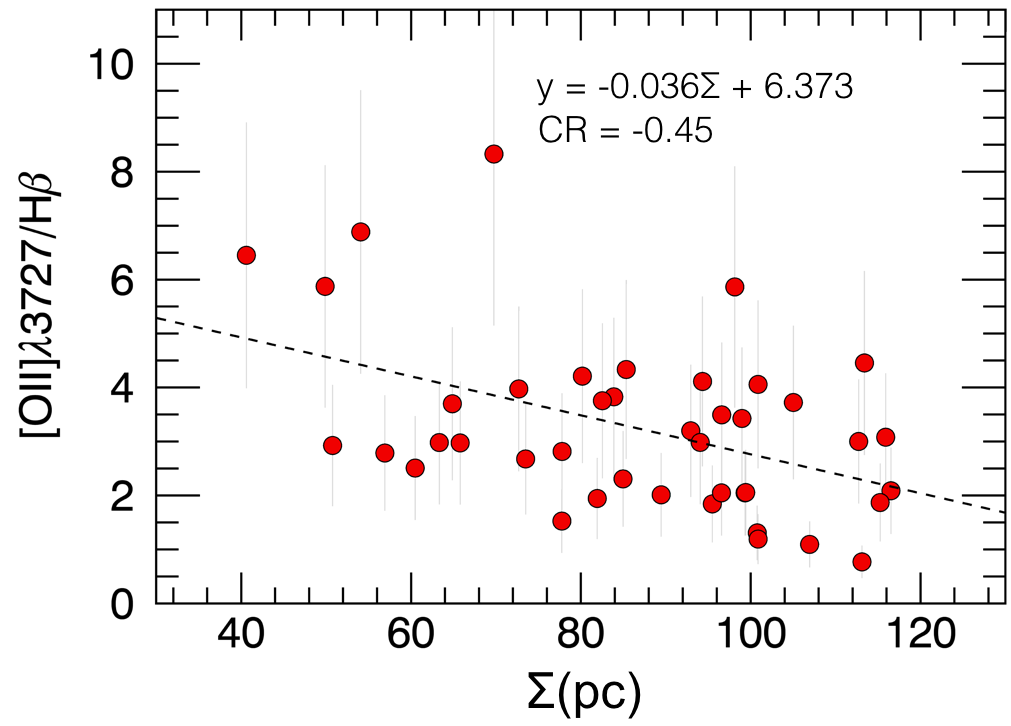} 
    \caption{Emision lines ratios for the Confirmed SNRs 
    as a function of $\Sigma$. The parameter $\Sigma$, a measurement of the SNR radius,  corresponds to the FWHM of the SNRs pseudo-Voight profile.  The dashed lines display a linear fit through the data. The fit parameters are given for each plot, along with the correlation coefficient ($CR$).
    %- as-tu regardé OII/OIII? : oui, CR = 0.09. pas de correlation
}
    \label{fig:sizerel}
\end{figure}

A measurement of the SNR radius $\Sigma$ is taken as the FWHM of the pseudo-Voight profile fitted for each region based on their [S\II]$\lambda$6716+[S\II]$\lambda$6731 emission  (\S~\ref{sec:auto}) considering that it is a good tracer for shocks. The spatial resolution obtained in this study (0.8$^{\prime\prime}$) allows us to identify SNR candidates with $\Sigma_{{\rm limit}} \ge 28.5$~pc. In our sample of Confirmed SNRs, the smallest one has $\Sigma \simeq 40$~pc, above this limit. As the size of the remnants may be an indication of its evolution \citep{2010MNRAS.407.1301B}, we have compared $\Sigma$ with the various emission lines ratios observed. In Figure~\ref{fig:sizerel}, we show the relation between $\Sigma$ and different emission line ratios: [S\II]/H$\alpha$, [N\II]/H$\alpha$, [S\II]$\lambda$6716/[S\II]$\lambda$6731,  [O\III]/H$\beta$, and also [O\II]/H$\beta$. Although we can superpose a linear fit to the data, revealing a negative slope for all ratios except [S\II]$\lambda$6716/[S\II]$\lambda$6731 where a positive slope is seen, the scatter is rather large (particularly when the weaker H$\beta$ line is involved) and the probability of a significant fit (as given in each plot) is rather small. Nevertheless, the trend displayed by the slopes is interesting and rather consistent with a smaller velocity (when [O\III]$\lambda$5007 is weaker) and a lower density (when [S\II]$\lambda$6716/[S\II]$\lambda$6731 is larger) in older SNR (i.e. when they are more extended according to $\Sigma$). 

%Figure 18

We also looked for a relation between $\Sigma$ and the position of the remnant withing the galaxy, wondering if the galactic environment has an affect on this parameter. More precisely, we draw a plot of $\Sigma$ as a function of the galactocentric distance, but we did not find any significant relation.
We also compared the average value of $\Sigma$ for SNRs located in different structures, like the inner ring, arms, and inter arms regions,  but again we did not find any important difference.

\begin{figure}
 \centering
   \includegraphics[width=0.7\columnwidth]{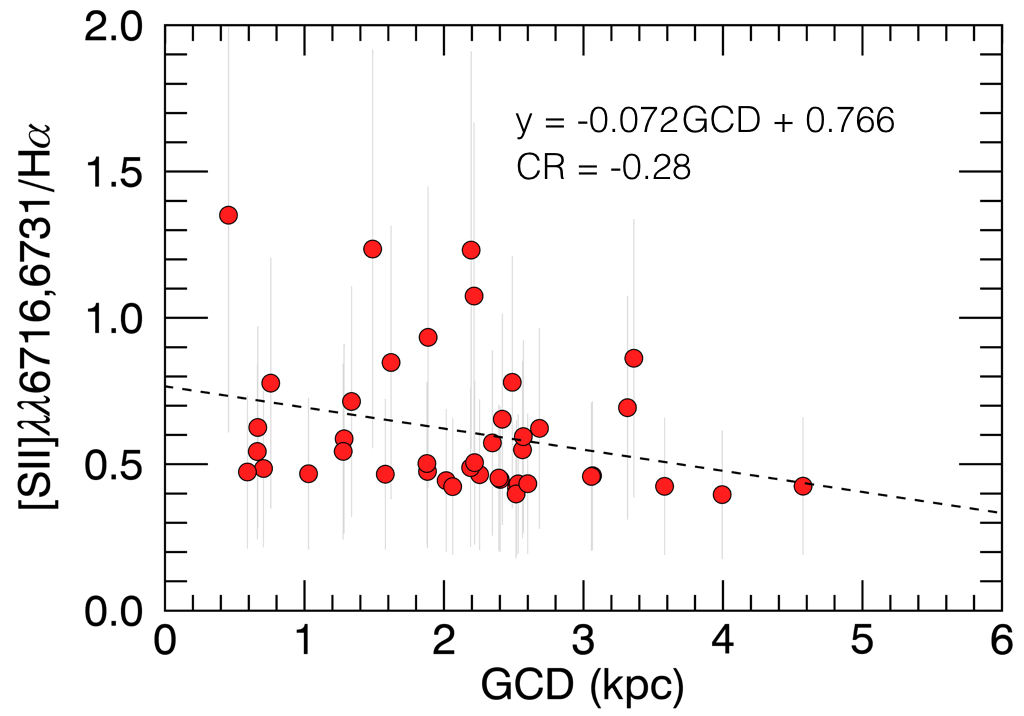}
    \includegraphics[width=0.7\columnwidth]{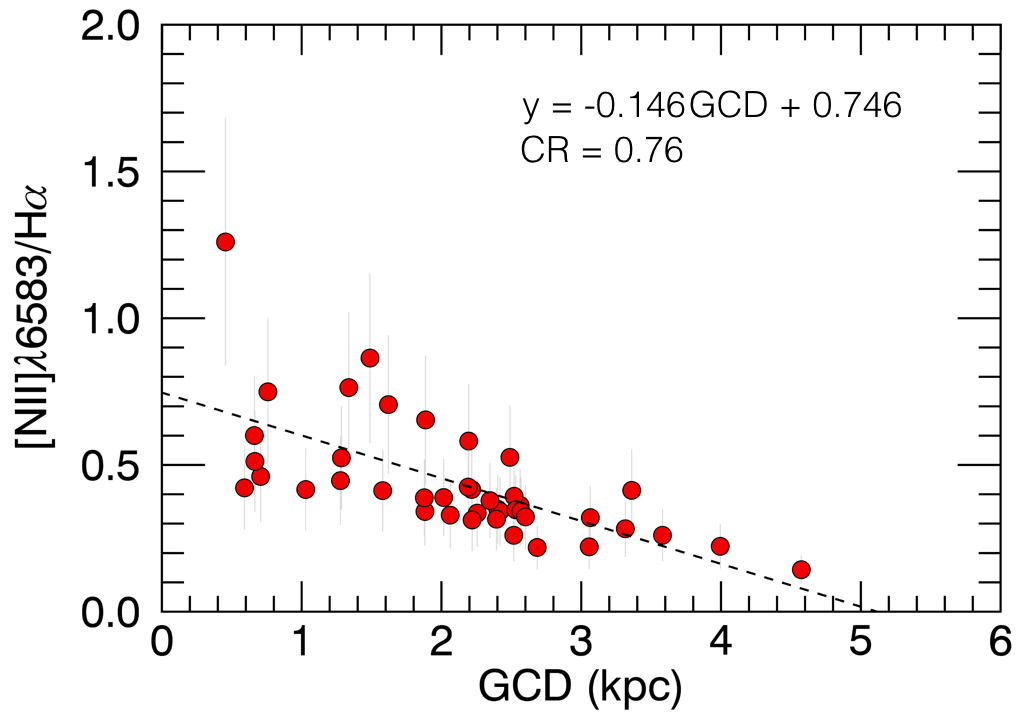}
    \includegraphics[width=0.7\columnwidth]{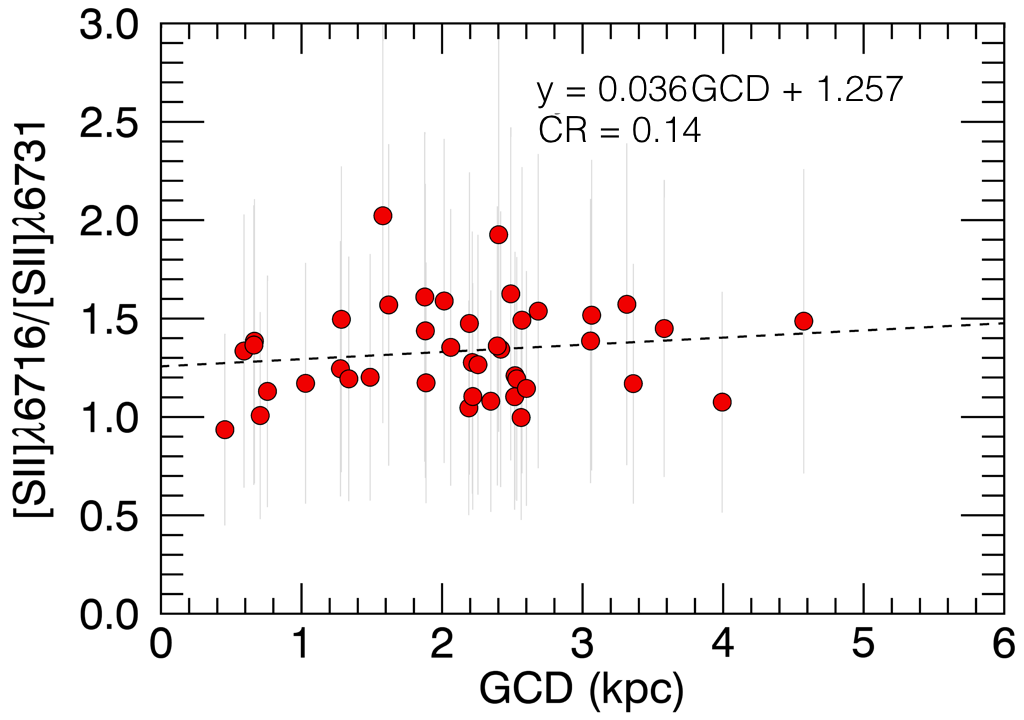}
    \includegraphics[width=0.7\columnwidth]{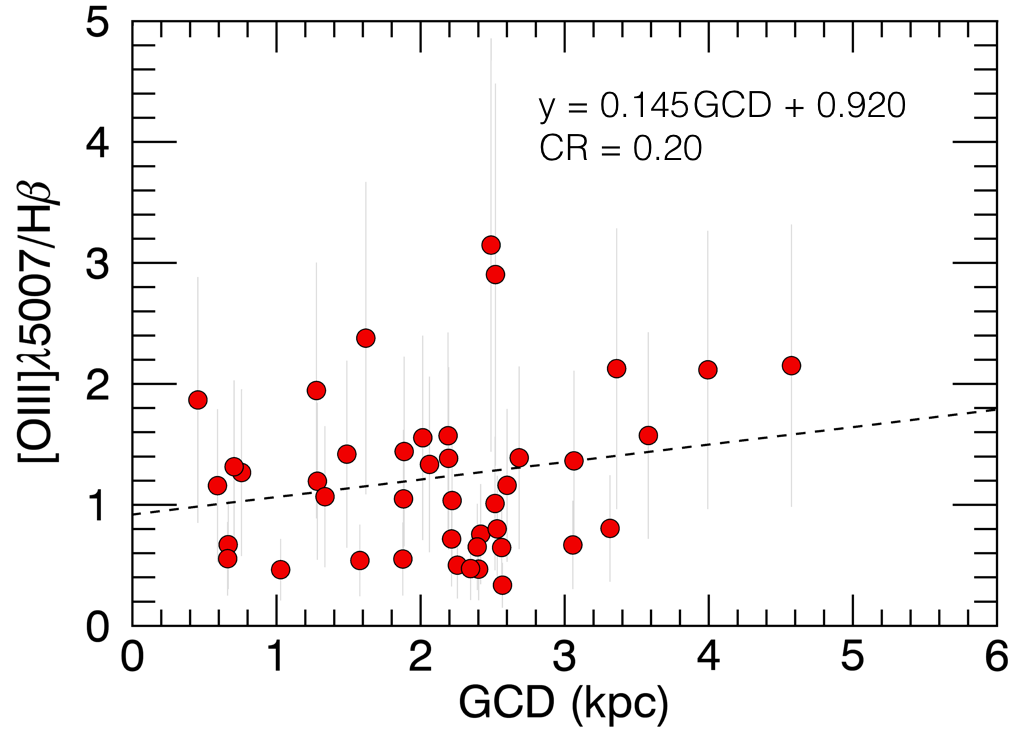}
    \includegraphics[width=0.7\columnwidth]{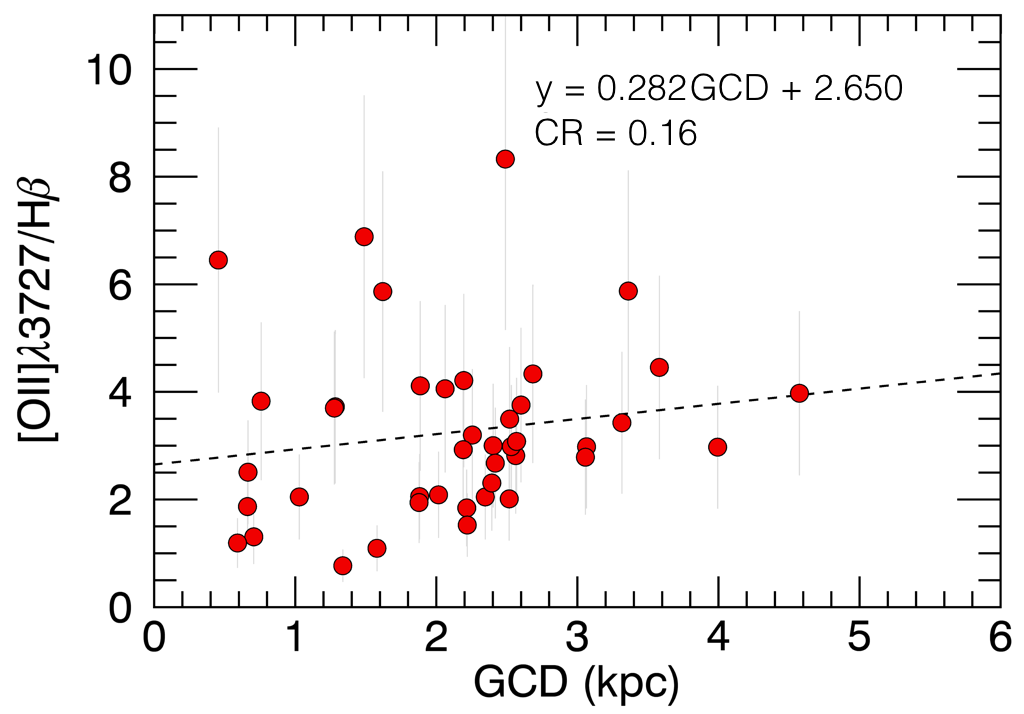}	  	
	\caption{Emission line ratios for the Confirmed SNRs 
	as a function of the galactocentric distance. 
    The dashed lines display a linear fit through the data. The fit parameters are given for each plot, along with the correlation coefficient ($CR$).
    }
    \label{fig:GCD-relation}
\end{figure}

Finally, we directly investigated the galactocentric variation of the different emission line ratios. As shown in Figure~\ref{fig:GCD-relation}, 
a negative gradient is seen for [S\II]/H$\alpha$ and [N\II]/H$\alpha$, while the other three ratios, [S\II]$\lambda$6716/[S\II]$\lambda$6731, [O\III]/H$\beta$ and [O\II]/H$\beta$, display a positive gradient. The plot with [N\II]/H$\alpha$ has less scatter and a higher probability, and may therefore suggests a higher metallicity for the SNRs near the galaxy centre.
This result is agreement with the strong abundance gradient for H\I\ regions found by \citep{1981PASP...93..273M} and confirmed by \citep{1988PASP..100.1428V}. 

%%%%%%%%%%%%%%%%%%%%%%%%%%%%%%%%%%%%%%%%%%%%%%%%%%
%%%%%%%%%%%%%%%%%%%%%%%%%%%%%%%%%%%%%%%%%%%%%%%%%%
\section{Conclusions}
%%%%%%%%%%%%%%%%%%%%%%%%%%%%%%%%%%%%%%%%%%%%%%%%%%

In this paper, we used the high spectral and spatial resolution data obtained with the iFTS SITELLE at the CFHT to study a sample of SNRs in the nearby galaxy NGC\,3344. Our conclusions are as follows:

\noindent 1) We report the first identification and confirmation of a sample of SNRs in NGC\,3344. Our systematic analysis using 
criteria (related to the  physics and morphology of the emission regions), reveals 129~SNR candidates which are classified in three categories: Confirmed (42), Probable (45), and Less likely (42) SNRs. 

\noindent 2) We present a self-consistent spectroscopic analysis, exploiting all the emission lines available with SITELLE and using the Sabbadin plots and BPT diagrams, to confirm the shock-heated nature of the ionization mechanism in the SNR candidates. Eventually, the [OI]$\lambda$6300 line should be added to this analysis, allowing to consider (i) the [OI]$\lambda$6300 emission as another tracer of SNRs and (ii) the BPT-OI diagram to complete the available list of diagrams confirming the mechanisms of ionization in the optical range.

\noindent 3) We have considered two methods to subtract the galaxy background spectrum from each emission region. 
One of these methods consisted in creating a 
local galaxy background (LGB) by considering a median spectrum in a small annulus centred on a region. For the other method we created a global galaxy background (GGB) for each region by considering in a ring centred on the galaxy, with a radius equal to the region galactocentric distance.
Although the LGB and GGB spectrum of a same region are similar in general (with more distinction for regions in the galaxy inner ring), we have adopted the GGB method because it may better represent, with less dispersion, the global stellar population in the disk along with the DIG component, which are quite important in this galaxy.

\noindent 4) After the background subtraction, the galaxy internal extinction seemed rather small, but this may require further investigation as the nature of the background is complex. 

\noindent 5) We have compared shock models from \citet{2008ApJS..178...20A}  with our emission line ratios obtained for the Confirmed SNRs. A metallicity ranging between LMC and 2$\times$solar is then prescribed for these objects. A comparison with the shock models of  \citet{1984ApJ...276..653D} reveal an O/N abundance between 6 and 12 and Z(O) varying from 1.5$\times$10$^{-4}$ to 6$\times$10$^{-4}$.

\noindent 6) We do not find a strong trend between the SNR $\Sigma$ parameter, i.e. a measurement of the radius of the remnants, and their emission lines ratios, except maybe that extended ones (i.e. older ones) have a lower density (i.e. higher [S\II]$\lambda$6716/[S\II]$\lambda$6731) and a lower velocity (i.e. weaker [O\III]$\lambda$5007 lines).
There are no obvious relation between $\Sigma$ and the SNR environments in the galaxy (i.e. their galactocentric radius and position inside the inner ring, arms, or between the arms). But we find a significant gradient withing the galaxy for the line ratio [N\II]/H$\alpha$. 
These correlations may indicate a metallicity gradient among
the SNR population, along with some evolutionary effect (age and shock velocity).

Although SITELLE is a powerful tool for the systematic identification and optical study of SNRs, multi-wavelength observations will also be necessary to have a complete picture of these objects, mainly to improve our understanding of their physics and evolutionary phase, to gain a detailed description of their morphology and environment, and to evaluate their impact on the galaxy evolution. A study using near-infrared data obtained recently with the [Fe\II]\,1.64\,$\mu$m filter (MOIRCS/Subaru), X-rays archived data (Chandra), and high spatial resolution optical images (HST) is ongoing to complete our work on the SNRs population in NGC\,3344 (Moumen et al. in prep.). Furthermore, the SITELLE data are being reanalyzed to describe the H\II\ regions population of NGC\,3344, along with its recent star formation history and in relation with the SNRs described here (Moumen et al. in prep.).

%%%%%%%%%%%%%%%%%%%%%%%%%%%%%%%%%%%%%%%%%%%%%%%%%%
%%%%%%%%%%%%%%%%%%%%%%%%%%%%%%%%%%%%%%%%%%%%%%%%%%
\section*{Acknowledgements}
%%%%%%%%%%%%%%%%%%%%%%%%%%%%%%%%%%%%%%%%%%%%%%%%%%

This research is based on observations obtained at the Canada-France-Hawaii Telescope (CFHT) which is operated from the summit of Mauna Kea by the National Research Council of Canada, the Institut National des Sciences de l’Univers of the Centre National de la Recherche Scientifique of France, and the University of Hawaii. The observations at the Canada-France-Hawaii Telescope were performed with care and respect from the summit of Mauna Kea which is a significant cultural and historic site. The observations were obtained with SITELLE, a joint project between Universit\'e Laval, ABB-Bomem, Universit\'e de Montr\'eal and the CFHT with funding support from the Canada Foundation for Innovation (CFI), the National Sciences and Engineering Research Council of Canada (NSERC), Fonds de Recherche du Qu\'ebec - Nature et Technologies (FRQNT), and CFHT. LRN, CR, and LD are grateful to the Fonds de recherche du Qu\'ebec - Nature et Technologies (FRQNT) for individual and team financial support. CR and LD are grateful to the Natural Sciences and Engineering Research Council of Canada (NSERC). This research has made use of NASA’s Astrophysics Data System and of the VizieR catalogue access tool, CDS, Strasbourg, France.

%%%%%%%%%%%%%%%%%%%%%%%%%%%%%%%%%%%%%%%%%%%%%%%%%%
%%%%%%%%%%%%%%%%% REFERENCES %%%%%%%%%%%%%%%%%%

\bibliographystyle{mnras}
\bibliography{ref} % if your bibtex file is called example.bib

%\begin{thebibliography}{99}
%\end{thebibliography}

%%%%%%%%%%%%%%%%%%%%%%%%%%%%%%%%%%%%%%%%%%%%%%%%%%
%%%%%%%%%%%%%%%%% APPENDICES %%%%%%%%%%%%%%%%%%%%%
\appendix
%%%%%%%%%%%%%%%%%%%%%%%%%%%%%%%%%%%%%%%%%%%%%%%%%%

%%%%%%%%%%%%%%%%%%%%%%%%%%%%%%%%%%%%%%%%%%%%%%%%%%
\section{Flux maps}
%%%%%%%%%%%%%%%%%%%%%%%%%%%%%%%%%%%%%%%%%%%%%%%%%%
\label{Flux_maps_A}

\begin{figure*}
	\includegraphics[width=0.3\textwidth]{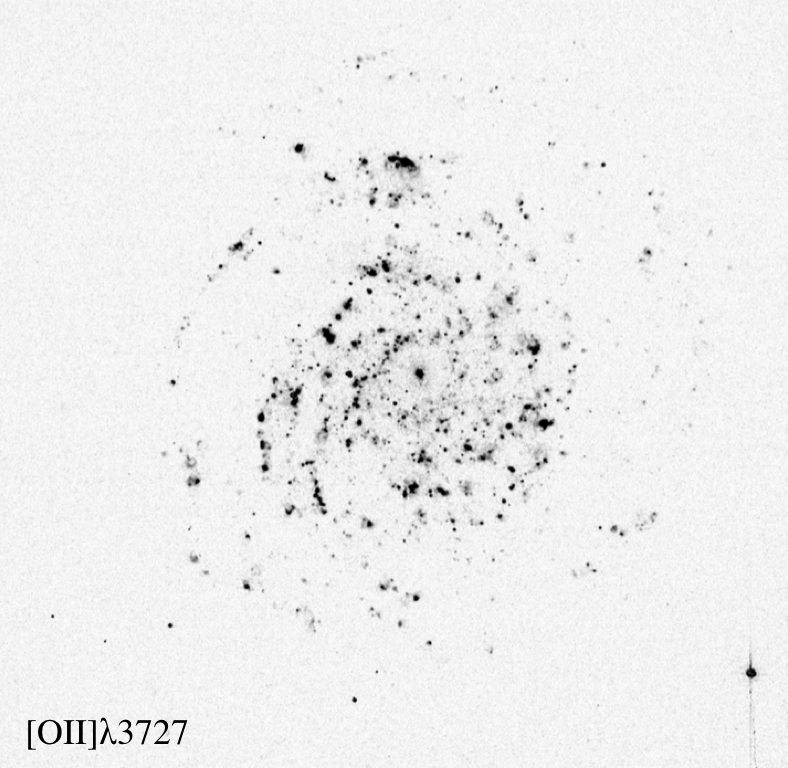}
	\includegraphics[width=0.3\textwidth]{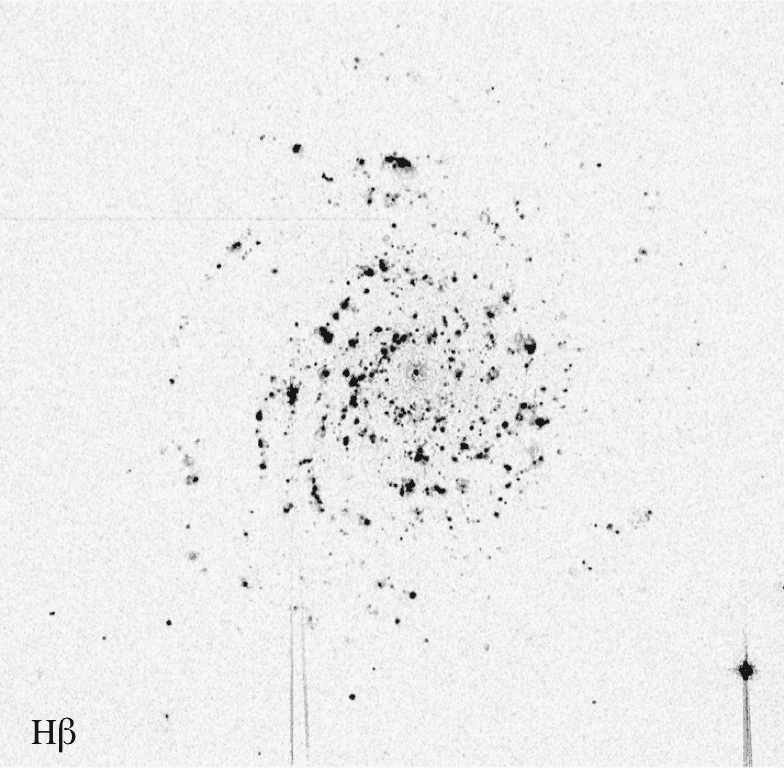}
	\includegraphics[width=0.3\textwidth]{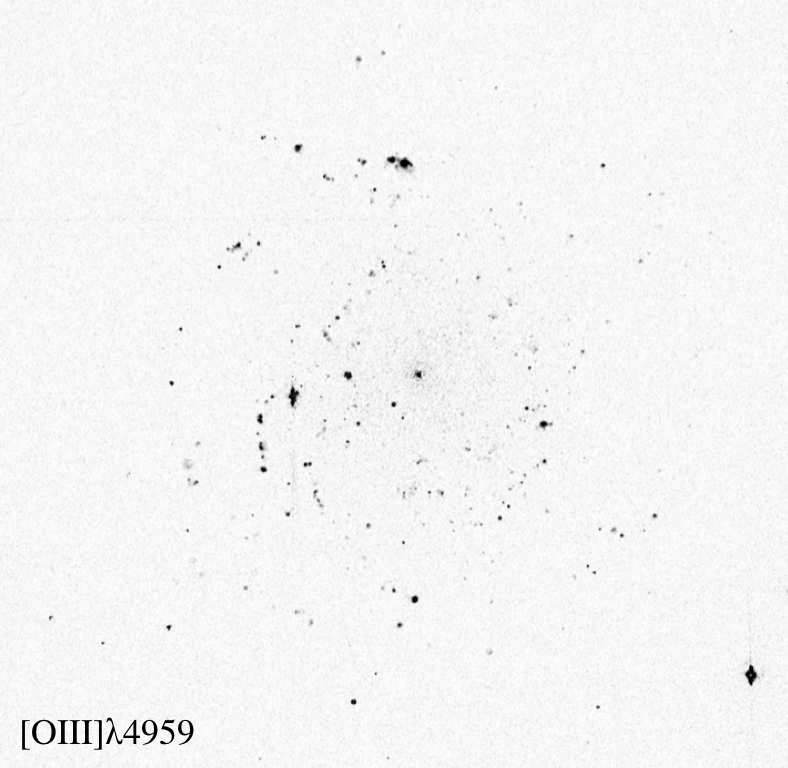}
	\includegraphics[width=0.3\textwidth]{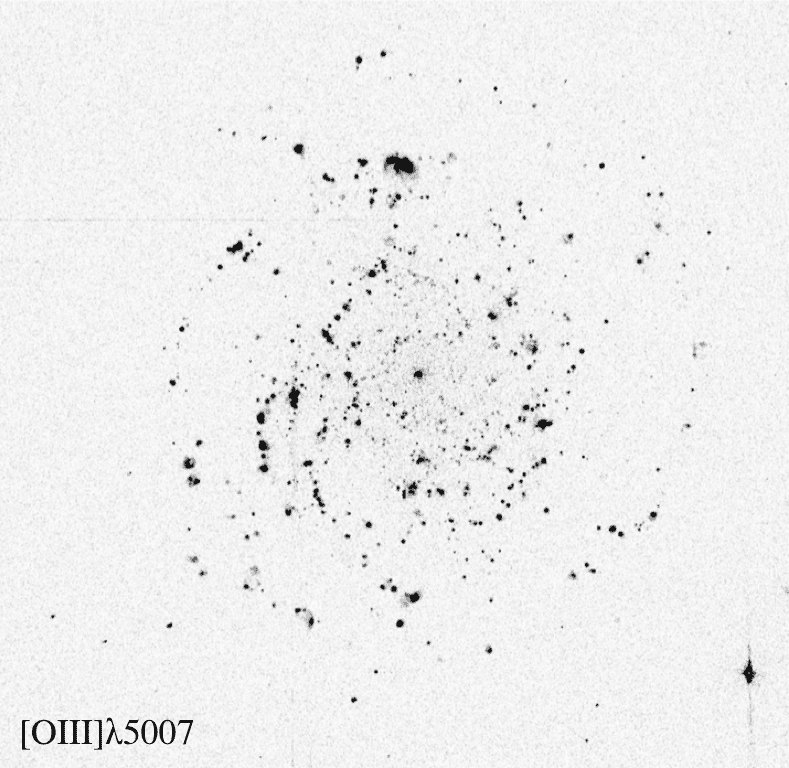}
	\includegraphics[width=0.3\textwidth]{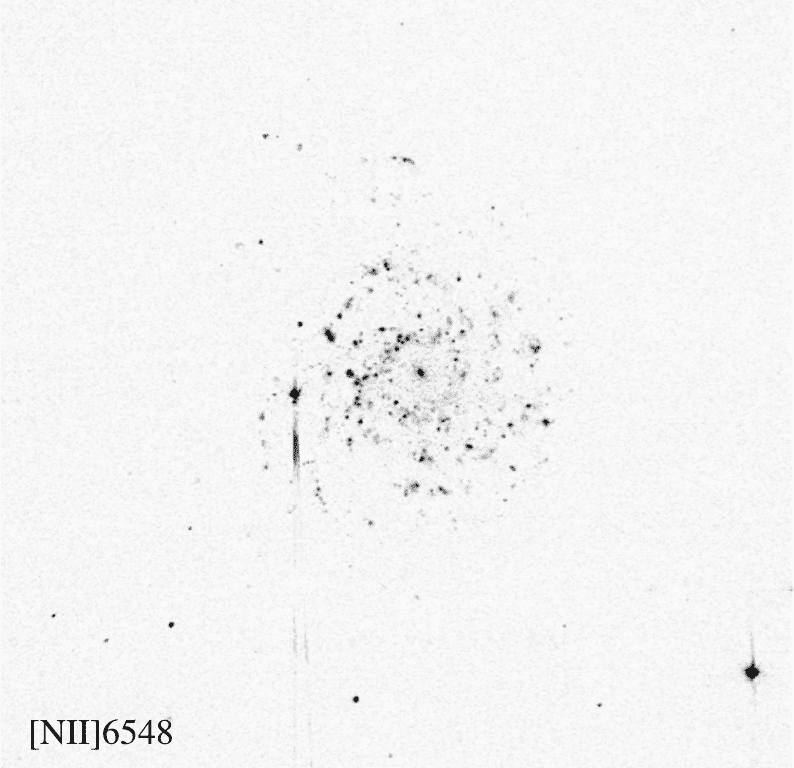}
	\includegraphics[width=0.3\textwidth]{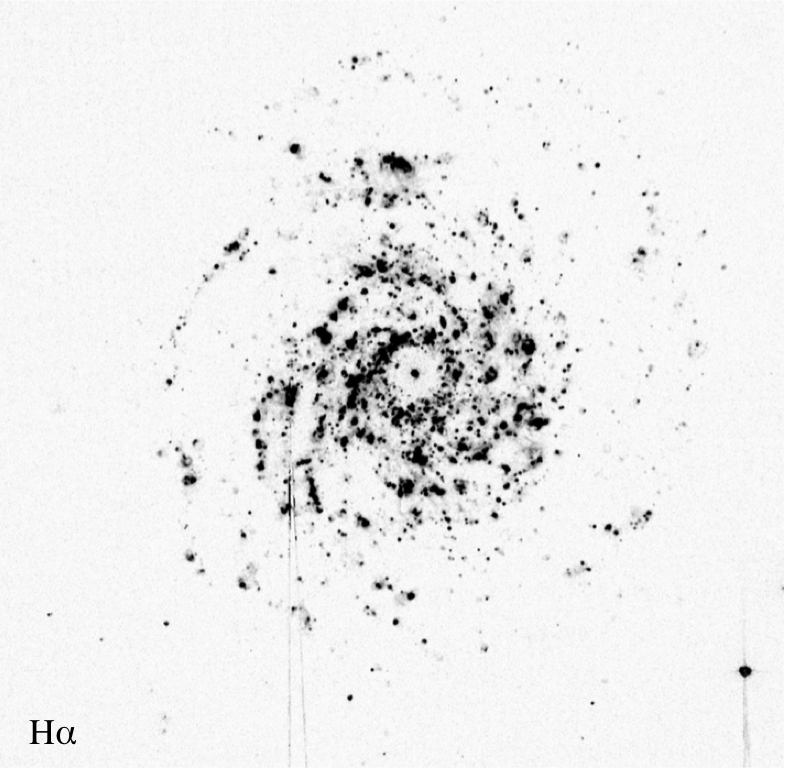}
	\includegraphics[width=0.3\textwidth]{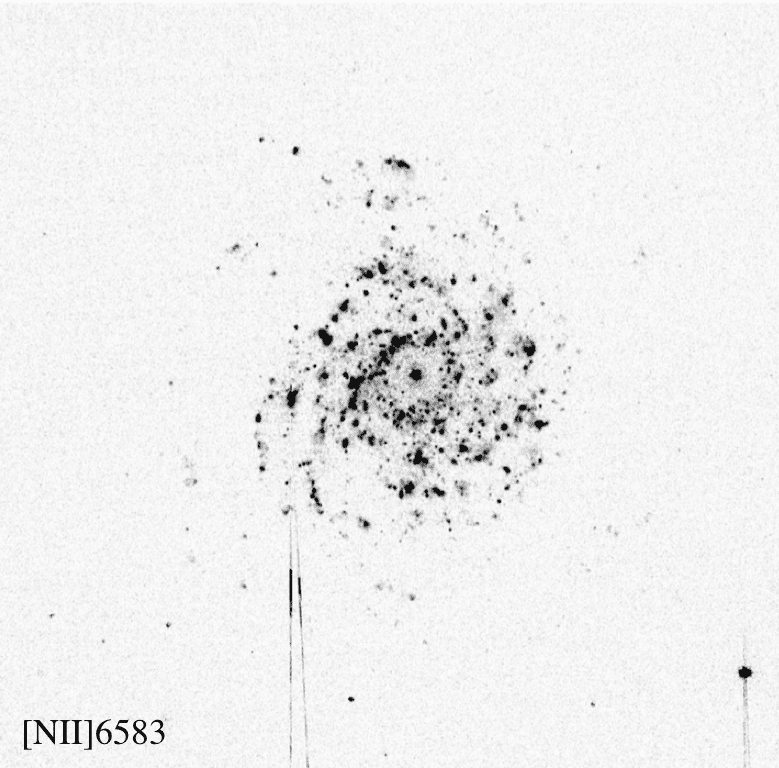}
	\includegraphics[width=0.3\textwidth]{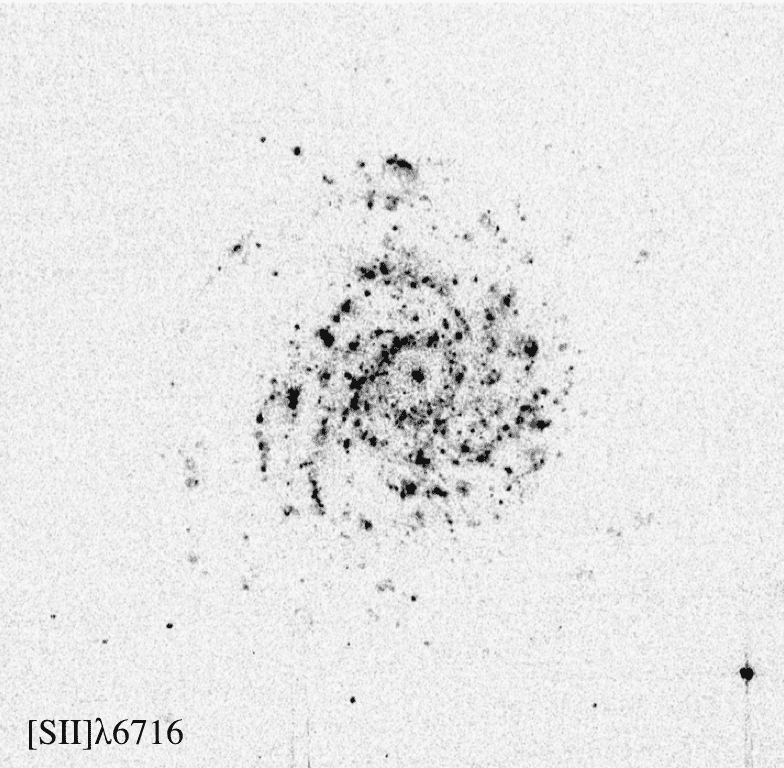}
	\includegraphics[width=0.3\textwidth]{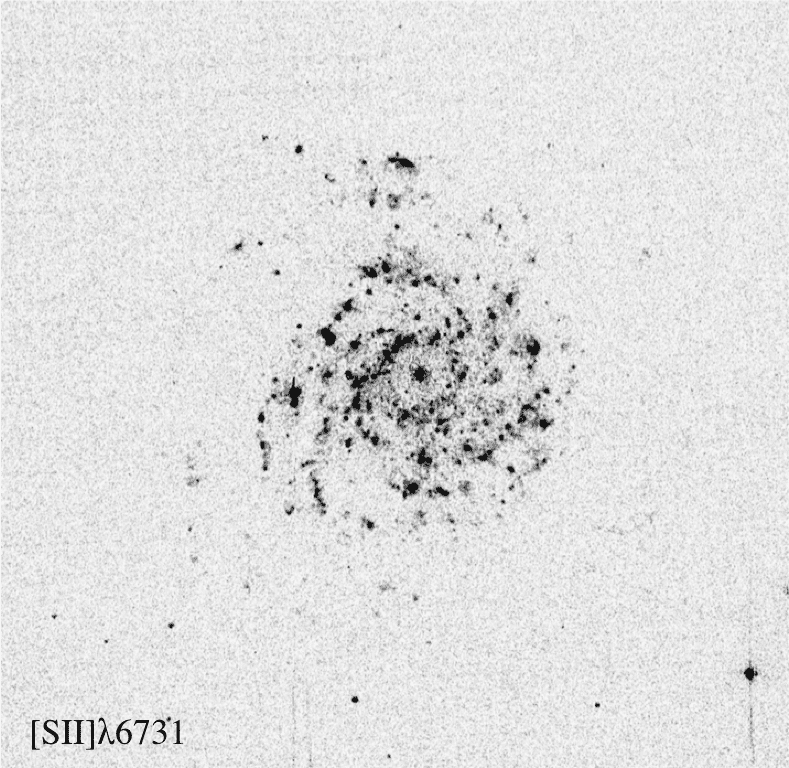}	
    \caption{Flux maps for all the emission lines from the SITELLE filters SN1, SN2, and SN3. 
    The flux has been measured for all the pixels.}
    \label{fig:figure_fluxmaps}
\end{figure*}

The flux map for all the emission lines seen in the three filters are shown in Figure~\ref{fig:figure_fluxmaps}. Emission lines from the same filter have been fitted simultaneously using the routine from ORCS (\S~\ref{sec:datareduction}) applied to all the spaxels individually. No subtraction of the galaxy background spectrum on the line of sight for each spaxels has been done here. All spaxels have been kept regardless of the signal-to-noise ratio measured for the lines.

% Figure A1

%%%%%%%%%%%%%%%%%%%%%%%%%%%%%%%%%%%%%%%%%%%%%%%%%%
\section{Catalogue of SNR Candidates in NGC\,3344}
%%%%%%%%%%%%%%%%%%%%%%%%%%%%%%%%%%%%%%%%%%%%%%%%%%
\label{cat_snr_cand}

\begin{figure*}
	\includegraphics[width=\textwidth]{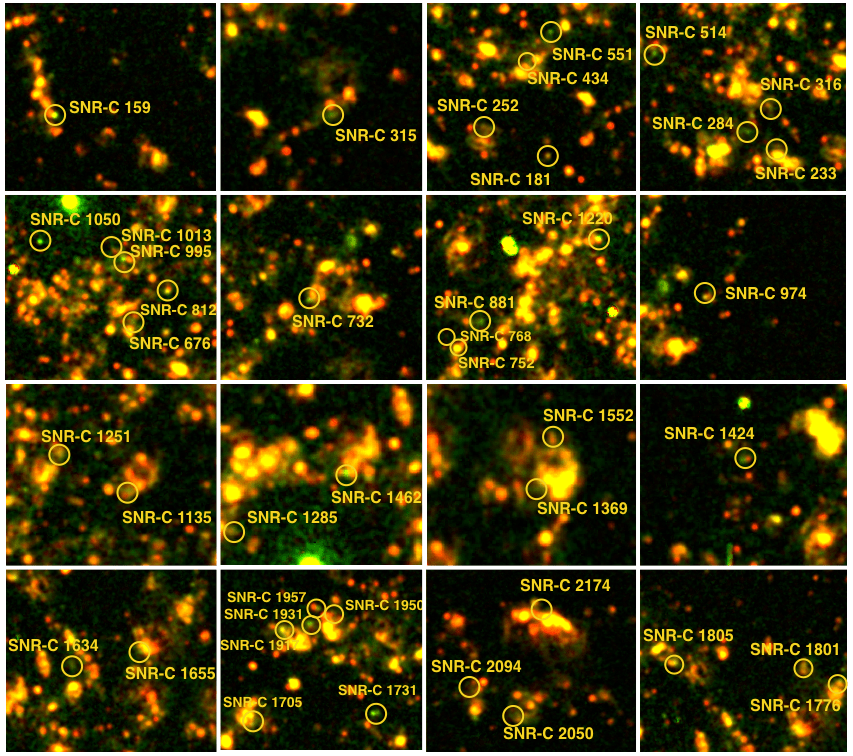}
    \caption{H$\alpha$ (red) + [S\II]$\lambda\lambda$6716,6731 (green) images of the 42 Confirmed SNRs. The size of each box is 2.9$^{\prime}$ $\times$ 2.6$^{\prime}$.  
    }
    \label{fig:figure_snrs2Ha}
\end{figure*}

This appendix presents, in Figure~\ref{fig:figure_snrs2Ha}, the H$\alpha$+[S\II] image of the NGC\,3344 with zooms on the 129~SNR candidates selected using the four criteria defined in Section~\ref{sec:iden-snr}. The SNR properties are given in Table~\ref{tab:example_table_snr}: identification number, coordinates, integrated H$\alpha$ flux 
and ([S\II]$\lambda$6716+[S\II]$\lambda$6731)/H$\alpha$ flux ratio 
(the flux uncertainties are given in 
Table~\ref{tab:table_emission_lines_flux}), region size ($\Sigma$), galactocentric distance (GCD), correlation coefficient ($R$) of the emission region pseudo-Voight profile, and velocity dispersion ($\sigma$). Table~\ref{tab:table_emission_lines_flux} gives, for the 129~SNR candidates, 
the integrated flux (within $\Sigma$) for all the emission lines measured in the SITELLE SN1, SN2, and SN3 filters,  after the subtraction of the global background.

% Figure B1

% Table B1
\begin{table*}
	\centering
	\caption{Catalogue of SNR Candidates in NGC\,3344}
	\label{tab:example_table_snr}
	\begin{tabular}{lcccccccc} % four columns, alignment for each
		\hline
		SNR-C ID & RA & DEC & flux $F_{\rm H_\alpha}$ & $\Sigma$ & GCD &$R$ & [S\II]/H$\alpha$ & $\sigma$ \\
		  & (J2000.0) & (J2000.0) & (erg cm$^{-2}$s$^{-1}$) & (pc) & (kpc) & (\%) & & km s$^{-1}$ \\
		  		\hline
		  &  &  & Confirmed SNRs &   &   &   &  & \\
		\hline\hline
		SNR-C159 & 10h43m36.57s & +24d53m44.48s & 3.12$\times$10$^{-16}$ & 50 & 3.36 & 94.5  & 0.86  $\pm$ 0.35 & 66.8 $\pm$ 3.4 \\
		SNR-C181 & 10h43m27.19s & +24d53m49.02s & 2.25$\times$10$^{-16}$ & 63 & 3.06 & 85.3  & 0.46  $\pm$ 0.18 & 51.5 $\pm$ 7.8 \\
		SNR-C233 & 10h43m30.21s & +24d53m56.61s & 7.05$\times$10$^{-16}$ & 112 & 2.40 & 80.0 & 0.45  $\pm$ 0.18 & 51 $\pm$ 12 \\
		SNR-C252 & 10h43m28.58s & +24d53m53.96s & 1.17$\times$10$^{-16}$ & 78 & 2.56 &  85.8 & 0.55  $\pm$ 0.22 & 47 $\pm$ 14 \\
		SNR-C284 & 10h43m30.92s & +24d54m02.07s & 6.82$\times$10$^{-17}$ & 96 & 2.21 & 81.8  & 1.07  $\pm$ 0.43 & 42 $\pm$ 20 \\
		SNR-C315 & 10h43m24.71s & +24d54m07.61s & 3.28$\times$10$^{-16}$ & 98 & 3.31 & 90.3  & 0.69  $\pm$ 0.27 & 40 $\pm$ 14 \\
		SNR-C316 & 10h43m30.37s & +24d54m09.99s & 6.53$\times$10$^{-16}$ & 116 & 2.01 & 80.1  & 0.44  $\pm$ 0.17 & 67.9 $\pm$ 8.4 \\
		SNR-C434 & 10h43m27.58s & +24d54m19.66s & 5.51$\times$10$^{-16}$ & 92 & 2.25 & 78.7  & 0.46  $\pm$ 0.19 & 35 $\pm$ 15 \\
		SNR-C514 & 10h43m33.16s & +24d54m28.93s & 1.97$\times$10$^{-16}$ & 98 & 1.62 &  75.6  & 0.85  $\pm$ 0.34 & 53 $\pm$ 12 \\
		SNR-C551 & 10h43m27.05s & +24d54m29.56s & 1.29$\times$10$^{-16}$ & 80 & 2.19 & 82.8  & 1.23  $\pm$ 0.49 & 52 $\pm$ 17 \\
		SNR-C676 & 10h43m29.66s & +24d54m40.61s & 5.40$\times$10$^{-16}$ & 106 & 1.29 & 77.6  & 0.59  $\pm$ 0.23 & 37 $\pm$ 12 \\
		SNR-C732 & 10h43m25.71s & +24d54m42.72s & 3.04$\times$10$^{-16}$ & 74 & 2.42 & 82.1  & 0.65  $\pm$ 0.26 & 31 $\pm$ 12 \\
		SNR-C752 & 10h43m36.33s & +24d54m48.06s & 9.09$\times$10$^{-16}$ & 50 & 2.19 & 97.0  & 0.49  $\pm$ 0.20 & 58.2 $\pm$ 2.1 \\
		SNR-C768 & 10h43m36.44s & +24d54m49.39s & 8.03$\times$10$^{-17}$ & 78 & 2.22 & 74.3  & 0.51  $\pm$ 0.20 & 70 $\pm$ 15 \\
		SNR-C812 & 10h43m28.83s & +24d54m49.74s & 4.21$\times$10$^{-16}$ & 64 & 1.28 &  92.9  & 0.54  $\pm$ 0.22 & 55.5 $\pm$ 5.5 \\
		SNR-C881 & 10h43m35.71s & +24d54m56.61s & 1.76$\times$10$^{-16}$ & 94 & 1.89 & 84.2  & 0.93  $\pm$ 0.38 & 59 $\pm$ 11 \\
		SNR-C974 & 10h43m23.58s & +24d54m57.56s & 3.56$\times$10$^{-16}$ & 56 & 3.06 &  90.1  & 0.46  $\pm$ 0.18 & 37.4 $\pm$ 6.9\\
		SNR-C995 & 10h43m29.86s & +24d55m00.86s & 5.70$\times$10$^{-17}$ & 84 & 0.76 & 85.5  & 0.78  $\pm$ 0.31 & 53.5 $\pm$ 7.1 \\
		SNR-C1013 & 10h43m29.93s & +24d55m02.53s & 1.17$\times$10$^{-16}$ & 100 & 0.71 & 82.1  & 0.49  $\pm$ 0.20 & 31 $\pm$ 16 \\
		SNR-C1050 & 10h43m31.86s & +24d55m07.49s & 1.04$\times$10$^{-16}$ & 40 & 0.45 & 94.6  & 1.35  $\pm$ 0.55 & 92.8 $\pm$ 5.4 \\
		SNR-C1135 & 10h43m27.16s & +24d55m12.87s &  5.26$\times$10$^{-16}$ & 106 & 1.58 & 72.8  & 0.47  $\pm$ 0.18 & -- \\
		SNR-C1220 & 10h43m32.81s & +24d55m22.85s & 1.09$\times$10$^{-15}$ & 60 & 0.66 & 94.1  & 0.63  $\pm$ 0.25 & 40.5 $\pm$ 5.3 \\
		SNR-C1251 & 10h43m28.50s & +24d55m24.13s & 1.36$\times$10$^{-15}$ & 100 & 1.03 & 96.3  & 0.46  $\pm$ 0.08 & -- \\
		SNR-C1285 & 10h43m32.46s & +24d55m29.55s & 9.86$\times$10$^{-16}$ & 100 & 0.59 & 81.1  & 0.47  $\pm$ 0.19 & 47 $\pm$ 11 \\
		SNR-C1369 & 10h43m25.17s & +24d55m34.56s & 1.59$\times$10$^{-16}$ & 96 & 2.35 & 75.8  & 0.57  $\pm$ 0.23 & -- \\
		SNR-C1424 & 10h43m37.69s & +24d55m43.61s & 4.32$\times$10$^{-16}$ & 86 & 2.68 & 93.3  & 0.62  $\pm$ 0.25 & 35 $\pm$ 13 \\
		SNR-C1462 & 10h43m30.47s & +24d55m42.82s & 9.53$\times$10$^{-16}$ & 116 & 0.66 & 53.8  & 0.54  $\pm$ 0.22 & 30.9 $\pm$ 7.3 \\
		SNR-C1552 & 10h43m24.89s & +24d55m46.50s & 1.31$\times$10$^{-16}$ & 96 &  2.52 & 85.0  & 0.42  $\pm$ 0.17 & --  \\
		SNR-C1634 & 10h43m28.81s & +24d55m57.15s &  2.41$\times$10$^{-16}$ & 114 & 1.34 & 80.2  & 0.71  $\pm$ 0.29 & --\\
		SNR-C1655 & 10h43m27.15s & +24d56m01.06s & 6.56$\times$10$^{-16}$ & 100 & 1.88 & 63.0  & 0.48  $\pm$ 0.19 & -- \\
		SNR-C1705 & 10h43m34.84s & +24d56m10.24s & 1.10$\times$10$^{-15}$ & 100 & 2.06 & 86.8  & 0.42  $\pm$ 0.17 & -- \\
		SNR-C1731 & 10h43m31.77s & +24d56m12.32s & 2.21$\times$10$^{-16}$ & 54 & 1.49 & 92.0  &  1.24  $\pm$ 0.50 & 71.9 $\pm$ 6.6 \\
		SNR-C1776 & 10h43m25.94s & +24d56m18.16s & 1.57$\times$10$^{-15}$ & 94 & 2.53 & 91.2  & 0.43  $\pm$ 0.17 & 39.1 $\pm$ 6.8 \\
		SNR-C1801 & 10h43m26.84s & +24d56m24.38s & 4.06$\times$10$^{-16}$ & 84 & 2.39 & 95.3  & 0.45  $\pm$ 0.18 & --\\
		SNR-C1805 & 10h43m30.11s & +24d56m26.65s & 5.00$\times$10$^{-16}$ & 82 & 1.88 & 87.2  & 0.50  $\pm$ 0.20 & -- \\
		SNR-C1917 & 10h43m34.00s & +24d56m40.89s & 1.27$\times$10$^{-15}$ & 116 & 2.57 & 67.9  & 0.59  $\pm$ 0.24 & --\\
		SNR-C1931 & 10h43m33.28s & +24d56m42.57s & 7.91$\times$10$^{-16}$ & 70 & 2.49 & 94.8  & 0.78  $\pm$ 0.31 & -- \\
		SNR-C1950 & 10h43m32.75s & +24d56m46.27s & 5.07$\times$10$^{-17}$ & 90 & 2.52 & 75.4  & 0.40  $\pm$ 0.16 & --\\
		SNR-C1957 & 10h43m33.10s & +24d56m47.72s & 4.58$\times$10$^{-16}$ & 82 & 2.60 & 83.2  & 0.43  $\pm$ 0.17 & -- \\
		SNR-C2050 & 10h43m32.40s & +24d57m26.17s & 1.99$\times$10$^{-15}$ & 114 & 3.58 & 89.7  & 0.42  $\pm$ 0.17 & --\\
		SNR-C2094 & 10h43m33.50s & +24d57m37.32s & 9.74$\times$10$^{-16}$ & 66 & 3.99 & 69.7  & 0.40  $\pm$ 0.16 & -- \\
		SNR-C2174 & 10h43m31.65s & +24d58m02.88s & 2.62$\times$10$^{-16}$ & 72 &  4.57 & 61.4  & 0.43  $\pm$ 0.17 & -- \\
		  		\hline
		  &  &  & Probable SNRs &   &   &   & &  \\
		\hline\hline
		SNR-C6 & 10h43m31.09s & +24d52m00.80s & 1.26$\times$10$^{-16}$ & 100 & 5.64 & 79.2 & 0.40  $\pm$ 0.16 & 57 $\pm$ 14 \\
		SNR-C178 & 10h43m29.79s & +24d53m49.64s & 6.81$\times$10$^{-16}$ & 86 & 2.63 & 93.0 & 0.40  $\pm$ 0.16 & 44.7 $\pm$ 6.7 \\
		SNR-C236 & 10h43m31.33s & +24d53m57.36s & 3.11$\times$10$^{-16}$ & 94 & 2.34 & 65.4  & 0.45  $\pm$ 0.18 & 54 $\pm$ 12 \\
		SNR-C244 & 10h43m32.07s & +24d53m58.93s & 5.73$\times$10$^{-16}$ & 86 &  2.31 & 72.4 & 0.45  $\pm$ 0.18 & 47.4 $\pm$ 8.3 \\
		SNR-C262 & 10h43m37.06s & +24d54m02.83s & 5.44$\times$10$^{-16}$ & 114 & 3.11 & 85.7  & 0.43  $\pm$ 0.17 & 41.2 $\pm$ 8.2 \\
		SNR-C332 & 10h43m27.17s & +24d54m11.08s & 5.10$\times$10$^{-16}$ & 108 & 2.55 & 82.2  & 0.41  $\pm$ 0.16 & 40 $\pm$ 10 \\
		SNR-C375 & 10h43m24.82s & +24d54m14.14s & 2.34$\times$10$^{-16}$ & 118 & 3.17 & 81.9  & 0.45  $\pm$ 0.18 & 46 $\pm$ 12 \\
		SNR-C380 & 10h43m28.28s & +24d54m15.70s & 3.77$\times$10$^{-16}$ & 118 & 2.18 & 76.3  & 0.40  $\pm$ 0.16 & 24 $\pm$ 16 \\
		SNR-C538 & 10h43m33.75s & +24d54m31.11s & 6.51$\times$10$^{-16}$ & 92 & 1.69 & 95.4  & 0.41  $\pm$ 0.16 & 37.7 $\pm$ 8.0 \\
		SNR-C570 & 10h43m40.02s & +24d54m36.51s & 1.61$\times$10$^{-16}$ & 80 & 3.64 & 87.2  & 0.59  $\pm$ 0.24 & --\\
		SNR-C601 & 10h43m29.80s & +24d54m34.48s & 5.14$\times$10$^{-16}$ & 84 & 1.42 & 90.8  & 0.50  $\pm$ 0.21 & 35 $\pm$ 15 \\
		SNR-C624 & 10h43m28.88s & +24d54m36.42s & 6.00$\times$10$^{-16}$ & 70 &  1.55 & 95.9  & 0.46  $\pm$ 0.18 & 31.4 $\pm$ 9.0 \\
		SNR-C668 & 10h43m32.15s & +24d54m40.58s & 1.81$\times$10$^{-13}$ & 80 &  1.17 &  87.3  & 0.47  $\pm$ 0.19 & 33 $\pm$ 22 \\
		SNR-C683 & 10h43m25.58s & +24d54m39.42s & 3.24$\times$10$^{-16}$ & 104 & 2.51 & 61.8  & 0.47  $\pm$ 0.19 & -- \\
		SNR-C699 & 10h43m36.33s & +24d54m44.80s & 7.45$\times$10$^{-16}$ & 80 & 2.23 & 95.2  & 0.40  $\pm$ 0.16 & 47.3 $\pm$ 8.7 \\
		SNR-C721 & 10h43m36.38s & +24d54m46.45s & 1.10$\times$10$^{-15}$ & 96 & 2.23 & 92.5  & 0.45  $\pm$ 0.18 & 47.0 $\pm$ 6.3 \\
		SNR-C733 & 10h43m25.04s & +24d54m42.48s & 8.80$\times$10$^{-16}$ & 80 & 2.66 & 60.6  & 0.46  $\pm$ 0.19 & 30 $\pm$ 15 \\
	\end{tabular}
\end{table*}

% Table B1 suite
\begin{table*}
	\centering
	\contcaption{}
	\label{tab:example_table_snr_2}
	\begin{tabular}{lcccccccc} % four columns, alignment for each
			  		\hline
		SNR-C738 & 10h43m25.19s & +24d54m42.86s & 5.21$\times$10$^{-17}$ & 116 & 2.60 & 70.3  & 0.70  $\pm$ 0.28 & -- \\
		SNR-C847 & 10h43m35.04s & +24d54m54.39s & 3.51$\times$10$^{-16}$ & 108 & 1.67 & 90.9  & 0.44  $\pm$ 0.17 & 51.4 $\pm$ 8.7 \\
		SNR-C968 & 10h43m34.31s & +24d55m01.60s & 2.73$\times$10$^{-16}$ & 94 &  1.33 &  76.9  & 0.50  $\pm$ 0.20 & -- \\	
		SNR-C1088 & 10h43m34.10s & +24d55m11.95s & 9.17$\times$10$^{-16}$ & 94 & 1.18 & 52.6  & 0.56  $\pm$ 0.23 & 46 $\pm$ 18 \\
		SNR-C1092 & 10h43m34.27s & +24d55m12.33s & 5.86$\times$10$^{-16}$ & 94 & 1.24 & 77.0  & 0.52  $\pm$ 0.21 & -- \\
		SNR-C1144 & 10h43m38.89s & +24d55m18.02s & 3.90$\times$10$^{-16}$ & 72 & 3.04 & 93.6  & 0.56 $\pm$ 0.22 & --\\
		SNR-C1180 & 10h43m28.65s & +24d55m17.34s & 3.24$\times$10$^{-16}$ & 108 & 0.98 & 63.9  & 0.47  $\pm$ 0.19 & -- \\
		SNR-C1217 & 10h43m36.51s & +24d55m24.28s & 5.46$\times$10$^{-16}$ & 92 & 2.12 & 77.9  & 0.47  $\pm$ 0.18 & 41 $\pm$ 11 \\
		SNR-C1242 & 10h43m37.03s & +24d55m27.08s & 1.08$\times$10$^{-15}$ & 68 & 2.33 & 76.7  & 0.54  $\pm$ 0.22 & 34.3 $\pm$ 6.4 \\
		SNR-C1415 & 10h43m31.74s & +24d55m40.70s & 1.01$\times$10$^{-15}$ & 92 & 0.62 & 78.5  & 0.41  $\pm$ 0.17 & 24 $\pm$ 18 \\
		SNR-C1429 & 10h43m25.50s & +24d55m39.26s & 1.33$\times$10$^{-16}$ & 110 & 2.24 & 67.6  & 0.44  $\pm$ 0.18 & 28 $\pm$ 17) \\
		SNR-C1432 & 10h43m24.54s & +24d55m38.90s & 1.24$\times$10$^{-15}$ & 92 & 2.60 & 92.5  & 0.42  $\pm$ 0.17 & 26.0 $\pm$ 8.0 \\
		SNR-C1574 & 10h43m36.64s & +24d55m53.01s & 3.49$\times$10$^{-16}$ & 102 & 2.39 & 89.8  & 0.44  $\pm$ 0.17 & 21 $\pm$ 19 \\
		SNR-C1584 & 10h43m36.50s & +24d55m53.92s & 3.66$\times$10$^{-16}$ & 94 & 2.34 & 94.2  & 0.43  $\pm$ 0.17 & 49.8 $\pm$ 9.3 \\
		SNR-C1588 & 10h43m36.61s & +24d55m54.62s & 4.71$\times$10$^{-16}$ & 94 & 2.39 & 82.2  & 0.44  $\pm$ 0.18 & 45.1 $\pm$ 8.7 \\
		SNR-C1606 & 10h43m36.47s & +24d55m55.53s & 8.96$\times$10$^{-17}$ & 82 & 2.35 & 87.7  & 0.47  $\pm$ 0.18 & 41 $\pm$ 10 \\
		SNR-C1637 & 10h43m26.97s & +24d55m57.41s & 1.97$\times$10$^{-16}$ & 110 & 1.88 & 55.9  & 0.42  $\pm$ 0.16 & -- \\
		SNR-C1712 & 10h43m27.27s & +24d56m07.63s & 3.34$\times$10$^{-16}$ & 72 &  1.96 & 93.1  & 0.43  $\pm$ 0.17 & -- \\
		SNR-C1727 & 10h43m33.61s & +24d55m15.23s & 9.34$\times$10$^{-16}$ & 80 &  2.12 & 95.4  & 0.40  $\pm$ 0.16 & -- \\
		SNR-C1769 & 10h43m26.14s & +24d56m16.62s & 6.32$\times$10$^{-16}$ & 104 & 2.45 & 87.5  & 0.42  $\pm$ 0.17 & -- \\
		SNR-C1799 & 10h43m25.86s & +24d56m22.68s & 1.42$\times$10$^{-16}$ & 80 & 2.64 & 75.4  & 0.77  $\pm$ 0.31 & 45 $\pm$ 16 \\
		SNR-C1809 & 10h43m31.47s & +24d56m27.52s & 3.61$\times$10$^{-16}$ & 104 & 1.89 & 79.1  & 0.60  $\pm$ 0.24 & --\\
		SNR-C1801 & 10h43m26.84s & +24d56m24.38s & 4.06$\times$10$^{-16}$ & 84 & 2.39 & 95.3  & 0.45  $\pm$ 0.18 & --\\
		SNR-C1805 & 10h43m30.11s & +24d56m26.65s & 5.00$\times$10$^{-16}$ & 82 & 1.88 & 87.2  & 0.50  $\pm$ 0.20 & -- \\
		SNR-C1845 & 10h43m30.55s & +24d56m31.42s & 7.74$\times$10$^{-16}$ & 68 & 1.99 & 91.4  & 0.48  $\pm$ 0.19 & 34.2 $\pm$ 3.4 \\
		SNR-C1873 & 10h43m31.96s & +24d56m34.57s & 7.08$\times$10$^{-16}$ & 104 & 2.12 & 80.9  & 0.41  $\pm$ 0.16 & --\\
		SNR-C1929 & 10h43m32.52s & +24d56m41.62s & 1.54$\times$10$^{-16}$ & 96 & 2.36 & 67.7  & 0.48  $\pm$ 0.19 & 44 $\pm$ 19 \\
		SNR-C2021 & 10h43m26.54s & +24d57m06.88s & 1.44$\times$10$^{-16}$ & 70 & 3.41 & 78.8  & 0.46  $\pm$ 0.18 & --\\
		SNR-C2048 & 10h43m32.28s & +24d57m25.14s & 5.23$\times$10$^{-16}$ & 94 &  3.55 & 86.6  & 0.69  $\pm$ 0.27 & -- \\
		\hline
		  &  &  &  Less likely SNRs &   &   &   &  & \\
		\hline\hline	
		SNR-C29 & 10h43m33.15s & +24d52m43.64s & 4.00$\times$10$^{-16}$ & 116 & 4.45 & 79.8 & 0.48  $\pm$ 0.19 & 56 $\pm$ 15 \\
		SNR-C88 & 10h43m33.90s & +24d53m27.55s & 1.06$\times$10$^{-15}$ & 100 & 3.31 & 83.9  & 0.44  $\pm$ 0.18 & 52.4 $\pm$ 8.9\\
		SNR-C158 & 10h43m36.86s & +24d53m44.60s & 2.79$\times$10$^{-16}$ & 120 & 3.42 & 85.1 & 0.41  $\pm$ 0.16 & 42 $\pm$ 11 \\
		SNR-C224 & 10h43m29.04s & +24d53m55.53s & 4.18$\times$10$^{-16}$ & 110 & 2.57 & 76.6  & 0.47  $\pm$ 0.19 & 37 $\pm$ 13 \\
		SNR-C241 & 10h43m32.93s & +24d53m58.94s & 7.22$\times$10$^{-17}$ & 98 & 2.37 &  89.2 & 0.63  $\pm$ 0.26 & 46 $\pm$ 21 \\
		SNR-C260 & 10h43m28.75s & +24d53m58.99s & 2.80$\times$10$^{-16}$ & 86 & 2.52 & 87.3  & 0.47  $\pm$ 0.19 & 51 $\pm$ 11 \\
		SNR-C301 & 10h43m30.81s & +24d54m07.22s & 3.32$\times$10$^{-16}$ & 92 & 2.07 & 76.1  & 0.62  $\pm$ 0.25 & 47.3 $\pm$ 9.9\\
		SNR-C317 & 10h43m43.72s & +24d54m15.61s & 3.65$\times$10$^{-16}$ & 114 & 5.19 & 82.2  & 0.47  $\pm$ 0.19 & 57 $\pm$ 15\\
		SNR-C333 & 10h43m30.44s & +24d54m12.61s & 2.45$\times$10$^{-16}$ & 88 & 1.94 & 68.0  & 0.64  $\pm$ 0.26 & 51 $\pm$ 15 \\
		SNR-C390 & 10h43m31.91s & +24d54m17.71s & 4.26$\times$10$^{-16}$ & 98 &  1.78 & 87.2  & 0.43  $\pm$ 0.18 & 48.6 $\pm$ 9.3 \\
		SNR-C440 & 10h43m36.78s & +24d54m24.17s & 5.37$\times$10$^{-16}$ & 90 & 2.66 &  78.9  & 0.63  $\pm$ 0.25 & 34 $\pm$ 17 \\
		SNR-C474 & 10h43m27.69s & +24d54m22.97s & 1.64$\times$10$^{-16}$ & 94 & 2.15 & 82.7  & 0.66  $\pm$ 0.27 & 54 $\pm$ 12 \\
		SNR-C543 & 10h43m30.43s & +24d54m30.16s & 4.79$\times$10$^{-16}$ & 102 & 1.45 & 92.1  & 0.47  $\pm$ 0.19 & 48 $\pm$ 14 \\
		SNR-C564 & 10h43m33.41s & +24d54m32.92s & 5.26$\times$10$^{-16}$ & 68 & 1.57 & 80.5  & 0.60  $\pm$ 0.24 & 61 $\pm$ 15 \\
		SNR-C566 & 10h43m30.52s & +24d54m32.14s & 2.32$\times$10$^{-16}$ & 110 & 1.39 & 55.1  & 0.57  $\pm$ 0.23 & 52 $\pm$ 11 \\
		SNR-C634 & 10h43m36.15s & +24d54m40.17s & 1.62$\times$10$^{-16}$ & 78 & 2.22 & 87.9  & 0.74  $\pm$ 0.29 & 57 $\pm$ 16 \\
		SNR-C638 & 10h43m34.36s & +24d54m39.79s & 1.56$\times$10$^{-15}$ & 92 & 1.66 &  86.3  & 0.43  $\pm$ 0.17 & 51.9 $\pm$ 8.5 \\
		SNR-C654 & 10h43m29.52s & +24d54m38.60s & 5.32$\times$10$^{-16}$ & 88 &  1.36 &  87.6  & 0.85  $\pm$ 0.34 & 41 $\pm$ 23 \\
		SNR-C672 & 10h43m32.32s & +24d54m41.29s & 1.68$\times$10$^{-16}$ & 102 & 1.18 & 81.0  & 0.43  $\pm$ 0.17 & 39 $\pm$ 19 \\
		SNR-C756 & 10h43m23.85s & +24d54m43.35s & 2.68$\times$10$^{-16}$ & 72 & 3.08 & 81.8  & 0.72  $\pm$ 0.29 & 46 $\pm$ 11 \\
		SNR-C805 & 10h43m26.68s & +24d54m48.27s & 9.18$\times$10$^{-17}$ & 106 & 2.00 & 71.4  & 0.82  $\pm$ 0.33 & -- \\
		SNR-C816 & 10h43m33.90s & +24d54m52.00s & 5.53$\times$10$^{-16}$ & 106 & 1.32 & 67.0  & 0.51  $\pm$ 0.20 & 46 $\pm$ 12 \\
		SNR-C929 & 10h43m26.16s & +24d54m55.89s & 1.41$\times$10$^{-16}$ & 106 & 2.09 & 79.1  & 0.91  $\pm$ 0.37 & --\\
		SNR-C944 & 10h43m30.13s & +24d54m58.34s & 4.48$\times$10$^{-15}$ & 114 & 0.75 &  72.2  & 0.72  $\pm$ 0.29 & 36 $\pm$ 16 \\
		SNR-C994 & 10h43m30.20s & +24d55m01.00s & 3.95$\times$10$^{-17}$ & 104 & 0.67 & 75.2  &  1.32  $\pm$ 0.53 & 42 $\pm$ 23 \\
		SNR-C997 & 10h43m24.65s & +24d54m58.89s & 1.27$\times$10$^{-16}$ & 84 & 2.63 & 56.0  & 0.97  $\pm$ 0.39 & -- \\
		SNR-C1012 & 10h43m33.66s & +24d55m03.95s & 5.64$\times$10$^{-16}$ & 110 & 1.08 & 75.7  & 0.65  $\pm$ 0.26 & 50 $\pm$ 14 \\
		SNR-C1213 & 10h43m28.60s & +24d55m20.90s & 2.30$\times$10$^{-15}$ & 102 & 0.99 & 68.4  & 0.41  $\pm$ 0.16 & 25 $\pm$ 15 \\
		SNR-C1829 & 10h43m31.37s & +24d56m30.10s & 4.64$\times$10$^{-16}$ & 88 & 1.96 & 54.5  & 0.61  $\pm$ 0.24 & -- \\
		SNR-C1844 & 10h43m31.10s & +24d56m31.64s & 3.70$\times$10$^{-16}$ & 78 & 1.99 & 80.3  & 0.51  $\pm$ 0.21 & 35 $\pm$ 19 \\
		SNR-C1861 & 10h43m32.42s & +24d56m33.43s & 5.16$\times$10$^{-17}$ & 92 & 2.13 & 76.3  & 0.92  $\pm$ 0.37 & -- \\
		SNR-C1887 & 10h43m31.62s & +24d56m36.06s & 2.33$\times$10$^{-16}$ & 90 & 2.14 & 92.5  & 0.40  $\pm$ 0.16 & 31 $\pm$ 18 \\
		SNR-C1922 & 10h43m33.81s & +24d56m41.47s & 3.67$\times$10$^{-16}$ & 84 & 2.55 & 87.5  & 0.49  $\pm$ 0.20 & -- \\
	\end{tabular}
\end{table*}

% Table B1 suite
\begin{table*}
	\centering
	\contcaption{}
	\label{tab:example_table_snr_3}
	\begin{tabular}{lcccccccc} % four columns, alignment for each
		SNR-C1940 & 10h43m31.53s & +24d56m43.51s & 5.54$\times$10$^{-16}$ & 118 & 2.34 & 63.7  & 0.42  $\pm$ 0.17 & -- \\
		SNR-C1971 & 10h43m31.54s & +24d56m51.35s & 1.04$\times$10$^{-15}$ & 90 & 2.56 & 86.0  & 0.50  $\pm$ 0.20 & -- \\
		SNR-C2051 & 10h43m32.02s & +24d57m26.02s & 1.42$\times$10$^{-16}$ & 88 &  3.55 & 88.8  & 0.72  $\pm$ 0.29 & -- \\
		SNR-C2053 & 10h43m33.45s & +24d57m27.22s & 7.33$\times$10$^{-16}$ & 72 & 3.71 & 92.2  & 0.58  $\pm$ 0.23 & 25 $\pm$ 17 \\
		SNR-C2058 & 10h43m32.06s & +24d57m27.99s & 1.59$\times$10$^{-16}$ & 86 & 3.61 & 89.5  & 0.50  $\pm$ 0.20 & 39 $\pm$ 15 \\
		SNR-C2085 & 10h43m32.21s & +24d57m34.22s & 8.94$\times$10$^{-17}$ & 90 & 3.80 & 93.5  & 0.59  $\pm$ 0.23 & -- \\
		SNR-C2086 & 10h43m31.90s & +24d57m34.42s & 5.02$\times$10$^{-17}$ & 86 & 3.78 & 88.5  & 0.40  $\pm$ 0.16 & 30 $\pm$ 16 \\
		SNR-C2090 & 10h43m30.44s & +24d57m34.81s & 1.20$\times$10$^{-16}$ & 90 & 3.76 & 82.2  & 0.53  $\pm$ 0.21 & -- \\
		SNR-C2161 & 10h43m29.77s & +24d57m58.23s & 6.34$\times$10$^{-18}$ & 90 & 4.43 & 71.5  & 1.02  $\pm$ 0.40 & -- \\
		SNR-C2179 & 10h43m31.98s & +24d58m05.27s & 6.07$\times$10$^{-16}$ & 70 & 4.65 & 93.5  & 0.43  $\pm$ 0.17 & 28.6 $\pm$ 7.6 \\
		\hline
	\end{tabular}
\end{table*}

% Table B2
\begin{landscape}
\begin{table*}
	\centering
	\caption{Emission lines fluxes of the SNR Candidates in NGC\,3344 corrected for the background contribution}
	\label{tab:table_emission_lines_flux}
	  \begin{adjustbox}{max width=\textwidth}
	\begin{tabular}{lccccccccc} % four columns, alignment for each
		\hline
		SNR-ID & [O\II]$\lambda$3727   & H$\beta$ & [O\III]$\lambda$4959 & [O\III]$\lambda$5007 & [N\II]$\lambda$6548   & H$\alpha$ & [N\II]$\lambda$6583   &  [S\II]$\lambda$6716   &  [S\II]$\lambda$6731 \\
		\hline\hline
        SNR-C6  & 1.32e-16 $\pm$   4.0e-17 &    4.52e-17 $\pm$   1.0e-17  &     9.86e-18 $\pm$   8.4e-18  &     3.64e-17 $\pm$   1.8e-17  &     1.31e-17 $\pm$   7.7e-18  &     1.26e-16 $\pm$   2.3e-17  &     2.70e-17 $\pm$   7.6e-18  &     2.85e-17 $\pm$   8.9e-18  &    2.13e-17 $\pm$   8.9e-18 \\
        
        SNR-C29 & 6.46e-16 $\pm$   1.9e-16 &    1.37e-16 $\pm$   3.2e-17  &     --  &     1.31e-16 $\pm$   6.4e-17  &     --  &     4.00e-16 $\pm$   7.2e-17  &     7.53e-17 $\pm$   2.1e-17  &     1.15e-16 $\pm$   3.6e-17  &    7.67e-17 $\pm$   3.2e-17 \\
        
        SNR-C88  & 7.52e-16 $\pm$   2.3e-16 &    3.40e-16 $\pm$   7.9e-17  &     --  &     1.76e-16 $\pm$   8.6e-17  &   --  &     1.06e-15 $\pm$   1.9e-16  &     2.63e-16 $\pm$   7.4e-17  &     2.51e-16 $\pm$   7.8e-17  &    2.17e-16 $\pm$   9.0e-17 \\
        
        SNR-C158 &  2.38e-16 $\pm$   7.2e-17 &    1.03e-16 $\pm$   2.4e-17  &     3.20e-17 $\pm$   2.7e-17  &     8.52e-17 $\pm$   4.2e-17  &    --  &     2.79e-16 $\pm$   5.0e-17  &     8.03e-17 $\pm$   2.2e-17  &     6.61e-17 $\pm$   2.1e-17  &    4.91e-17 $\pm$   2.0e-17 \\
        
        SNR-C159 &  6.52e-16 $\pm$   2.0e-16 &    1.11e-16 $\pm$   2.6e-17  &     7.09e-17 $\pm$   6.0e-17  &     2.36e-16 $\pm$   1.2e-16  &     5.34e-17 $\pm$   3.1e-17  &     3.12e-16 $\pm$   5.6e-17  &     1.29e-16 $\pm$   3.6e-17  &     1.45e-16 $\pm$   4.5e-17  &    1.24e-16 $\pm$   5.1e-17 \\
        
        SNR-C178 &  6.31e-16 $\pm$   1.9e-16 &    2.27e-16 $\pm$   5.3e-17  &     --  &    --  &     5.71e-17 $\pm$   3.4e-17  &     6.81e-16 $\pm$   1.2e-16  &     2.05e-16 $\pm$   5.7e-17  &     1.68e-16 $\pm$   5.2e-17  &    1.07e-16 $\pm$   4.4e-17 \\
        
        SNR-C181 &  2.36e-16 $\pm$   7.1e-17 &    7.92e-17 $\pm$   1.8e-17  &     2.73e-17 $\pm$   2.3e-17  &     1.08e-16 $\pm$   5.3e-17  &     2.66e-17 $\pm$   1.6e-17  &     2.25e-16 $\pm$   4.1e-17  &     7.21e-17 $\pm$   2.0e-17  &     6.25e-17 $\pm$   1.9e-17  &    4.12e-17 $\pm$   1.7e-17 \\
        
        SNR-C224 & 2.77e-16 $\pm$   8.3e-17 &    1.45e-16 $\pm$   3.4e-17  &     --  &     --  &     -- &     4.18e-16 $\pm$   7.5e-17  &     1.57e-16 $\pm$   4.4e-17  &     1.10e-16 $\pm$   3.4e-17  &    8.46e-17 $\pm$   3.5e-17 \\
        
        SNR-C233 &  7.05e-16 $\pm$   2.1e-16 &    2.35e-16 $\pm$   5.9e-17  &     --  &     1.10e-16 $\pm$   5.4e-17  &     6.46e-17 $\pm$   3.8e-17  &     7.05e-16 $\pm$   1.3e-16  &     2.46e-16 $\pm$   6.9e-17  &     2.08e-16 $\pm$   6.5e-17  &    1.08e-16 $\pm$   4.5e-17 \\
        
        SNR-C236 &  3.60e-16 $\pm$   1.1e-16 &    1.04e-16 $\pm$   2.4e-17  &     --  &     -- &     4.53e-17 $\pm$   2.7e-17  &     3.11e-16 $\pm$   5.6e-17  &     1.11e-16 $\pm$   3.1e-17  &     7.60e-17 $\pm$   2.4e-17  &    6.53e-17 $\pm$   2.7e-17 \\
        
        SNR-C241 & -- &    2.64e-17 $\pm$   6.1e-18  &    --  &     --  &    --  &     7.22e-17 $\pm$   1.3e-17  &     2.29e-17 $\pm$   6.4e-18  &     2.12e-17 $\pm$   6.6e-18  &    2.41e-17 $\pm$   1.0e-17 \\
        
        SNR-C244 &   3.81e-16 $\pm$   1.1e-16 &    1.98e-16 $\pm$   4.6e-17  &     --  &     1.26e-16 $\pm$   6.2e-17  &     4.67e-17 $\pm$   2.8e-17  &     5.73e-16 $\pm$   1.0e-16  &     1.76e-16 $\pm$   4.9e-17  &     1.42e-16 $\pm$   4.4e-17  &    1.18e-16 $\pm$   4.9e-17 \\
        
        SNR-C252 &  1.14e-16 $\pm$   3.4e-17 &    4.05e-17 $\pm$   9.4e-18  &     --  &     2.62e-17 $\pm$   1.3e-17  &     2.05e-17 $\pm$   1.2e-17  &     1.17e-16 $\pm$   2.1e-17  &     4.24e-17 $\pm$   1.2e-17  &     3.21e-17 $\pm$   1.0e-17  &    3.22e-17 $\pm$   1.3e-17 \\
        
        SNR-C260 &    1.67e-16 $\pm$   5.0e-17 &    9.99e-17 $\pm$   2.3e-17  &     --  &     3.48e-17 $\pm$   1.7e-17  &    --  &     2.80e-16 $\pm$   5.0e-17  &     8.54e-17 $\pm$   2.4e-17  &     7.52e-17 $\pm$   2.3e-17  &    5.58e-17 $\pm$   2.3e-17 \\
        
        SNR-C262 &   2.80e-16 $\pm$   8.4e-17 &    1.90e-16 $\pm$   4.4e-17  &     --  &     7.91e-17 $\pm$   3.9e-17  &     3.04e-17 $\pm$   1.8e-17  &     5.44e-16 $\pm$   9.8e-17  &     1.31e-16 $\pm$   3.7e-17  &     1.37e-16 $\pm$   4.2e-17  &    9.61e-17 $\pm$   4.0e-17 \\
        
        SNR-C284 &  4.57e-17 $\pm$   1.4e-17 &    2.48e-17 $\pm$   5.8e-18  &     --  &     1.78e-17 $\pm$   8.7e-18  &     1.23e-17 $\pm$   7.2e-18  &     6.82e-17 $\pm$   1.2e-17  &     2.84e-17 $\pm$   8.0e-18  &     4.11e-17 $\pm$   1.3e-17  &    3.22e-17 $\pm$   1.3e-17 \\
        
        SNR-C301  & 4.28e-16 $\pm$   1.3e-16 &    1.11e-16 $\pm$   2.6e-17  &    --  &     1.21e-16 $\pm$   5.9e-17  &    --  &     3.32e-16 $\pm$   6.0e-17  &     1.49e-16 $\pm$   4.2e-17  &     1.12e-16 $\pm$   3.5e-17  &    9.50e-17 $\pm$   3.9e-17 \\
        
        SNR-C315 &  3.94e-16 $\pm$   1.2e-16 &    1.15e-16 $\pm$   2.7e-17  &     4.13e-17 $\pm$   3.5e-17  &     9.26e-17 $\pm$   4.5e-17  &     3.96e-17 $\pm$   2.3e-17  &     3.28e-16 $\pm$   5.9e-17  &     9.27e-17 $\pm$   2.6e-17  &     1.39e-16 $\pm$   4.3e-17  &    8.84e-17 $\pm$   3.7e-17 \\
        
        SNR-C316 &  4.59e-16 $\pm$   1.4e-16 &    2.20e-16 $\pm$   5.1e-17  &     1.27e-16 $\pm$   1.1e-16  &     3.42e-16 $\pm$   1.7e-16  &     1.05e-16 $\pm$   6.2e-17  &     6.53e-16 $\pm$   1.2e-16  &     2.54e-16 $\pm$   7.1e-17  &     1.78e-16 $\pm$   5.5e-17  &    1.12e-16 $\pm$   4.6e-17 \\
        
        SNR-C317 &  5.46e-16 $\pm$   1.6e-16 &    1.23e-16 $\pm$   2.9e-17  &     9.15e-17 $\pm$   7.8e-17  &     3.13e-16 $\pm$   1.5e-16  &     --  &     3.65e-16 $\pm$   6.6e-17  &     5.83e-17 $\pm$   1.6e-17  &     8.37e-17 $\pm$   2.6e-17  &    8.89e-17 $\pm$   3.7e-17 \\
        
        SNR-C332 &  2.69e-16 $\pm$   8.1e-17 &    1.74e-16 $\pm$   4.1e-17  &     --  &     1.22e-16 $\pm$   6.0e-17  &     5.20e-17 $\pm$   3.1e-17  &     5.10e-16 $\pm$   9.2e-17  &     1.31e-16 $\pm$   3.7e-17  &     1.39e-16 $\pm$   4.3e-17  &    7.22e-17 $\pm$   3.0e-17 \\
        
        SNR-C333 &   2.57e-16 $\pm$   7.7e-17 &    8.22e-17 $\pm$   1.9e-17  &     --  &     7.35e-17 $\pm$   3.6e-17  &    --  &     2.45e-16 $\pm$   4.4e-17  &     9.52e-17 $\pm$   2.7e-17  &     7.19e-17 $\pm$   2.2e-17  &    8.60e-17 $\pm$   3.6e-17 \\
        
        SNR-C375 &  2.04e-16 $\pm$   6.1e-17 &    8.47e-17 $\pm$   2.0e-17  &     --  &     3.80e-17 $\pm$   1.9e-17  &     2.10e-17 $\pm$   1.2e-17  &     2.34e-16 $\pm$   4.2e-17  &     6.69e-17 $\pm$   1.9e-17  &     5.64e-17 $\pm$   1.7e-17  &    4.87e-17 $\pm$   2.0e-17 \\ 
        
        SNR-C380 &   2.89e-16 $\pm$   8.7e-17 &    1.28e-16 $\pm$   3.0e-17  &     --  &     4.38e-17 $\pm$   2.1e-17  &     5.11e-17 $\pm$   3.0e-17  &     3.77e-16 $\pm$   6.8e-17  &     1.29e-16 $\pm$   3.6e-17  &     9.01e-17 $\pm$   2.8e-17  &    6.12e-17 $\pm$   2.5e-17 \\
        
        SNR-C390 &  --  &    1.49e-16 $\pm$   3.5e-17  &    --  &     --  &    --  &     4.26e-16 $\pm$   7.7e-17  &     1.31e-16 $\pm$   3.7e-17  &     8.64e-17 $\pm$   2.7e-17  &    9.71e-17 $\pm$   4.0e-17 \\
        
        SNR-C434 &   5.66e-16 $\pm$   1.7e-16 &    1.77e-16 $\pm$   4.1e-17  &   --  &     8.87e-17 $\pm$   4.3e-17  &     6.53e-17 $\pm$   3.8e-17  &     5.51e-16 $\pm$   9.9e-17  &     1.85e-16 $\pm$   5.2e-17  &     1.43e-16 $\pm$   4.4e-17  &    1.13e-16 $\pm$   4.7e-17 \\
        
        SNR-C440 &  5.05e-16 $\pm$   1.5e-16 &    --  &     --  &     --  &    -- &     5.37e-16 $\pm$   9.7e-17  &     1.55e-16 $\pm$   4.3e-17  &     2.06e-16 $\pm$   6.4e-17  &    1.32e-16 $\pm$   5.5e-17 \\
        
        SNR-C474 &    1.14e-16 $\pm$   3.4e-17 &    5.74e-17 $\pm$   1.3e-17  &     --  &    --  &     --  &     1.64e-16 $\pm$   2.9e-17  &     6.19e-17 $\pm$   1.7e-17  &     5.75e-17 $\pm$   1.8e-17  &    5.09e-17 $\pm$   2.1e-17 \\
        
        SNR-C514 &   4.72e-16   $\pm$ 8.5e-17 &   7.89e-17  $\pm$  1.8e-17 &   8.07e-17   $\pm$ 1.8e-17 &   1.86e-16   $\pm$ 3.4e-17 &   3.76e-17  $\pm$  1.6e-17 &   2.22e-16  $\pm$  4.1e-17 &   1.57e-16  $\pm$  3.0e-17  &  1.15e-16  $\pm$  2.4e-17 &   7.32e-17  $\pm$  1.9e-17 \\
        
        SNR-C538 &  -- &    2.23e-16 $\pm$   5.2e-17  &    --  &     --  &     8.77e-17 $\pm$   5.2e-17  &     6.51e-16 $\pm$   1.2e-16  &     2.44e-16 $\pm$   6.8e-17  &     1.50e-16 $\pm$   4.7e-17  &    1.15e-16 $\pm$   4.7e-17 \\
        
        SNR-C543 &  3.10e-16 $\pm$   9.3e-17 &    1.54e-16 $\pm$   3.6e-17  &     --  &     --  &     --  &     4.79e-16 $\pm$   8.6e-17  &     1.42e-16 $\pm$   4.0e-17  &     1.26e-16 $\pm$   3.9e-17  &    9.93e-17 $\pm$   4.1e-17 \\
        
        SNR-C551 &   1.84e-16 $\pm$   5.5e-17 &    4.37e-17 $\pm$   1.0e-17  &    --  &     6.05e-17 $\pm$   3.0e-17  &     3.33e-17 $\pm$   2.0e-17  &     1.29e-16 $\pm$   2.3e-17  &     7.50e-17 $\pm$   2.1e-17  &     9.47e-17 $\pm$   2.9e-17  &    6.42e-17 $\pm$   2.7e-17 \\
        
        SNR-C564 &   -- &    --  &     --  &     --  &    --  &     5.26e-16 $\pm$   9.5e-17  &     2.13e-16 $\pm$   6.0e-17  &     1.81e-16 $\pm$   5.6e-17  &    1.33e-16 $\pm$   5.5e-17 \\
        
        SNR-C566 &  -- &    8.32e-17 $\pm$   1.9e-17  &    --  &     --  &     1.93e-17 $\pm$   1.1e-17  &     2.32e-16 $\pm$   4.2e-17  &     7.20e-17 $\pm$   2.0e-17  &     7.31e-17 $\pm$   2.3e-17  &    5.92e-17 $\pm$   2.5e-17 \\
        
        SNR-C570 &  2.17e-16 $\pm$   6.5e-17 &    5.50e-17 $\pm$   1.3e-17  &    --  &    --  &     2.10e-17 $\pm$   1.2e-17  &     1.61e-16 $\pm$   2.9e-17  &     6.00e-17 $\pm$   1.7e-17  &     5.55e-17 $\pm$   1.7e-17  &    4.00e-17 $\pm$   1.7e-17 \\
        
        SNR-C601 &   3.78e-16 $\pm$   1.1e-16 &    1.70e-16 $\pm$   4.0e-17  &    --  &    --  &     7.26e-17 $\pm$   4.3e-17  &     5.14e-16 $\pm$   9.3e-17  &     1.96e-16 $\pm$   5.5e-17  &     1.19e-16 $\pm$   3.7e-17  &    1.40e-16 $\pm$   5.8e-17 \\
        
        SNR-C624 &   3.75e-16 $\pm$   1.1e-16 &    2.05e-16 $\pm$   4.8e-17  &     --  &    --  &     6.34e-17 $\pm$   3.7e-17  &     6.00e-16 $\pm$   1.1e-16  &     2.05e-16 $\pm$   5.7e-17  &     1.69e-16 $\pm$   5.3e-17  &    1.10e-16 $\pm$   4.6e-17 \\
        
        SNR-C634 &  2.73e-16 $\pm$   8.2e-17 &   --  &   --  &     6.28e-17 $\pm$   3.1e-17  &    -- &     1.62e-16 $\pm$   2.9e-17  &     5.69e-17 $\pm$   1.6e-17  &     7.51e-17 $\pm$   2.3e-17  &    4.50e-17 $\pm$   1.9e-17 \\
        
        SNR-C638 &  --   &    4.88e-16 $\pm$   1.1e-16  &     --  &    --  &   --  &     1.56e-15 $\pm$   2.8e-16  &     3.94e-16 $\pm$   1.1e-16  &     3.71e-16 $\pm$   1.1e-16  &    2.92e-16 $\pm$   1.2e-16 \\
        
        SNR-C654 & -- &    --  &    --  &   --  &    -- &     5.32e-16 $\pm$   9.6e-17  &     3.11e-16 $\pm$   8.7e-17  &     2.72e-16 $\pm$   8.4e-17  &    1.82e-16 $\pm$   7.5e-17 \\
        
        SNR-C668 &  --  &    --  &    --  &     2.39e-13 $\pm$   1.2e-13  &     2.66e-14 $\pm$   1.6e-14  &     1.81e-13 $\pm$   3.3e-14  &     7.80e-14 $\pm$   2.2e-14  &     4.11e-14 $\pm$   1.3e-14  &    4.31e-14 $\pm$   1.8e-14 \\
        
        SNR-C672 &  -- &    5.96e-17 $\pm$   1.4e-17  &    --  &    --  &    --  &     1.68e-16 $\pm$   3.0e-17  &     6.72e-17 $\pm$   1.9e-17  &     3.84e-17 $\pm$   1.2e-17  &    3.36e-17 $\pm$   1.4e-17 \\
        
        SNR-C676 &   6.29e-16 $\pm$   1.9e-16 &    1.69e-16 $\pm$   3.9e-17  &    -- &     2.02e-16 $\pm$   9.8e-17  &     1.07e-16 $\pm$   6.3e-17  &     5.40e-16 $\pm$   9.7e-17  &     2.83e-16 $\pm$   7.9e-17  &     1.90e-16 $\pm$   5.9e-17  &    1.27e-16 $\pm$   5.2e-17 \\
        
        SNR-C683 &    -- &    1.16e-16 $\pm$   2.7e-17  &     --  &     --  &     3.44e-17 $\pm$   2.0e-17  &     3.24e-16 $\pm$   5.9e-17  &     7.77e-17 $\pm$   2.2e-17  &     8.97e-17 $\pm$   2.8e-17  &    6.34e-17 $\pm$   2.6e-17 \\
        
        SNR-C699 &  9.28e-16 $\pm$   2.8e-16 &    2.34e-16 $\pm$   5.5e-17  &     --  &     --  &     1.27e-16 $\pm$   7.5e-17  &     7.45e-16 $\pm$   1.3e-16  &     3.02e-16 $\pm$   8.5e-17  &     1.76e-16 $\pm$   5.5e-17  &    1.20e-16 $\pm$   5.0e-17 \\
        
        SNR-C721 &  1.03e-15 $\pm$   3.1e-16 &    3.66e-16 $\pm$   8.5e-17  &     8.38e-17 $\pm$   7.1e-17  &     1.87e-16 $\pm$   9.1e-17  &     1.73e-16 $\pm$   1.0e-16  &     1.10e-15 $\pm$   2.0e-16  &     3.57e-16 $\pm$   1.0e-16  &     2.85e-16 $\pm$   8.9e-17  &    2.13e-16 $\pm$   8.8e-17 \\
        
        SNR-C732 &   2.89e-16 $\pm$   8.7e-17 &    1.08e-16 $\pm$   2.5e-17  &    --  &     8.18e-17 $\pm$   4.0e-17  &     4.86e-17 $\pm$   2.9e-17  &     3.04e-16 $\pm$   5.5e-17  &     1.04e-16 $\pm$   2.9e-17  &     1.14e-16 $\pm$   3.5e-17  &    8.48e-17 $\pm$   3.5e-17 \\
        
        SNR-C733 &   1.03e-15 $\pm$   3.1e-16 &    2.83e-16 $\pm$   6.6e-17  &    --  &     --  &     6.80e-17 $\pm$   4.0e-17  &     8.80e-16 $\pm$   1.6e-16  &     2.89e-16 $\pm$   8.1e-17  &     2.14e-16 $\pm$   6.6e-17  &    1.93e-16 $\pm$   8.0e-17 \\
        
        SNR-C738 &  3.10e-17 $\pm$   9.3e-18 &    1.96e-17 $\pm$   4.6e-18  &    --  &     --  &     8.31e-18 $\pm$   4.9e-18  &     5.21e-17 $\pm$   9.4e-18  &     1.67e-17 $\pm$   4.7e-18  &     2.04e-17 $\pm$   6.3e-18  &    1.60e-17 $\pm$   6.6e-18 \\
        
        SNR-C752 & 9.42e-16 $\pm$   2.8e-16 &    3.22e-16 $\pm$   7.5e-17  &     1.64e-16 $\pm$   1.4e-16  &     5.06e-16 $\pm$   2.5e-16  &     1.43e-16 $\pm$   8.5e-17  &     9.09e-16 $\pm$   1.6e-16  &     3.86e-16 $\pm$   1.1e-16  &     2.27e-16 $\pm$   7.1e-17  &    2.17e-16 $\pm$   9.0e-17 \\
        
        SNR-C756 &  4.60e-16 $\pm$   1.4e-16 &    9.01e-17 $\pm$   2.1e-17  &    --  &     8.76e-17 $\pm$   4.3e-17  &     --  &     2.68e-16 $\pm$   4.8e-17  &     7.44e-17 $\pm$   2.1e-17  &     1.01e-16 $\pm$   3.1e-17  &    9.15e-17 $\pm$   3.8e-17 \\
        
        SNR-C768 &   4.30e-17 $\pm$   1.3e-17 &    2.82e-17 $\pm$   6.6e-18  &     2.01e-17 $\pm$   1.7e-17  &     2.92e-17 $\pm$   1.4e-17  &     1.55e-17 $\pm$   9.1e-18  &     8.03e-17 $\pm$   1.4e-17  &     2.51e-17 $\pm$   7.0e-18  &     2.13e-17 $\pm$   6.6e-18  &    1.93e-17 $\pm$   8.0e-18 \\
        
        SNR-C805 &  -- &    --  &    --  &    --  &     --  &     9.18e-17 $\pm$   1.7e-17  &     3.76e-17 $\pm$   1.0e-17  &     4.09e-17 $\pm$   1.3e-17  &    3.44e-17 $\pm$   1.4e-17 \\
        
        SNR-C812 &  5.40e-16 $\pm$   1.6e-16 &    1.46e-16 $\pm$   3.4e-17  &     8.91e-17 $\pm$   7.6e-17  &     2.84e-16 $\pm$   1.4e-16  &     5.64e-17 $\pm$   3.3e-17  &     4.21e-16 $\pm$   7.6e-17  &     1.88e-16 $\pm$   5.3e-17  &     1.27e-16 $\pm$   4.0e-17  &    1.02e-16 $\pm$   4.2e-17 \\
        
        SNR-C816 &  --  &    1.78e-16 $\pm$   4.1e-17  &     8.51e-17 $\pm$   7.2e-17  &     1.13e-16 $\pm$   5.5e-17  &     --  &     5.53e-16 $\pm$   1.0e-16  &     2.17e-16 $\pm$   6.1e-17  &     1.75e-16 $\pm$   5.4e-17  &    1.06e-16 $\pm$   4.4e-17 \\
        
        SNR-C847 & -- &    1.20e-16 $\pm$   2.8e-17  &     --  &     --  &     4.83e-17 $\pm$   2.8e-17  &     3.51e-16 $\pm$   6.3e-17  &     1.23e-16 $\pm$   3.4e-17  &     9.00e-17 $\pm$   2.8e-17  &    6.43e-17 $\pm$   2.7e-17 \\
        
        SNR-C881 &   2.50e-16 $\pm$   7.5e-17 &    6.08e-17 $\pm$   1.4e-17  &   --  &     8.76e-17 $\pm$   4.3e-17  &     3.94e-17 $\pm$   2.3e-17  &     1.76e-16 $\pm$   3.2e-17  &     1.15e-16 $\pm$   3.2e-17  &     8.87e-17 $\pm$   2.8e-17  &    7.56e-17 $\pm$   3.1e-17 \\
        
        SNR-C929 & -- &    4.70e-17 $\pm$   1.1e-17  &     --  &     9.40e-17 $\pm$   4.6e-17  &     --  &     1.41e-16 $\pm$   2.5e-17  &     5.19e-17 $\pm$   1.5e-17  &     6.51e-17 $\pm$   2.0e-17  &    6.29e-17 $\pm$   2.6e-17 \\
        
        SNR-C944 &  -- &    --  &     --  &    --  &     -- &     4.48e-15 $\pm$   8.1e-16  &     2.56e-15 $\pm$   7.2e-16  &     1.65e-15 $\pm$   5.1e-16  &    1.57e-15 $\pm$   6.5e-16 \\
        
        SNR-C968 &  -- &    9.56e-17 $\pm$   2.2e-17  &    -- &     -- &     4.98e-17 $\pm$   2.9e-17  &     2.73e-16 $\pm$ 4.9e-17  &     9.64e-17 $\pm$   2.7e-17  &     8.54e-17 $\pm$   2.6e-17  &    5.02e-17 $\pm$   2.1e-17 \\
        
        SNR-C974 &  3.51e-16 $\pm$   1.1e-16 &    1.26e-16 $\pm$   2.9e-17  &    --  &     8.42e-17 $\pm$   4.1e-17  &     2.99e-17 $\pm$   1.7e-17  &     3.56e-16 $\pm$   6.4e-17  &     7.87e-17 $\pm$   2.2e-17  &     9.48e-17 $\pm$   2.9e-17  &    6.84e-17 $\pm$   2.8e-17 \\
        
        SNR-C994 & -- &    --  &     --  &    -- &     -- &     3.95e-17 $\pm$   7.1e-18  &     1.53e-17 $\pm$   4.3e-18  &     2.60e-17 $\pm$   8.1e-18  &    2.61e-17 $\pm$   1.1e-17 \\
					
	\end{tabular}
	\end{adjustbox}
\end{table*}
\end{landscape}

\begin{landscape}
\begin{table*}
	\centering
	\contcaption{}
	\label{tab:table_emission_lines_flux_2}
	  \begin{adjustbox}{max width=\textwidth}
	\begin{tabular}{lccccccccc} % four columns, alignment for each
 SNR-C995  & 7.31e-16 $\pm$   2.2e-16 &    1.91e-16 $\pm$   4.5e-17  &     --  &     2.42e-16 $\pm$   1.2e-16  &     1.26e-16 $\pm$   7.4e-17  &     5.70e-16 $\pm$   1.0e-16  &     4.27e-16 $\pm$   1.2e-16  &     2.35e-16 $\pm$   7.3e-17  &    2.08e-16 $\pm$   8.6e-17 \\
 
 SNR-C997  & 1.22e-16 $\pm$   3.7e-17 &    4.65e-17 $\pm$   1.1e-17  &   --  &     -- &     --  &     1.27e-16 $\pm$   2.3e-17  &     5.30e-17 $\pm$   1.5e-17  &     6.98e-17 $\pm$   2.2e-17  &    5.39e-17 $\pm$   2.2e-17 \\
 
 SNR-C1012  & -- &    1.84e-16 $\pm$   4.3e-17  &    --  &     --  &     --  &     5.64e-16 $\pm$   1.0e-16  &     2.18e-16 $\pm$   6.1e-17  &     2.22e-16 $\pm$   6.9e-17  &    1.43e-16 $\pm$   5.9e-17 \\
 
 SNR-C1013  & 5.68e-17 $\pm$   1.7e-17 &    4.35e-17 $\pm$   1.0e-17  &     1.71e-17 $\pm$   1.5e-17  &     5.72e-17 $\pm$   2.8e-17  &     2.23e-17 $\pm$   1.3e-17  &     1.17e-16 $\pm$   2.1e-17  &     5.39e-17 $\pm$   1.5e-17  &     2.85e-17 $\pm$   8.9e-18  &    2.83e-17 $\pm$   1.2e-17 \\
 
 SNR-C1050  &  2.38e-16 $\pm$   7.2e-17 &    3.69e-17 $\pm$   8.6e-18  &     1.94e-17 $\pm$   1.6e-17  &     6.89e-17 $\pm$   3.4e-17  &     3.98e-17 $\pm$   2.3e-17  &     1.04e-16 $\pm$   1.9e-17  &     1.31e-16 $\pm$   3.7e-17  &     6.79e-17 $\pm$   2.1e-17  &    7.26e-17 $\pm$   3.0e-17 \\
 
 SNR-C1088  &  -- &    --  &     --  &     --  &     1.34e-16 $\pm$   7.9e-17  &     9.17e-16 $\pm$   1.7e-16  &     4.17e-16 $\pm$   1.2e-16  &     2.69e-16 $\pm$   8.4e-17  &    2.47e-16 $\pm$   1.0e-16 \\
 
 SNR-C1092 &  5.84e-16 $\pm$   1.8e-16 &    1.84e-16 $\pm$   4.3e-17  &     --  &     -- &     9.04e-17 $\pm$   5.3e-17  &     5.86e-16 $\pm$   1.1e-16  &     2.72e-16 $\pm$   7.6e-17  &     1.68e-16 $\pm$   5.2e-17  &    1.38e-16 $\pm$   5.7e-17 \\
 
 SNR-C1135  & --  &    1.84e-16 $\pm$   4.3e-17  &     --  &     9.93e-17 $\pm$   4.8e-17  &     9.40e-17 $\pm$   5.5e-17  &     5.26e-16 $\pm$   9.5e-17  &     2.17e-16 $\pm$   6.1e-17  &     1.64e-16 $\pm$   5.1e-17  &    8.11e-17 $\pm$   3.4e-17 \\
 
 SNR-C1144  &  5.18e-16 $\pm$   1.6e-16 &    1.26e-16 $\pm$   2.9e-17  &    --  &    --  &     8.26e-17 $\pm$   4.8e-17  &     3.90e-16 $\pm$   7.0e-17  &     1.42e-16 $\pm$   4.0e-17  &     1.24e-16 $\pm$   3.8e-17  &    9.33e-17 $\pm$   3.9e-17 \\
 
 SNR-C1180  &  2.02e-16 $\pm$   6.1e-17 &    1.09e-16 $\pm$   2.5e-17  &  --  &    --  &     4.63e-17 $\pm$   2.7e-17  &     3.24e-16 $\pm$   5.8e-17  &     1.13e-16 $\pm$   3.2e-17  &     8.06e-17 $\pm$   2.5e-17  &    7.12e-17 $\pm$   2.9e-17 \\
 
 SNR-C1213  & --  &   --  &     --  &    -- &    --  &     2.30e-15 $\pm$   4.1e-16  &     8.53e-16 $\pm$   2.4e-16  &     5.13e-16 $\pm$   1.6e-16  &    4.19e-16 $\pm$   1.7e-16 \\
 
 SNR-C1217  & 2.70e-16 $\pm$   8.1e-17 &    1.86e-16 $\pm$   4.3e-17  &    --  &     9.09e-17 $\pm$   4.4e-17  &     6.32e-17 $\pm$   3.7e-17  &     5.46e-16 $\pm$   9.8e-17  &     1.45e-16 $\pm$   4.1e-17  &     1.59e-16 $\pm$   4.9e-17  &    9.67e-17 $\pm$   4.0e-17 \\
 
 SNR-C1220  &  8.90e-16 $\pm$   2.7e-16 &    3.55e-16 $\pm$   8.3e-17  &     --  &     2.38e-16 $\pm$   1.2e-16  &     1.96e-16 $\pm$   1.2e-16  &     1.09e-15 $\pm$   2.0e-16  &     5.58e-16 $\pm$   1.6e-16  &     3.96e-16 $\pm$   1.2e-16  &    2.86e-16 $\pm$   1.2e-16 \\
 
 SNR-C1242  &  1.03e-15 $\pm$   3.1e-16 &    3.62e-16 $\pm$   8.4e-17  &     --  &     --  &     1.32e-16 $\pm$   7.8e-17  &     1.08e-15 $\pm$   1.9e-16  &     4.21e-16 $\pm$   1.2e-16  &     3.42e-16 $\pm$   1.1e-16  &    2.46e-16 $\pm$   1.0e-16 \\
 
 SNR-C1251  &  9.09e-16 $\pm$   2.7e-16 &    4.44e-16 $\pm$   1.0e-16  &    --  &     2.06e-16 $\pm$   1.0e-16  &     2.02e-16 $\pm$   1.2e-16  &     1.36e-15 $\pm$   2.5e-16  &     5.66e-16 $\pm$   1.6e-16  &     3.43e-16 $\pm$   1.1e-16  &    2.93e-16 $\pm$   1.2e-16 \\
 
 SNR-C1285  &  -- &    3.09e-16 $\pm$   7.2e-17  &    --  &     3.58e-16 $\pm$   1.7e-16  &     1.63e-16 $\pm$   9.6e-17  &     9.86e-16 $\pm$   1.8e-16  &     4.16e-16 $\pm$   1.2e-16  &     2.67e-16 $\pm$   8.3e-17  &    2.00e-16 $\pm$   8.3e-17 \\
 
 SNR-C1369  &  1.13e-16 $\pm$   3.4e-17 &    5.52e-17 $\pm$   1.3e-17  &     --  &     2.61e-17 $\pm$   1.3e-17  &     3.48e-17 $\pm$   2.0e-17  &     1.59e-16 $\pm$   2.9e-17  &     6.02e-17 $\pm$   1.7e-17  &     4.73e-17 $\pm$   1.5e-17  &    4.38e-17 $\pm$   1.8e-17 \\
 
 SNR-C1415  & -- &    3.23e-16 $\pm$   7.5e-17  &     --  &     --  &     1.03e-16 $\pm$   6.1e-17  &     1.01e-15 $\pm$   1.8e-16  &     4.72e-16 $\pm$   1.3e-16  &     1.91e-16 $\pm$   5.9e-17  &    2.22e-16 $\pm$   9.2e-17 \\
 
 SNR-C1424  &  6.24e-16 $\pm$   1.9e-16 &    1.44e-16 $\pm$   3.3e-17  &    --  &     2.00e-16 $\pm$   9.8e-17  &     4.60e-17 $\pm$   2.7e-17  &     4.32e-16 $\pm$   7.8e-17  &     9.45e-17 $\pm$   2.6e-17  &     1.63e-16 $\pm$   5.1e-17  &    1.06e-16 $\pm$   4.4e-17 \\
 
 SNR-C1429  &  6.27e-17 $\pm$   1.9e-17 &    4.66e-17 $\pm$   1.1e-17  &     --  &     1.69e-17 $\pm$   8.3e-18  &     1.89e-17 $\pm$   1.1e-17  &     1.33e-16 $\pm$   2.4e-17  &     3.29e-17 $\pm$   9.2e-18  &     2.68e-17 $\pm$   8.3e-18  &    3.20e-17 $\pm$   1.3e-17 \\
 
 SNR-C1432  &  1.17e-15 $\pm$   3.5e-16 &    4.18e-16 $\pm$   9.8e-17  &    --  &     1.22e-16 $\pm$   6.0e-17  &     1.23e-16 $\pm$   7.3e-17  &     1.24e-15 $\pm$   2.2e-16  &     3.78e-16 $\pm$   1.1e-16  &     3.13e-16 $\pm$   9.7e-17  &    2.07e-16 $\pm$   8.6e-17 \\
 
 SNR-C1462  & 5.96e-16 $\pm$   1.8e-16 &    3.19e-16 $\pm$   7.4e-17  &     --  &     1.77e-16 $\pm$   8.6e-17  &     1.57e-16 $\pm$   9.3e-17  &     9.53e-16 $\pm$   1.7e-16  &     5.72e-16 $\pm$   1.6e-16  &     2.99e-16 $\pm$   9.3e-17  &    2.19e-16 $\pm$   9.1e-17 \\
 
 SNR-C1552  & 1.60e-16 $\pm$   4.8e-17 &    4.58e-17 $\pm$   1.1e-17  &     4.29e-17 $\pm$   3.7e-17  &     1.33e-16 $\pm$   6.5e-17  &     1.56e-17 $\pm$   9.2e-18  &     1.31e-16 $\pm$   2.4e-17  &     5.15e-17 $\pm$   1.4e-17  &     3.00e-17 $\pm$   9.3e-18  &    2.48e-17 $\pm$   1.0e-17 \\
 
 SNR-C1574  &  2.36e-16 $\pm$   7.1e-17 &    1.22e-16 $\pm$   2.8e-17  &     --  &     --  &     4.83e-17 $\pm$   2.8e-17  &     3.49e-16 $\pm$   6.3e-17  &     1.03e-16 $\pm$   2.9e-17  &     9.73e-17 $\pm$   3.0e-17  &    5.61e-17 $\pm$   2.3e-17 \\
 
 SNR-C1584  & 2.93e-16 $\pm$   8.8e-17 &    1.22e-16 $\pm$   2.8e-17  &    --  &     4.56e-17 $\pm$   2.2e-17  &     3.03e-17 $\pm$   1.8e-17  &     3.66e-16 $\pm$   6.6e-17  &     1.02e-16 $\pm$   2.9e-17  &     8.76e-17 $\pm$   2.7e-17  &    6.94e-17 $\pm$   2.9e-17 \\
 
 SNR-C1588  &  3.04e-16 $\pm$   9.1e-17 &    1.68e-16 $\pm$   3.9e-17  &     --  &    --  &     3.68e-17 $\pm$   2.2e-17  &     4.71e-16 $\pm$   8.5e-17  &     1.33e-16 $\pm$   3.7e-17  &     1.13e-16 $\pm$   3.5e-17  &    9.33e-17 $\pm$   3.9e-17 \\
 
 SNR-C1606  &  4.37e-17 $\pm$   1.3e-17 &    3.28e-17 $\pm$   7.6e-18  &    --  &     --  &     1.02e-17 $\pm$   6.0e-18  &     8.96e-17 $\pm$   1.6e-17  &     2.20e-17 $\pm$   6.2e-18  &     2.51e-17 $\pm$   7.8e-18  &    1.66e-17 $\pm$   6.9e-18 \\
 
 SNR-C1634  &  -- &    8.16e-17 $\pm$   1.9e-17  &    --  &     8.71e-17 $\pm$   4.3e-17  &     6.79e-17 $\pm$   4.0e-17  &     2.41e-16 $\pm$   4.4e-17  &     1.84e-16 $\pm$   5.2e-17  &     9.37e-17 $\pm$   2.9e-17  &    7.85e-17 $\pm$   3.3e-17 \\
 
 SNR-C1637  &  1.08e-16 $\pm$   3.3e-17 &    7.02e-17 $\pm$   1.6e-17  &    -- &     --  &     2.36e-17 $\pm$   1.4e-17  &     1.97e-16 $\pm$   3.5e-17  &     6.52e-17 $\pm$   1.8e-17  &     4.88e-17 $\pm$   1.5e-17  &    3.30e-17 $\pm$   1.4e-17 \\
 
 SNR-C1655  & 4.58e-16 $\pm$   1.4e-16 &    2.23e-16 $\pm$   5.2e-17  &     7.66e-17 $\pm$   6.5e-17  &     2.34e-16 $\pm$   1.1e-16  &     7.23e-17 $\pm$   4.3e-17  &     6.56e-16 $\pm$   1.2e-16  &     2.24e-16 $\pm$   6.3e-17  &     1.84e-16 $\pm$   5.7e-17  &    1.28e-16 $\pm$   5.3e-17 \\
 
 SNR-C1705  &  1.42e-15 $\pm$   4.3e-16 &    3.50e-16 $\pm$   8.2e-17  &     1.44e-16 $\pm$   1.2e-16  &     4.67e-16 $\pm$   2.3e-16  &     1.13e-16 $\pm$   6.7e-17  &     1.10e-15 $\pm$   2.0e-16  &     3.62e-16 $\pm$   1.0e-16  &     2.68e-16 $\pm$   8.3e-17  &    1.98e-16 $\pm$   8.2e-17 \\
 
 SNR-C1712  &  1.68e-16 $\pm$   5.1e-17 &    1.16e-16 $\pm$   2.7e-17  &     -- &     --  &     2.97e-17 $\pm$   1.8e-17  &     3.34e-16 $\pm$   6.0e-17  &     1.02e-16 $\pm$   2.8e-17  &     8.03e-17 $\pm$   2.5e-17  &    6.42e-17 $\pm$   2.7e-17 \\
 
  SNR-C1727  &  8.58e-16 $\pm$   2.6e-16 &    3.10e-16 $\pm$   7.2e-17  &   --  &     1.04e-16 $\pm$   5.1e-17  &     8.75e-17 $\pm$   5.2e-17  &     9.34e-16 $\pm$   1.7e-16  &     2.76e-16 $\pm$   7.7e-17  &     2.19e-16 $\pm$   6.8e-17  &    1.51e-16 $\pm$   6.2e-17 \\

 SNR-C1731  &  5.19e-16 $\pm$   1.6e-16 &    7.54e-17 $\pm$   1.8e-17  &     --  &     1.07e-16 $\pm$   5.2e-17  &     6.99e-17 $\pm$   4.1e-17  &     2.21e-16 $\pm$   4.0e-17  &     1.91e-16 $\pm$   5.4e-17  &     1.49e-16 $\pm$   4.6e-17  &    1.24e-16 $\pm$   5.1e-17 \\
 
 SNR-C1769   & 4.60e-16 $\pm$   1.4e-16 &    2.12e-16 $\pm$   4.9e-17  &    --  &     6.26e-17 $\pm$   3.1e-17  &     6.59e-17 $\pm$   3.9e-17  &     6.32e-16 $\pm$   1.1e-16  &     2.00e-16 $\pm$   5.6e-17  &     1.61e-16 $\pm$   5.0e-17  &    1.07e-16 $\pm$   4.4e-17 \\
 
 SNR-C1776  & 1.50e-15 $\pm$   4.5e-16 &    5.03e-16 $\pm$   1.2e-16  &   --  &     4.03e-16 $\pm$   2.0e-16  &     2.26e-16 $\pm$   1.3e-16  &     1.57e-15 $\pm$   2.8e-16  &     5.45e-16 $\pm$   1.5e-16  &     3.70e-16 $\pm$   1.1e-16  &    3.10e-16 $\pm$   1.3e-16 \\
 
 SNR-C1799  &  3.45e-16 $\pm$   1.0e-16 &   --  &    --  &     1.12e-16 $\pm$   5.4e-17  &     2.78e-17 $\pm$   1.6e-17  &     1.42e-16 $\pm$   2.6e-17  &     6.69e-17 $\pm$   1.9e-17  &     6.14e-17 $\pm$   1.9e-17  &    4.76e-17 $\pm$   2.0e-17 \\
 
 SNR-C1801  & 3.09e-16 $\pm$   9.3e-17 &    1.34e-16 $\pm$   3.1e-17  &   --  &     8.74e-17 $\pm$   4.3e-17  &     6.58e-17 $\pm$   3.9e-17  &     4.06e-16 $\pm$   7.3e-17  &     1.28e-16 $\pm$   3.6e-17  &     1.06e-16 $\pm$   3.3e-17  &    7.79e-17 $\pm$   3.2e-17 \\
 
 SNR-C1805  &  3.44e-16 $\pm$   1.0e-16 &    1.77e-16 $\pm$   4.1e-17  &    -- &     9.77e-17 $\pm$   4.8e-17  &     5.37e-17 $\pm$   3.2e-17  &     5.00e-16 $\pm$   9.0e-17  &     1.94e-16 $\pm$   5.4e-17  &     1.55e-16 $\pm$   4.8e-17  &    9.63e-17 $\pm$   4.0e-17 \\
 
 SNR-C1809  & -- &    --  &    -- &    -- &     9.06e-17 $\pm$   5.3e-17  &     3.61e-16 $\pm$   6.5e-17  &     1.41e-16 $\pm$   3.9e-17  &     1.11e-16 $\pm$   3.5e-17  &    1.04e-16 $\pm$   4.3e-17 \\
 
 SNR-C1829  & -- &    1.51e-16 $\pm$   3.5e-17  &    --  &     --  &   --  &     4.64e-16 $\pm$   8.4e-17  &     2.03e-16 $\pm$   5.7e-17  &     1.61e-16 $\pm$   5.0e-17  &    1.23e-16 $\pm$   5.1e-17 \\
 
 SNR-C1844  &  --  &    1.17e-16 $\pm$   2.7e-17  &     --  &    --  &   --  &     3.70e-16 $\pm$   6.7e-17  &     9.21e-17 $\pm$   2.6e-17  &     1.04e-16 $\pm$   3.2e-17  &    8.52e-17 $\pm$   3.5e-17 \\
 
 SNR-C1845  &  5.24e-16 $\pm$   1.6e-16 &    2.83e-16 $\pm$   6.6e-17  &     --  &     5.36e-17 $\pm$   2.6e-17  &     1.10e-16 $\pm$   6.5e-17  &     7.74e-16 $\pm$   1.4e-16  &     2.76e-16 $\pm$   7.7e-17  &     2.15e-16 $\pm$   6.7e-17  &    1.59e-16 $\pm$   6.6e-17 \\
 
 SNR-C1861  &  -- &   --  &   --  &     --  &     --  &     5.16e-17 $\pm$   9.3e-18  &     --  &     2.68e-17 $\pm$   8.3e-18  &    2.06e-17 $\pm$   8.6e-18 \\
 
 SNR-C1873  &  1.14e-16 $\pm$   3.4e-17 &    2.41e-16 $\pm$   5.6e-17  &    --  &     8.25e-17 $\pm$   4.0e-17  &     6.04e-17 $\pm$   3.6e-17  &     7.08e-16 $\pm$   1.3e-16  &     2.02e-16 $\pm$   5.7e-17  &     1.80e-16 $\pm$   5.6e-17  &    1.13e-16 $\pm$   4.7e-17  \\
 
 SNR-C1887  & -- &    7.68e-17 $\pm$   1.8e-17  &    --  &    --  &     2.41e-17 $\pm$   1.4e-17  &     2.33e-16 $\pm$   4.2e-17  &     7.58e-17 $\pm$   2.1e-17  &     5.26e-17 $\pm$   1.6e-17  &    4.02e-17 $\pm$   1.7e-17 \\
 
 SNR-C1917  & 1.33e-15 $\pm$   4.0e-16 &    4.32e-16 $\pm$   1.0e-16  &   -- &     1.45e-16 $\pm$   7.1e-17  &     1.98e-16 $\pm$   1.2e-16  &     1.27e-15 $\pm$   2.3e-16  &     4.35e-16 $\pm$   1.2e-16  &     4.52e-16 $\pm$   1.4e-16  &    3.03e-16 $\pm$   1.3e-16 \\
 
 SNR-C1922  & 2.66e-16 $\pm$   8.0e-17 &    1.25e-16 $\pm$   2.9e-17  &     --  &     --  &     --  &     3.67e-16 $\pm$   6.6e-17  &     1.29e-16 $\pm$   3.6e-17  &     1.04e-16 $\pm$   3.2e-17  &    7.62e-17 $\pm$   3.2e-17 \\
 
 SNR-C1929  & 1.99e-16 $\pm$   6.0e-17 &    5.06e-17 $\pm$   1.2e-17  &    --  &     --  &     2.84e-17 $\pm$   1.7e-17  &     1.54e-16 $\pm$   2.8e-17  &     5.56e-17 $\pm$   1.6e-17  &     3.71e-17 $\pm$   1.1e-17  &    3.62e-17 $\pm$   1.5e-17 \\
 
 SNR-C1931  & 2.09e-15 $\pm$   6.3e-16 &    2.51e-16 $\pm$   5.9e-17  &     1.65e-16 $\pm$   1.4e-16  &     7.90e-16 $\pm$   3.9e-16  &     1.80e-16 $\pm$   1.1e-16  &     7.91e-16 $\pm$   1.4e-16  &     4.16e-16 $\pm$   1.2e-16  &     3.82e-16 $\pm$   1.2e-16  &    2.35e-16 $\pm$   9.7e-17 \\
 
 SNR-C1940  & -- &    1.70e-16 $\pm$   4.0e-17  &    --  &     --  &    --  &     5.54e-16 $\pm$   1.0e-16  &     1.30e-16 $\pm$   3.6e-17  &     1.18e-16 $\pm$   3.7e-17  &    1.15e-16 $\pm$   4.8e-17 \\
 
 SNR-C1950  &  3.62e-17 $\pm$   1.1e-17 &    1.80e-17 $\pm$   4.2e-18  &    --  &     1.82e-17 $\pm$   8.9e-18  &     1.00e-17 $\pm$   5.9e-18  &     5.07e-17 $\pm$   9.1e-18  &     1.32e-17 $\pm$   3.7e-18  &     1.06e-17 $\pm$   3.3e-18  &    9.61e-18 $\pm$   4.0e-18 \\
 
 SNR-C1957  & 5.82e-16 $\pm$   1.7e-16 &    1.55e-16 $\pm$   3.6e-17  &     7.48e-17 $\pm$   6.4e-17  &     1.80e-16 $\pm$   8.8e-17  &     4.08e-17 $\pm$   2.4e-17  &     4.58e-16 $\pm$   8.3e-17  &     1.48e-16 $\pm$   4.2e-17  &     1.06e-16 $\pm$   3.3e-17  &    9.26e-17 $\pm$   3.8e-17 \\
 
 SNR-C1971  &  1.93e-15 $\pm$   5.8e-16 &    3.02e-16 $\pm$   7.0e-17  &     3.79e-16 $\pm$   3.2e-16  &     4.10e-16 $\pm$   2.0e-16  &     --  &     1.04e-15 $\pm$   1.9e-16  &     3.28e-16 $\pm$   9.2e-17  &     2.54e-16 $\pm$   7.9e-17  &    2.66e-16 $\pm$   1.1e-16 \\
 
 SNR-C2021  & 7.13e-17 $\pm$   2.1e-17 &    5.38e-17 $\pm$   1.2e-17  &    --  &     2.48e-17 $\pm$   1.2e-17  &     1.35e-17 $\pm$   8.0e-18  &     1.44e-16 $\pm$   2.6e-17  &     3.53e-17 $\pm$   9.9e-18  &     3.88e-17 $\pm$   1.2e-17  &    2.70e-17 $\pm$   1.1e-17 \\
 
 SNR-C2048  & 1.43e-15 $\pm$   4.3e-16 &   --  &     --     &     2.55e-16 $\pm$   1.2e-16  &     1.10e-16 $\pm$   6.5e-17  &     5.23e-16 $\pm$   9.4e-17  &     1.65e-16 $\pm$   4.6e-17  &     2.01e-16 $\pm$   6.2e-17  &    1.58e-16 $\pm$   6.5e-17 \\
 
 SNR-C2050  &  2.86e-15 $\pm$   8.6e-16 &    6.42e-16 $\pm$   1.5e-16  &     3.93e-16 $\pm$   3.3e-16  &     1.01e-15 $\pm$   4.9e-16  &     2.35e-16 $\pm$   1.4e-16  &     1.99e-15 $\pm$   3.6e-16  &     5.18e-16 $\pm$   1.4e-16  &     5.00e-16 $\pm$   1.6e-16  &    3.45e-16 $\pm$   1.4e-16 \\
 
 SNR-C2051  &   2.08e-16 $\pm$   6.3e-17 &    4.81e-17 $\pm$   1.1e-17  &   --  &     4.13e-17 $\pm$   2.0e-17  &    --  &     1.42e-16 $\pm$   2.6e-17  &    --  &     5.10e-17 $\pm$   1.6e-17  &    5.06e-17 $\pm$   2.1e-17 \\
 
 SNR-C2053  &  1.10e-15 $\pm$   3.3e-16 &    2.41e-16 $\pm$   5.6e-17  &     --  &     2.40e-16 $\pm$   1.2e-16  &    --  &     7.33e-16 $\pm$   1.3e-16  &     2.38e-16 $\pm$   6.7e-17  &     2.43e-16 $\pm$   7.5e-17  &    1.84e-16 $\pm$   7.6e-17 \\
 
 SNR-C2058  & 2.11e-16 $\pm$   6.4e-17 &    5.53e-17 $\pm$   1.3e-17  &    --  &     3.13e-17 $\pm$   1.5e-17  &    --  &     1.59e-16 $\pm$   2.9e-17  &     3.68e-17 $\pm$   1.0e-17  &     4.69e-17 $\pm$   1.5e-17  &    3.20e-17 $\pm$   1.3e-17 \\
 
 SNR-C2085  & 9.13e-17 $\pm$   2.7e-17 &    3.18e-17 $\pm$   7.4e-18  &     1.19e-17 $\pm$   1.0e-17  &     2.05e-17 $\pm$   1.0e-17  &    --  &     8.94e-17 $\pm$   1.6e-17  &     1.28e-17 $\pm$   3.6e-18  &     3.24e-17 $\pm$   1.0e-17  &    2.00e-17 $\pm$   8.3e-18 \\
 
 SNR-C2086  & 4.67e-17 $\pm$   1.4e-17 &    1.85e-17 $\pm$   4.3e-18  &     --  &     3.61e-17 $\pm$   1.8e-17  &    --  &     5.02e-17 $\pm$   9.1e-18  &     1.13e-17 $\pm$   3.2e-18  &     1.06e-17 $\pm$   3.3e-18  &    9.68e-18 $\pm$   4.0e-18 \\
 
 SNR-C2090  & 8.47e-17 $\pm$   2.5e-17 &    4.15e-17 $\pm$   9.7e-18  &     -- &    --  &    --  &     1.20e-16 $\pm$   2.2e-17  &     1.57e-17 $\pm$   4.4e-18  &     3.60e-17 $\pm$   1.1e-17  &    2.76e-17 $\pm$   1.1e-17 \\
 
 SNR-C2094   & -- &    2.92e-16 $\pm$   6.8e-17  &     1.95e-16 $\pm$   1.7e-16  &     6.18e-16 $\pm$   3.0e-16  &     1.26e-16 $\pm$   7.4e-17  &     9.74e-16 $\pm$   1.8e-16  &     2.17e-16 $\pm$   6.1e-17  &     2.00e-16 $\pm$   6.2e-17  &    1.86e-16 $\pm$   7.7e-17 \\
 
 SNR-C2161   & -- &    2.64e-18 $\pm$   6.2e-19  &     --  &     --  &    -- &     6.34e-18 $\pm$   1.1e-18  &     1.62e-18 $\pm$   4.5e-19  &     4.03e-18 $\pm$   1.2e-18  &    2.41e-18 $\pm$   1.0e-18 \\
 
 SNR-C2174   &  3.64e-16 $\pm$   1.1e-16 &    9.16e-17 $\pm$   2.1e-17  &     --  &     1.97e-16 $\pm$   9.6e-17  &     2.99e-17 $\pm$   1.8e-17  &     2.62e-16 $\pm$   4.7e-17  &     3.75e-17 $\pm$   1.0e-17  &     6.66e-17 $\pm$   2.1e-17  &    4.48e-17 $\pm$   1.9e-17 \\
 
 SNR-C2179   & 5.48e-16 $\pm$   1.6e-16 &    2.16e-16 $\pm$   5.0e-17  &     8.51e-17 $\pm$   7.2e-17  &     2.04e-16 $\pm$   1.0e-16  &     --  &     6.07e-16 $\pm$   1.1e-16  &     9.61e-17 $\pm$   2.7e-17  &     1.46e-16 $\pm$   4.5e-17  &    1.14e-16 $\pm$   4.7e-17 \\
	\hline				
	\end{tabular}
	\end{adjustbox}
\end{table*}
\end{landscape}


%%%%%%%%%%%%%%%%%%%%%%%%%%%%%%%%%%%%%%%%%%%%%%%%%%

% Don't change these lines
\bsp	% typesetting comment
\label{lastpage}
\end{document}